\documentclass[iicol]{sn-jnl}

\usepackage{graphicx}%
\usepackage{multirow}%
\usepackage{amsmath,amssymb,amsfonts}%
\usepackage{amsthm}%
\usepackage{mathrsfs}%
\usepackage[title]{appendix}%
\usepackage{xcolor, soul}%
\usepackage{textcomp}%
\usepackage{manyfoot}%
\usepackage{booktabs}%
\usepackage{algorithm}%
\usepackage{algorithmicx}%
\usepackage{algpseudocode}%
\usepackage{listings}%

\usepackage{subcaption}
\usepackage{multirow}
\sethlcolor{lightgray} 
\usepackage{bm}

\newtheorem{example}{Example}

\newcommand*\diff{\mathop{}\!\mathrm{d}}

\begin{document}

\title[Article Title]{Sampling from Conditional Distributions of Simplified Vines}

\author*[1]{\fnm{Ariane} \sur{Hanebeck}}\email{ariane.hanebeck@tum.de}

\author[2]{\fnm{Özge} \sur{\c{S}ahin}}\email{O.Sahin@tudelft.nl}

\author[1]{\fnm{Petra} \sur{Havlíčková}}

\author[3]{\fnm{Claudia} \sur{Czado}}\email{czado@tum.de}

\affil*[1]{School of Computation, Information and Technology, Technical University of Munich, Munich, Germany}

\affil[2]{Delft Institute of Applied Mathematics, Delft University of Technology, Delft, the Netherlands}

\affil[3]{School of Computation, Information and Technology \& Munich Data Science Institute, Technical University of Munich, Munich, Germany}

\abstract{Simplified vine copulas are flexible tools over standard multivariate distributions for modeling and understanding different dependence properties in high-dimensional data. Their conditional distributions are of utmost importance, from statistical learning to graphical models. However, the conditional densities of vine copulas and, thus, vine distributions cannot be obtained in closed form without integration for all possible sets of conditioning variables.
We propose a Markov Chain Monte Carlo based approach of using Hamiltonian Monte Carlo to sample from any conditional distribution of arbitrarily specified simplified vine copulas and thus vine distributions. We show its accuracy through simulation studies and analyze data of multiple maize traits such as flowering times, plant height, and vigor. Use cases from predicting traits to estimating conditional Kendall's tau are presented.}


\maketitle 

\section{Introduction}\label{sec:Int}

Conditional distributions are often of interest for different tasks, such as prediction in regression and classification problems, as well as conditional independence tests required for structure learning. Thus, among others, they have applications in finance, genomics, and physics \cite{hastie2009elements}.

 Regular (R-) vine distributions are a flexible class of multivariate distributions beyond normality and allow for asymmetrical (tail) dependence structures in data \cite{Aas2009,joeDependenceModelingCopulas2014,
czadoAnalyzingDependentData2019} using the copula approach of \cite{sklarFonctionsRepartitionDimensions1959}. They are 
constructed using only bivariate copulas as building 
blocks, which are glued together by conditioning arguments. A so-called vine tree structure of linked trees introduced by \cite{bedfordProbabilityDensityDecomposition2001} identifies the conditioning variables, and edges in the trees are connected with bivariate (pair) copulas. Boundary cases of the regular vines are D-vines (each tree is a path) and C-vines (each tree is a star). To insure tractable statistical inference in higher dimensional problems with $d>5$ variables, a simplifying assumption (see \cite{stoeber2013simplified}, \cite{mroz2021simplifying},  \cite{kurz2022testing}, \cite{nagler2024simplified}, and \cite{czado2022vine} for definition and further discussion) has to be made. In particular, \cite{mroz2021simplifying} show that this subclass is not dense in the class of all distributions and construct examples illustrating this. However they fail to give real data applications of their examples. Further, \cite{kurz2022testing} show in their real data applications that in 3 out of their 4 investigated examples the simplifying assumption cannot be rejected using their testing approach. Therefore, we follow the quote: \textit{Essentially, all models are wrong} by George Box and focus on simplified vine copulas, which remain computationally tractable in high dimensions. However, we recommend that in applications the simplifying assumption to be checked using the tools in \cite{kurz2022testing} and \cite{derumignyTestsSimplifyingAssumption2017}. Finally, fitting non-simplified vines using the approaches of \cite{acar2012beyond} and \cite{vatter2018generalized} are too costly computationally and require larger sample sizes. The restriction of using a simplified vine is further mitigated by using a large catalog of parametric pair copula families as well as allowing for nonparametric ones as suggested by \cite{naglerNonparametricEstimationSimplified2017}. Simplified R-, D-, and C-vines are widely used for high-dimensional multivariate applications. Often the \texttt{R} package \texttt{rvinecopulib} of \cite{nagler2022rvinecopulib} is utilized in many domains \cite{czado2022vine}.

Here, we consider the use of univariate and multivariate conditional distributions associated with a joint simplified vine distribution of the response(s) and the covariates to construct more flexible nonlinear non-Gaussian regression models with non-additive error structures. Such an approach fits into the class of distributional regression models as discussed, for example, in \cite{klein2024distributional}. Additionally, we restrict to the case of continuous response and covariate variables.

Obtaining arbitrary conditional distributions associated with simplified regular vine distributions and sampling from them are big challenges since only certain conditional densities of a vine distribution are available without integration. In the case of a univariate regression problem, we need to insure that the response is contained in a leaf node of each tree in the regular vine tree structure.  This can be achieved in two ways. In the first approach propagated by \cite{chang2019prediction}, a simplified regular vine is fitted to the covariates, and the response variable is added in such a way that the response variable occurs only in leaf nodes at each tree level. This fits the dependence structure among the covariates optimally but does not optimize with regard to the conditional distribution of the response given the covariates, which is the focus of interest. Therefore, \cite{krausDVineCopulaBased2017} restricted themselves to D-vine models for the joint distribution of the response and covariates, where they assume that the nodes involving the response are a leaf node in each path tree. This allowed them to construct a forward selection algorithm of the covariates. \cite{tepegjozova2022nonparametric} extended this approach to C-vines. Here the response variable has to be never involved in the central nodes of the C-vine to allow for a conditional density of the response given the covariates to be available without integration. They also extended the greedy forward selection algorithm to a two-step ahead-looking approach for the selection of the next covariate and allowed for nonparametric pair copulas. Finally, \cite{sahin2022high} proposed sparse variable selection algorithms in the context of the D-vine regression model of \cite{krausDVineCopulaBased2017} for high-dimensional regression problems. To allow for more general R-vine based regressions \cite{herrmann2018regular} developed a heuristic algorithm based on using partial correlations of the transformed data with standard normal margins. A similar heuristic approach based on partial correlations was followed in
\cite{zhu2021simplified}.

Lately, \cite{tepegjozovaBivariateVineCopula2023} introduced and applied bivariate conditional distribution estimation using vines, allowing for a symmetrical treatment of the response variables.  Another vine based approach for bivariate conditional distributions was followed by \cite{zhu2021simplified} with an asymmetrical treatment of the response variables. However, all approaches mentioned above restrict the structure of the vine to have a simple expression for the corresponding conditional densities without any integration.

While integration can be used to estimate the conditional densities for the multivariate response case of the cardinality one or two, it is not efficient for the higher cardinality case. Therefore, we propose a sampling approach based on Markov Chain Monte Carlo (MCMC) to sample from conditional distributions of simplified vines without restricting the vine structure, using the programming language \texttt{Stan} \cite{carpenterStanProbabilisticProgramming2017}. It uses the No-U-Turn Sampler \cite{hoffmanNoUTurnSamplerAdaptively2014}, an extension of the Hamiltonian Monte Carlo Algorithm (HMC). HMC is inspired by the Hamiltonian dynamics known in physics. It combines the potential energy related to the current position, e.g., parameter estimate, and the kinetic energy related to the momentum, e.g., exploration direction. Thus, HMC simulates trajectories in the joint space of position and momentum to move from one parameter to another. As a result, compared to the Metropolis Hastings algorithm, HMC provides faster convergence and a more accurate representation of complex probability distributions. 

Even though MCMC-based methods are computationally costly, they provide information on the full conditional distribution instead of being restricted to point estimates. Kernel density estimation using the MCMC samples can be utilized to estimate the full conditional density. 
Additionally, prediction intervals of the response for fixed
covariate values are constructed using the empirical quantiles of the MCMC samples. Furthermore, point estimates like modes, means, and medians of the conditional distribution are also available.
So far, \cite{brechmannConditionalCopulaSimulation2013} considered sampling from C-vines conditional on a single variable only.
While \cite{kahmAssessingSystemRelevance2014} has developed a tailor-made finely tuned Metropolis-Hastings MCMC algorithm to obtain samples from multivariate conditional distributions of C-vines, we utilize the faster HMC approach together with a No-U-Turn sampler to
achieve an easy and user-friendly implementation to obtain MCMC samples from conditional distributions of arbitrary vines. 
Given the flexibility of R-vine copulas and the importance and usage of conditional distributions, this solves a problem of high interest in many areas. 

Our simulation studies study the performance of our approach to univariate and bivariate conditional distributions of vine copulas in $d$ dimensions. The true densities (available or obtained by integration in small dimensions) and estimated densities using MCMC samples compare well.

Finally, we provide different use cases of our sampling method by analyzing multiple maize traits jointly. They include the estimation of univariate, bivariate, and trivariate responses conditioned on some covariates. In particular, it allows the prediction of maize traits at later growth stages, given the traits at the early stages, thanks to the point estimates obtained by our MCMC samples together with prediction intervals. Furthermore, we obtain estimates of the associated conditional Kendall's tau in the case of a bivariate trait as a response. We also propose a simple heuristic to select an appropriate vine structure for a univariate conditional distribution estimation task. We compare our estimates with the existing analytical methods and show the advantages of sampling by our proposal.

In summary, the paper makes the following contributions:
(1) it provides a general sampling-based framework to investigate conditional distributions from  arbitrarily specified simplified R-vines,
(2) the \textcolor{black}{code for this} framework is available on GitHub\footnote{github.com/ArianeHanebeck/Sampling\_Conditional\_Vines} and implemented in \texttt{Stan} for arbitrary setups allowing 
for conditional distributions of several variables given multivariate conditioning variables,
(3) this method can be used to estimate bivariate conditional Kendall's $\tau$ given more than one conditioning value without having to aggregate the effect of the covariates
into a univariate function as proposed in \cite{vatter2015generalized} and \cite{schellhase2018estimating}.

The paper is organized as follows: Section \ref{sec:Background} gives the necessary theoretical background, while Section \ref{Sec:Proposal} introduces our sampling proposal. Simulation studies are provided in Section \ref{sec:Sim}, and real data analyses are given in Section \ref{Sec:Data}. We provide our discussions, future research directions, and conclusions in Section \ref{Sec:Conc}.

\section{Background}\label{sec:Background}

\subsection{Introduction to Copulas and Vine Copulas}
\label{Sec:IntCopula}
We consider a $d$-dimensional random vector $\mathbf{X}=(X_1,\ldots,X_d)$ with joint distribution function $F$ and density $f$. Further, let $F_1,\ldots,F_d$ be the marginal distribution functions with densities $f_1,\ldots,f_d$. The copula approach allows to separate the dependence structure from the univariate margins. For this, $(X_1,\ldots,X_d)$ is transformed to $(U_1=F_1(X_1),\ldots,U_d=F_d(X_d)) \in [0,1]^d$ using probability integral transforms. The random variables $U_j, j=1,\ldots,d$, are then uniformly distributed on $[0,1]$. Since $F_1,\ldots, F_d$ can be arbitrary, different marginal distributions are allowed, contrary to the case of a multivariate Gaussian distribution.

We call $X_1,\ldots, X_d$ the variables on the x-scale (original scale) and $U_1,\ldots, U_d$ the variables on the u-scale (copula scale). The variables $(Z_1,\ldots,Z_d)=(\Phi^{-1}(U_1),\ldots,\Phi^{-1}(U_d))$ are on the so-called z-scale. Here, $\Phi^{-1}$ is the inverse of the standard normal distribution function. On this scale, signature shapes of contours arising from different bivariate copula families can be compared to empirical contours \cite[Section 3.8]{czadoAnalyzingDependentData2019}. For the bivariate Gaussian copula, elliptical contours on the z-scale are observed.

After the margins are standardized, resulting in data on the copula scale, the goal is to model the dependence structure using a copula. A $d$-dimensional copula $C$ is a multivariate distribution function on the $d$-dimensional hypercube $[0,1]^d$ with uniformly distributed margins. In the absolutely continuous case, its density $c$ is obtained by partial differentiation. In this case, 
Sklar's theorem \cite{sklarFonctionsRepartitionDimensions1959} allows the joint distribution function of $\mathbf{X}$ to be uniquely expressed as
\begin{equation}
    F(x_1,\ldots,X_d)=C(F_1(x_1),\ldots,F_d(x_d))
\end{equation}
with associated density
\begin{equation}
	\begin{aligned}
		f(x_1,\ldots,x_d)={}& c(F_1(x_1),\ldots,F_d(x_d))\cdot\\
        & \cdot f_1(x_1) \cdot \ldots \cdot f_d(x_d).
	\end{aligned}
\end{equation}

Three main classes of copulas are elliptical, Archimedean, and extreme-value copulas, differing in their construction. Different bivariate copula families allow for very different dependence structures (see \cite{czadoAnalyzingDependentData2019} for examples).

To measure the dependence between two random variables $X_1$ and $X_2$, Kendall's $\tau$ is most often used. \textcolor{black}{It is defined as \cite{kendall1938new, czado2022vine}
\begin{align*}
    \tau(X_1,X_2) = &P((X_{11}-X_{21})(X_{12}-X_{22})>0)\\
    -&P((X_{11}-X_{21})(X_{12}-X_{22})<0),
\end{align*}
where $(X_{11},X_{12})$ and $(X_{21},X_{22})$ are i.i.d copies of $(X_1,X_2)$.}
Kendall's $\tau$ can be expressed solely in terms of the associated copula and, therefore, does not depend on the marginal distribution. For some bivariate copulas, the range of Kendall's tau is restricted to $[0,1]$ or $[-1,0]$. To extend the range of dependence, counterclockwise rotations of the copula density can be used \cite[Section 3.6]{czadoAnalyzingDependentData2019}.

While the set of available parametric bivariate copula families can accommodate different dependence properties such as no, symmetric, and asymmetric tail dependence, this is not true for $d>2$. In particular the multivariate Gaussian copula has no tail dependence, while the multivariate t-copula has only symmetric tail dependence. Further, bivariate dependence patterns derived from a multivariate t copula are governed by the pairwise association parameter and the common degree parameter alone.
\textcolor{black}{Grouped-$t$ copulas \cite{daul_grouped_2003} generalize the standard multivariate $t$ model by partitioning the $d$ margins into $m$ disjoint groups, each sharing its own degrees-of-freedom parameter $\nu_g$.
For each pair of components within group $g$, the upper and lower tail dependence coefficients coincide and depend on the common $\nu_g$. This shows that only symmetric tail dependence can be modeled.}
In the case of multivariate Archimedean copulas, all pairwise dependence patterns are the same and determined by a single dependence parameter.
\textcolor{black}{Hierarchical or nested Archimedean copulas model multivariate dependence by arranging variables into a tree of nested clusters, so that any two variables in the same cluster share the same dependence strength, and more distant clusters have weaker dependence.  This parsimonious model with closed‐form density and sampling algorithms is suited to applications where variables naturally group. However, one cannot accommodate variation in pairwise dependence or asymmetric tail behavior across all pairs of variables. Further, to result in a valid $d$-dimensional copula, the generators must satisfy nesting conditions \cite{mcneil_sampling_2008, joe_parametric_1993}. Such conditions are not needed for a vine construction}. 

To overcome these restrictive patterns, vine copulas allow the construction of multivariate distributions in $d \geq 2$ out of bivariate copulas as building blocks. To couple the bivariate copulas, a conditioning argument is utilized, allowing to decompose the joint density of a random vector $(X_1,\ldots, X_d)$ \cite{bedfordProbabilityDensityDecomposition2001, joeDependenceModelingCopulas2014}.
From now on, for a vector \textcolor{black}{$\boldsymbol x \in R^d$ and set $D = \{i_1,\ldots,i_{|D|}\} \subset \{1,\ldots,d\}$}, $\boldsymbol x_D$ denotes the subvector $(x_{i_1},\ldots,x_{i_{|D|}})$.
For the decomposition, we require the following notations: For $D \subset \{1,\ldots,d\}\backslash \{i,j\}$, $C_{ij;D}(\cdot,\cdot;\mathbf{x}_D)$ denotes the copula associated with the bivariate conditional distribution of $(X_i,X_j)$ given $\mathbf{X}_D=\mathbf{x}_D$. The corresponding copula density is denoted by $c_{ij;D}(\cdot,\cdot;\mathbf{x}_D)$.
For $D \subset \{1,\ldots,d\}\backslash \{i\}$, $F_{i|D}(\cdot|\mathbf{x}_D)$ denotes the univariate conditional distribution function of $X_i$ given $\mathbf{X}_D=\mathbf{x}_D$. Furthermore, $C_{i|D}(\cdot|\mathbf{u}_D)$ denotes the univariate conditional distribution function of $U_i$ given $\mathbf{U}_D=\mathbf{u}_D$, where $\mathbf{U}_D$ corresponds to $\mathbf{X}_D$ transformed to the u-scale.

An arbitrary three-dimensional density (see Section 4.1 of \cite{czadoAnalyzingDependentData2019}) can be written as
\begin{align*}
    f(x_1, {}& x_2, x_3)=c_{13;2}(F_{1|2}(x_1|x_2),F_{3|2}(x_3|x_2);x_2) \cdot\\
    & \cdot c_{23}(F_2(x_2),F_3(x_3))\cdot c_{12}(F_1(x_1),F_2(x_2))\cdot\\
    &\cdot f_3(x_3) \cdot f_2(x_2) \cdot f_1(x_1).
\end{align*}
\textcolor{black}{Consider now the bivariate conditional copula density $c_{13;2}(\cdot,\cdot;x_2)$, which depends on the value $x_2$. Since each value of $x_2$ typically appears only once in the dataset, the copula cannot be reliably estimated from this single observation. Hence, to facilitate tractability in higher dimensions, the simplifying assumption (discussed in Section \ref{sec:Int}) is used.}
It neglects the dependence of a conditional copula $C_{ij;k}(\cdot,\cdot;x_k)$ on the conditioning value $x_k$ \cite{stoeber2013simplified, czado2022vine}.
Using this assumption, the above decomposition becomes the construction 
\begin{equation}\label{Eq:Decomp}
    \begin{aligned}
        f(x_1,x_2,x_3)= {}& c_{13;2}(F_{1|2}(x_1|x_2),F_{3|2}(x_3|x_2)) \cdot \\
        & \cdot c_{23}(F_2(x_2),F_3(x_3))\cdot\\
        &\cdot c_{12}(F_1(x_1),F_2(x_2))\cdot \\
        & \cdot f_3(x_3) \cdot f_2(x_2) \cdot f_1(x_1).
    \end{aligned}
\end{equation}
The dependence on the conditioning value $x_2$ is still captured by the arguments of $c_{13;2}(F_{1|2}(x_1|x_2),F_{3|2}(x_3|x_2))$.
On the copula scale, \autoref{Eq:Decomp} reduces to
\begin{equation*}
    \begin{aligned}
        c(u_1,u_2,u_3)={}&c_{13;2}(C_{1|2}(u_1|u_2),C_{3|2}(u_3|u_2)) \cdot \\
        & \cdot c_{23}(u_2,u_3) \cdot c_{12}(u_1,u_2).
    \end{aligned}
\end{equation*}

From now on we will make the simplifying assumption, thereby working with simplified vine distributions.

In higher dimensions, \cite{bedfordProbabilityDensityDecomposition2001} represent this density construction using a so-called vine tree structure. A regular (R-) vine tree sequence $\mathcal{V}$ on $d$ elements consists of $d-1$ linked trees $T_1,\ldots,T_{d-1}$, where $T_1$ is a tree with node set $N_1=\{1,\ldots,d\}$ and edge set $E_1$. In the next tree, $T_2$, the nodes are the edges of $T_1$, i.e., $N_2=E_1$. Two nodes $a$ and $b$ in $N_2$ can be connected by an edge in $E_2$ if they share a common node in Tree 1. This \textit{proximity condition} can be extended to the trees $T_3,\ldots,T_{d-1}$.

The set 
$A_e := \left\{ j \in N_1 \vert \exists e_1 \in E_1, \dots, \right.$ $\left.e_{i-1} \in E_{i-1}\text{ such that } j\in e_1 \in \dots \in e_{i-1} \in e \right\} 
$ is called the complete union for the edge $e =\{a,b\} \in E_i$.
Further, the conditioning set is defined as $D_e:=A_a \cap A_b$ and the conditioned sets consisting of two elements are $\mathcal{C}_{e,a} := A_a \backslash D_e, \mathcal{C}_{e,b} := A_b \backslash D_e$.
An edge $e=(\mathcal{C}_{e,a}, \mathcal{C}_{e,b}; D_e)$ is abbreviated by $e=(a(e),b(e);D(e))$.
With these notations, the R-vine distribution $(\mathcal{F}, \mathcal{V}, \mathcal{B})$ for a $d$-dimensional random vector $\mathbf{X}$ can be defined. Here, $\mathcal{F}=(F_1,\ldots,F_d)$ is the vector of continuous marginal distribution functions of $X_1,\ldots,X_d$. Further, $\mathcal{V}$ is an R-vine tree sequence described above. Lastly, let $\mathcal{B}=\{C_e|e \in E_i, i=1,\ldots,d-1\}$ be the associated set of bivariate copula families for each edge in the vine tree structure. For $e = \{a,b\} \in E_i, i=1,\ldots,d-1$, $C_e$ is the copula associated with the bivariate conditional distribution $X_{a(e)}$ and $X_{b(e)}$ given $\mathbf{X}_{D(e)}=\mathbf{x}_{D(e)}$. For further details, see 
\cite{joeDependenceModelingCopulas2014, czadoAnalyzingDependentData2019, czado2022vine}.
This allows the density of an R-vine copula to be written as the product of all pair copula densities in $\mathcal{B}$, which are evaluated at conditional distribution functions, i.e.,
\begin{equation}\label{eq:rvinedensityuscale}
	\begin{aligned}
		c(u_1{}&, \dots, u_d) = \\
        & \prod_{i=1}^{d-1} \prod_{e \in E_i} c_{a(e)b(e);D(e)}(C_{a(e) \vert D(e)}(u_{a(e)} \vert \mathbf{u}_{D(e)}),\\
        &  C_{b(e) \vert D(e)}(u_{b(e)} \vert \mathbf{u}_{D(e)})).
	\end{aligned}
\end{equation}
Special cases of R-vines are canonical (C-) and drawable (D-) vines. A C-vine is present if in each tree $T_i$, there is one root node $n \in N_i, i=1,\ldots,d-1$, such that $|\{e \in E_i|n \in e\}|=d-i$. Thus, all trees have a star shape. For a D-vine, $|\{e \in E_i|n \in e\}|\leq 2$ for each node $n \in N_i, i=1,\ldots,d-1$, resulting in a path shape for all trees in the vine tree structure. The order of the nodes in the first tree defines the D-vine completely.

\subsubsection*{Conditional Densities and Distributions of Vine Copulas}
Assume the random variables $\mathbf{U} = (U_1, \dots, U_d)$ on the copula scale follow a regular vine copula distribution, whose joint density is given in \autoref{eq:rvinedensityuscale}.
We define the subsets $\mathcal{C}_1 = \{ \mathcal{C}_{1,1}, \dots, \mathcal{C}_{1,k} \}$ and $\mathcal{C}_2 = \{ \mathcal{C}_{2,1}, \dots \mathcal{C}_{2,\ell} \}$ of $\mathcal{C}= \{ 1, \dots, d\}$ with $\mathcal{C}_1 \cap \mathcal{C}_2 = \emptyset$, $\mathcal{C}_1 \cup \mathcal{C}_2 = \mathcal{C}$, and cardinalities $ \vert \mathcal{C}_1 \vert = k$ and $\vert \mathcal{C}_2 \vert = \ell$, respectively. We want to analyze the distribution of $(\mathbf{U}_{\mathcal{C}_1} \vert \mathbf{U}_{\mathcal{C}_2} = \mathbf{u}_{\mathcal{C}_2})$ where $\mathbf{U}_{\mathcal{C}_1} = (U_{\mathcal{C}_{1,1}}, \dots, U_{\mathcal{C}_{1,k}})$ and $\mathbf{U}_{\mathcal{C}_2} = (U_{\mathcal{C}_{2,1}}, \dots, U_{\mathcal{C}_{2,\ell}})$, respectively.
In general, however, the corresponding conditional density can only be obtained via integration since not all required components are given directly by the representation of the vine. We demonstrate this problem in an example.

\begin{example}[Determining conditional density from a given vine tree structure]
\label{ex:condensproblem}
Consider the $3$-dimensional vine tree structure given in \autoref{fig:3dvine}. We can see that the density of $(U_1 \vert U_2 = u_2, U_3 = u_3)$ is of the form
\begin{equation}\label{eq:exampleconddensity1}
	\begin{aligned}
		{}&c_{1 \vert 23}(u_1 \vert u_2, u_3) = \frac{c_{123}(u_1, u_2, u_3)}{ c_{23}(u_2, u_3)}\\
        &= c_{13;2}(C_{1 \vert 2}(u_1 \vert u_2), C_{3 \vert 2}(u_3 \vert u_2))c_{12}(u_1, u_2).
	\end{aligned}
\end{equation}

\begin{figure}
    \centering
    \includegraphics[width=0.6\linewidth]{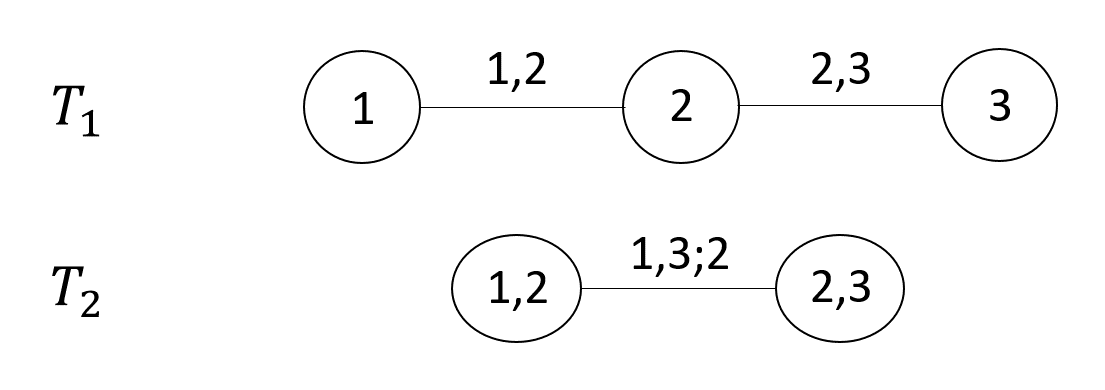}
    \caption{3-dimensional vine tree structure with copula density given by $c_{123}(u_1, u_2, u_3) = c_{13;2}(C_{1 \vert 2}(u_1 \vert u_2), C_{3 \vert 2}(u_3 \vert u_2))c_{12}(u_1, u_2)c_{23}(u_2, u_3)$.}
    \label{fig:3dvine}
\end{figure}

\noindent All components of this equation are known from the representation of the vine tree structure given in \autoref{fig:3dvine}. The density of $(U_3 \vert U_1 = u_1, U_2 = u_2)$ can be expressed similarly. On the other hand, the density of $(U_2 \vert U_1 = u_1, U_3 = u_3)$ is given by
\begin{equation}
	\begin{aligned}
		c_{2 \vert 13}{}&(u_2 \vert u_1, u_3) = \frac{c_{123}(u_1, u_2, u_3)}{ c_{13}(u_1, u_3)} \\
        &= \frac{c_{123}(u_1, u_2, u_3)}{\int_0^1 c_{123} (u_1, a_2, u_3) \diff a_2}.
	\end{aligned}
\end{equation}

\end{example}

For some vine tree structures, the conditional density of a single variable, given the remaining variables, can be obtained without integration. This includes all R-vine structures where the single variable is contained in any leaf node in all trees as shown in \cite{chang2019prediction,zhu2021simplified}.
As a special case, consider D-vines, where all trees are a path. Then, one can express the conditional density of a leaf node given all the other variables in the first tree in a closed form. Suppose we have a D-vine copula with the node order $1 - \ldots - d$  in the first tree, corresponding to the variables $U_1, \ldots, U_d$. Then, the conditional density $c_{1|2,\ldots, d}(u_1|u_2, \ldots, u_d)$ of $U_1$ given the others is
\begin{equation}
	\begin{aligned}
		{}&c_{1|2\ldots,d}(u_1|u_2,\ldots, u_d)=  \bigg[\! \prod_{j=3}^{d} c_{1, j;2, \ldots, j-1}\\
        &\big( C_{1 | 2, \ldots, j-1}(u_1 | u_{2}, \ldots, u_{j-1}),\\
        &C_{j | 2, \ldots, j-1}(u_{j} | u_2, \ldots, u_{j-1}\big)\bigg]  .
	\end{aligned}
\end{equation}
 Hence, if the outcome of interest is to estimate the conditional distribution/density of a given variable, i.e., response, a forward-selection method can be applied so that variables are added into the D-vine as far as they improve the conditional log-likelihood. Such ideas are implemented in \cite{krausDVineCopulaBased2017, sahin2022high} using D-vines and \cite{ tepegjozova2022nonparametric} using C-vines. 

Given the flexibility of simplified vine copulas, we aim to estimate a conditional distribution arising from any vine structure. To avoid integration, we propose a sampling approach, which is explained next.

\clearpage
\subsection{Introduction to the No-U-Turn Sampler and \texttt{Stan}}

We use the modeling and computation platform \texttt{Stan} \cite{standevelopmentteamStanModelingLanguage2023}. It applies the No-U-Turn Sampler (NUTS) \cite{hoffmanNoUTurnSamplerAdaptively2014} to sample from highly complex (multivariate) distributions. 

NUTS is an extension of Hamiltonian Monte Carlo (HMC), which is a Markov chain Monte Carlo (MCMC-) method with the goal of obtaining samples from a parametric distribution of interest.
In MCMC, random samples are chosen from a Markov chain with a limiting distribution equal to the desired distribution. In most cases, the desired distribution is high-dimensional, and MCMC is a widely used approximating approach for a (multivariate) distribution that is too complex for analytical investigation.
The idea of HMC is to construct an artificial physical system that describes the movement of a particle following the so-called Hamiltonian dynamics \cite{nealMcmcUsingHamiltonian2012, hoffmanNoUTurnSamplerAdaptively2014, betancourtConceptualIntroductionHamiltonian2018, thomasLearningHamiltonianMonte2020}. Intuitively, this particle moves on the curve of the negative logarithm of the target density. The movement is defined by the Hamiltonian dynamics, modeling the potential and kinetic energy of the particle. It is designed to visit the bottom of the curve, i.e., the regions of high-density values, more often than the other regions.

In each HMC iteration, $L$ discrete steps of size $\epsilon$ are performed to find a new proposal for a sample from the desired distribution.
The efficiency of HMC strongly depends on the choice of these tuning parameters $\epsilon$ and $L$. In \texttt{Stan}, the No-U-Turn sampler is implemented \cite{carpenterStanProbabilisticProgramming2017, hoffmanNoUTurnSamplerAdaptively2014}, choosing $L$ automatically, while $\epsilon$ can be adjusted by employing \textit{primal-dual averaging} \cite{nesterovPrimaldualSubgradientMethods2009}.
To eliminate the necessity for tuning $L$, steps are performed until the distance between the proposal parameters and the initial value does no longer increase, i.e., a U-turn is performed.


To study whether the samples of the Markov chain have converged to its limiting distribution,  we use the scale reduction factor $\hat{R}$ \cite[Chapter 11.4]{gelmanBayesianDataAnalysis2013} and the effective sample size (ESS) \cite{vehtariRankNormalizationFoldingLocalization2021}.

\section{Sampling Conditional Distributions of Simplified Vines}
\label{Sec:Proposal}
Our goal is to sample from and analyze conditional distributions associated with a specified vine copula. 
For a general given simplified R-vine copula, the density of $(\mathbf{U}_{\mathcal{C}_1} \vert \mathbf{U}_{\mathcal{C}_2} = \mathbf{u}_{\mathcal{C}_2})$ can either be expressed directly provided we have all the components available from the vine structure for $(\mathbf{U}_{\mathcal{C}_1}, \mathbf{U}_{\mathcal{C}_2})$ or by using integration
\begin{equation}
	\begin{aligned}
		{}&c_{\mathcal{C}_1 \vert \mathcal{C}_2 }(\mathbf{u}_{\mathcal{C}_1} \vert \mathbf{u}_{\mathcal{C}_2}) = \frac{c(\mathbf{u}_{\mathcal{C}_1}, \mathbf{u}_{\mathcal{C}_2})}{c(\mathbf{u}_{\mathcal{C}_2})}\\
        &= \frac{c(\mathbf{u})}{\int_{[0,1]^k}c(\mathbf{u}) \diff  u_{\mathcal{C}_{1,1}} \dots \diff  u_{\mathcal{C}_{1,k}}},
	\end{aligned}
\end{equation}
\noindent where $\mathbf{u}_{\mathcal{C}_1} \in [0,1]^k$, $\mathbf{u}_{\mathcal{C}_2} \in [0,1]^{\ell}$.

In our MCMC-based proposal, we take advantage of the fact that 
\begin{equation}
\label{eq:proportionalconditionaldensity}
c_{\mathcal{C}_1 \vert \mathcal{C}_2 }(\mathbf{u}_{\mathcal{C}_1} \vert \mathbf{u}_{\mathcal{C}_2}) = \frac{c(\mathbf{u}_{\mathcal{C}_1}, \mathbf{u}_{\mathcal{C}_2})}{c(\mathbf{u}_{\mathcal{C}_2})} \propto c(\mathbf{u}_{\mathcal{C}_1},\mathbf{u}_{\mathcal{C}_2}) = c(\mathbf{u})
\end{equation}
as a function of $\mathbf{u}_{\mathcal{C}_1}$ alone. Hence, for fixed values of $\mathbf{u}_{\mathcal{C}_2}$, we can draw MCMC samples from the distribution corresponding to the joint density $c(\mathbf{u})$ given in \autoref{eq:rvinedensityuscale} to obtain samples of $(\mathbf{U}_{\mathcal{C}_1} \vert \mathbf{U}_{\mathcal{C}_2} = \mathbf{u}_{\mathcal{C}_2})$.

\textcolor{black}{Using these samples, we can investigate any statistics based on the conditional distribution. As an example, we illustrate this for the conditional Kendall's $\tau$ in the data application in Section \ref{Sec:Data}, among other statistics.}

To obtain the MCMC samples, \texttt{Stan} is used, which can deal with proportional densities. More specifically, the \texttt{R} package \texttt{rstan} provides an \texttt{R} interface to \texttt{Stan}.
For further details, see \url{www.mc-stan.org}.

To represent our conditional statistical model in \texttt{rstan}, we use the logarithm of the joint density $c(\mathbf{u})$ as a function of $\mathbf{u}_{\mathcal{C}_1}$ alone as mentioned above. The implementation of this function is inspired by the function  \texttt{RVineLogLik} of the \texttt{R} package \texttt{VineCopula} of \cite{naglerVineCopulaStatisticalInference2023}.

\section{Simulation Study}\label{sec:Sim}

We test our proposed sampling algorithm using \texttt{Stan} in a large simulation study. Here, we show a small part of it and refer to \cite{havlickovaAnalysisConditionalVine2022} for further simulation scenarios and results.
The simulation study consists of two parts.
First, we consider univariate conditional distributions of $d$-dimensional simplified vine copulas, i.e., $\vert \mathcal{C}_1 \vert = k = 1$ and $\vert \mathcal{C}_2 \vert = \ell = d-1$. Second, bivariate conditional distributions are examined, i.e., $ \vert \mathcal{C}_1 \vert = k= 2$ and $\vert \mathcal{C}_2 \vert = \ell = d-2$.

\subsection{Case 1: Univariate Conditional Distribution}


\begin{table*}[ht]
    \centering
    \begin{tabular}{ |c|c|c|c|c| } 
        \hline
        Setup & Vine Structure & $d$ & Density availability & Conditioned, conditioning set \\ 
        \hline 
        1 & D & 3 & T & $\mathcal{C}_1 = \{ 1 \}$, $\mathcal{C}_2 = \{ 2,3 \}$  \\
        \hline
        2 & D & 3 & F & $\mathcal{C}_1 = \{ 2 \}$, $\mathcal{C}_2 = \{ 1,3 \}$  \\
        \hline
        3 & D & 5 & T & $\mathcal{C}_1 = \{ 1 \}$, $\mathcal{C}_2 = \{ 2,3,4,5 \}$  \\
        \hline
        4 & D & 5 & F & $\mathcal{C}_1 = \{ 2 \}$, $\mathcal{C}_2 = \{ 1,3,4,5 \}$  \\
        \hline
        5 & R & 5 & T & $\mathcal{C}_1 = \{ 5 \}$, $\mathcal{C}_2 = \{ 1,2,3,4 \}$  \\
        \hline
        6 & R & 5 & F & $\mathcal{C}_1 = \{ 2 \}$, $\mathcal{C}_2 = \{ 1,3,4,5 \}$  \\
        \hline
    \end{tabular}
\caption{Selected simulation setups when sampling from univariate conditional distributions of $d$-dimensional simplified vine copulas. Density availability means whether we can express the conditional density without integration. We show the chosen conditioning and conditioned variables in the conditioning and conditioned set.}
\label{tab:univarsetups}
\end{table*}

We consider D-vines in 3 and 5 dimensions and R-vines in 5 dimensions.
The setups are shown in \autoref{tab:univarsetups}.
For all setups, our proposal leads to very good results with regard to the Kolmogorov-Smirnov test for the uniformity of the copula data and the conditional density estimation of interest.  As the results are similar, we only show the details for Setup 6. The results for the other scenarios can be found in \cite{havlickovaAnalysisConditionalVine2022}. 

For all simulation setups, three sets of informative conditioning values are chosen, based on their distance from the central vector $\mathbf{u_c}:=(0.5, 0.5, \dots, 0.5) \in [0,1]^d$ in the copula space.
For this, sample $\mathbf{u}_k := (u_{k1}, u_{k2}, \dots, u_{kd})$, $k=1, \dots, K=1000$, using the \texttt{R} package \texttt{rvinecopulib} of \cite{nagler2022rvinecopulib}, and compute the Euclidean distance $e(\mathbf{u}_k)$ to $\mathbf{u_c}$, so that we obtain $\mathbf{e} = (e(\mathbf{u}_1), \dots, e(\mathbf{u}_K)) \in \mathbb{R}^K$. Let $\hat{q}_{\alpha}(\mathbf{e})$ be the empirical $\alpha$-quantile of $\mathbf{e}$. Then, for $\alpha=0.05$ (low), $\alpha=0.5$ (medium), and $\alpha=0.95$ (high), we find the sample iteration $k_{\alpha}$ such that $e(\mathbf{u}_{k_{\alpha}})$ is closest to $\hat{q}_{\alpha}(\mathbf{e})$. 

We are now interested in sampling from $U_{\mathcal{C}_1} \vert \mathbf{U}_{\mathcal{C}_2} = \mathbf{u}_{k_{\alpha}}^{\mathcal{C}_2}$, where $\mathbf{u}_{k_{\alpha}}^{\mathcal{C}_2}= (u_{k_{\alpha}}^{\mathcal{C}_{2,1}}, \dots, u_{k_{\alpha}}^{\mathcal{C}_{2,\ell}})$.
For sampling, we aim at sampling i.i.d. samples from the desired conditional distribution of sizes $R=1000$, $5000$, and $10000$ after burnin. We achieve this by drawing an MCMC sample of size $10000$, $50000$ or $100000$ after burnin using the HMC approach. Then, we use thinning by choosing every $10^{th}$ iteration to remove the autocorrelation present in the sampled Markov chain. We denote the sampled values from $U_{\mathcal{C}_1} \vert \mathbf{U}_{\mathcal{C}_2} = \mathbf{u}_{k_{\alpha}}^{\mathcal{C}_2}$ by $u^{(r)}(\mathbf{u}_{k_{\alpha}}^{\mathcal{C}_2})$, $r=1, \dots, R$.
Hence, for all simulation steps, we consider three sets of conditioning values derived from empirical quantiles and three different numbers of MCMC samples, leading to nine scenarios per simulation setup.

To measure how well the MCMC samples agree with the desired distribution, we apply the probability integral transform to the sampled values to obtain uniform data
$v^{(r)}(\mathbf{u}_{k_{\alpha}}^{\mathcal{C}_2}) := C_{U_{\mathcal{C}_1} \vert \mathbf{U}_{\mathcal{C}_2}}\left(u^{(r)}(\mathbf{u}_{k_{\alpha}}^{\mathcal{C}_2}) \big\vert \mathbf{u}_{k_{\alpha}}^{\mathcal{C}_2}\right)$
using the theoretical distribution function of $U_{\mathcal{C}_1} \vert \mathbf{U}_{\mathcal{C}_2}$, which is obtained by integration.  The Kolmogorov-Smirnov (KS) test is used to test whether the transformed data follows a uniform distribution by examining the p-values.
For each of the nine scenarios per simulation setup, $N=100$ repeated simulations are executed, resulting in $N$ Kolmogorov-Smirnov tests per scenario.

\subsubsection*{Setup 6: R-vine copula, d=5, Conditional Density Expressed with Integration}

The specification of the R-vine copula that we consider is given in \autoref{fig:univarsetup6fampar} in the Appendix. Thus, the conditional density of $(U_{2} \vert U_{1}= u_{1}, U_{3}= u_{3}, \dots, U_{5}= u_{5})$ is given by
\begin{equation*}
	\begin{aligned}
		{}&c_{2 \vert 1345}(u_2 \vert u_1, u_3, u_4, u_5) =\\
        & \frac{c_{12345} (u_1, u_2, u_3, u_4, u_5)}{\int_0^1 c_{12345} (u_1, a_2, u_3, u_4, u_5) \diff a_2}.
	\end{aligned}
\end{equation*}
The conditional distribution function is given by
\begin{equation*}
	\begin{aligned}
		{}&C_{2 \vert 1345}(u_2 \vert u_1, u_3, u_4, u_5) =\\
        & \int_0^{u_2} c_{2 \vert  1345}(a_2\vert u_1,u_3,u_4,u_5) \diff a_2.
	\end{aligned}
\end{equation*}

\begin{figure*}
     \centering
     \begin{subfigure}[b]{\textwidth}
         \centering
         \includegraphics[width=\linewidth]{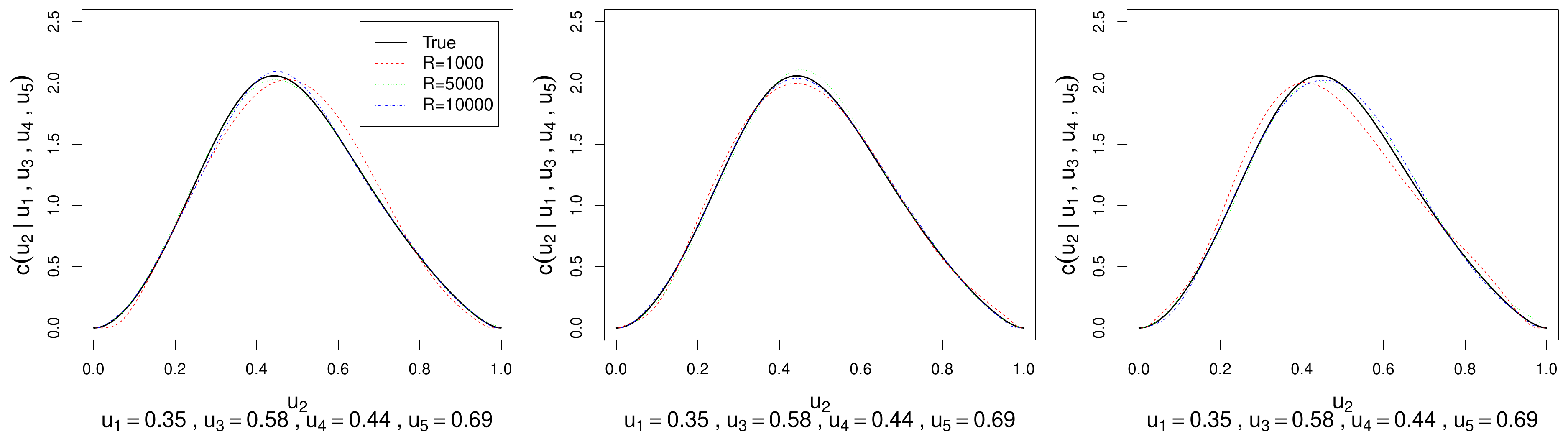}
     \end{subfigure}
     
     \begin{subfigure}[b]{\textwidth}
         \centering
         \includegraphics[width=\linewidth]{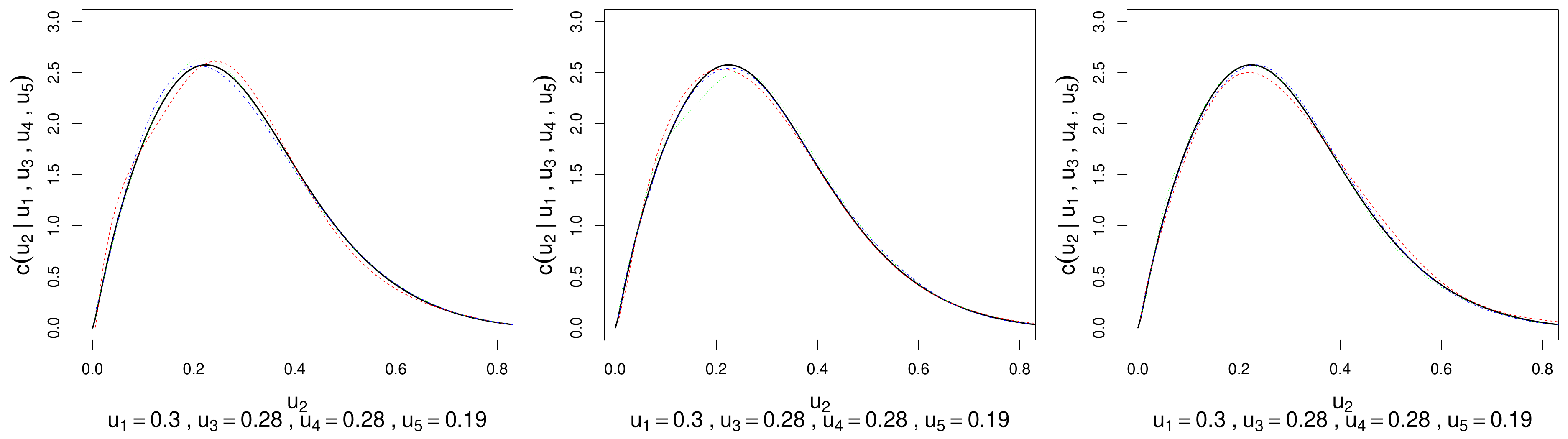}
     \end{subfigure}
     
     \begin{subfigure}[b]{\textwidth}
         \centering
         \includegraphics[width=\linewidth]{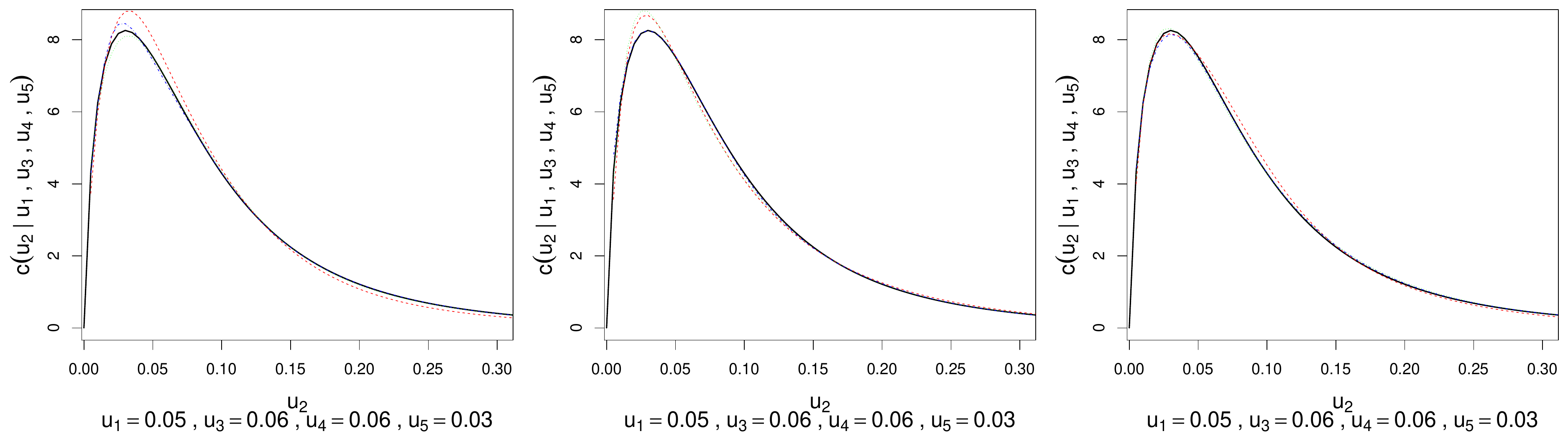}
     \end{subfigure}     
    \caption{Comparison of densities for simulation setup 6. For the
    vine specification in \autoref{fig:univarsetup6fampar}, each column shows the plot of one chosen iteration out of 100. The rows correspond to different conditioning values resulting in low ($\alpha=0.05$), medium ($\alpha=0.5$), and high ($\alpha=0.95$) (empirical) Euclidean distances to the u-data center $\bold{u}_c$.}
    \label{fig:univardenssetup6}
\end{figure*}

To compare the sampled values $u^{(r)}(\mathbf{u}_{k_{\alpha}}^{\mathcal{C}_2})$, $r=1, \dots, R$, to the theoretical conditional density of $(U_{\mathcal{C}_1} \vert \mathbf{U}_{\mathcal{C}_2} = \mathbf{u}_{k_{\alpha}}^{\mathcal{C}_2})$, we plot the theoretical density together with the kernel density estimates of the sampled values.
To avoid issues at the boundaries \cite{naglerNonparametricEstimationSimplified2017} of the support $[0,1]$, the kernel density estimation is performed on $(-\infty,\infty)$ (see Section \ref{Sec:TransOneDim} in the Appendix).
As an example, we show the comparison of the theoretical density to the estimated densities for three of the $N$ iterations corresponding to the columns in \autoref{fig:univardenssetup6}.
As expected, the fit improves when the sample size $R$ increases. From visual inspection, the estimated densities are very close to the theoretical density.

In \autoref{tab:results_setup6}, we give the percentage of the $N=100$ Kolmogorov-Smirnov tests for which the test rejects a uniform distribution at the significance level $\alpha=5\%$.
Further, the effective sample size divided by the number of samples $R$, and $\hat{R}$ is given.
The percentage of rejected tests is less than 7\%, which is a very good result. According to \cite{vehtariRankNormalizationFoldingLocalization2021}, the value of $\hat{R}$ should be lower than 1.01, which is always the case. Further, the effective sample size should be higher than 100. Since we show the effective sample size divided by $R$, it should be higher than $0.1, 0.02$, and $0.01$ for $R=1000, 5000$, and $10000$, which clearly always holds.

\begin{table*}
\centering
\begin{tabular}{|c|l|c|c|c|c|c|}
  \hline
  Conditioning & & Rejected  & \multicolumn{2}{|c|}{ESS} & \multicolumn{2}{|c|}{$\hat{R}$} \\
  values & & KS tests &  min & max & min & max \\ 
  \hline
  \multirow{3}{*}{low} & R=1000 & 4\% & 0.784 & 1.184 & 0.999 & 1.006 \\ 
  & R=5000 & 3\% & 0.852 & 1.054 & 1.000 & 1.002 \\ 
  & R=10000 & 6\% & 0.901 & 1.034 & 1.000 & 1.000 \\
   \hline
  \multirow{3}{*}{medium} & R=1000 & 1\% & 0.738 & 1.159 & 0.999 & 1.005 \\ 
  & R=5000 & 3\% & 0.869 & 1.077 & 1.000 & 1.002 \\ 
  & R=10000 & 3\% & 0.796 & 1.054 & 1.000 & 1.000 \\
   \hline
  \multirow{3}{*}{high} & R=1000 & 4\% & 0.775 & 1.174 & 0.999 & 1.008 \\ 
  & R=5000 & 5\% & 0.835 & 1.078 & 1.000 & 1.001 \\ 
  & R=10000 & 7\% & 0.890 & 1.053 & 1.000 & 1.001 \\
   \hline
\end{tabular}
\caption{For the nine scenarios of simulation Setup 6, we show the percentage of the rejected KS tests as well as the minimum and maximum values of ESS and $\hat{R}$.} 
\label{tab:results_setup6}
\end{table*}

\subsection{Case II: Bivariate Conditional Distribution}

\begin{table*}[ht]
    \centering
    \begin{tabular}{ |c|c|c|c|c| } 
        \hline
        Setup & Vine Structure & $d$ & Density availability & Conditioned, conditioning set \\ 
        \hline 
        1 & D & 3 & T & $\mathcal{C}_1 = \{ 1, 2 \}$, $\mathcal{C}_2 = \{ 3 \}$  \\
        \hline
        2 & D & 5 & F & $\mathcal{C}_1 = \{ 2, 4 \}$, $\mathcal{C}_2 = \{ 1,3,5 \}$  \\
        \hline
        3 & R & 5 & F & $\mathcal{C}_1 = \{ 2, 4 \}$, $\mathcal{C}_2 = \{ 1,3,5 \}$  \\
        \hline
    \end{tabular}
\caption{Selected simulation setups in sampling from bivariate conditional distributions of $d$-dimensional simplified vine copulas. Density availability means whether we can express the conditional density without integration, hence true or false. We show the chosen conditioning and conditioned variables in the conditioning and conditioned set.}
\label{tab:bivarsetups}
\end{table*}

The simulation study for bivariate conditional distributions is similar to the univariate one, with exceptions in the testing procedure.
We are again using D-vine copulas and an R-vine copula. The three simulation setups are in \autoref{tab:bivarsetups}. Here, we only present the results of Setup 3, while Setups 1 and 2 can be found in \cite{havlickovaAnalysisConditionalVine2022}.
For every setup, nine scenarios are executed by choosing $R=1000,5000$, $10000$, and three sets of conditioning values based on their Euclidean distance to the central vector. Here, the conditioning values are chosen less extreme, i.e., $\alpha=0.05 \text{ (low)}, \alpha=0.5 \text{ (medium)}$ and $\alpha=0.85 \text{ (high)}$.
The sampled values from $\mathbf{U}_{\mathcal{C}_1} \vert \mathbf{U}_{\mathcal{C}_2} = \mathbf{u}_{k_{\alpha}}^{\mathcal{C}_2}$ are denoted by $\mathbf{u}^{(r)}(\mathbf{u}_{k_{\alpha}}^{\mathcal{C}_2})=(u^{(r)}_{1}(\mathbf{u}_{k_{\alpha}}^{\mathcal{C}_2}),u^{(r)}_{2}(\mathbf{u}_{k_{\alpha}}^{\mathcal{C}_2}))$, $r=1, \dots, R$.

To measure the goodness-of-fit in the bivariate conditional case, we use the Rosenblatt transformation \textcolor{black}{\cite{rosenblattRemarksNonparametricEstimates1956}}: \textit{Let $(U_1,U_2)$ be a random vector with distribution function $C(u_1,u_2)$ and $V_1=C_1(U_1), V_2=C_{2|1}(U_2|U_1)$. Then, $V_1$ and $V_2$ are independent and uniformly distributed on $[0,1]$.}

We apply the transformation to the samples $\mathbf{u}^{(r)}(\mathbf{u}_{k_{\alpha}}^{\mathcal{C}_2})$, $r=1, \dots, R$, and obtain
\begin{align*}
v^{(r)}_{1,m} {}&= C_{U_{\mathcal{C}_{11}} \vert \mathbf{U}_{\mathcal{C}_2}}\left(u^{(r)}_{1}(\mathbf{u}_{k_{\alpha}}^{\mathcal{C}_2}) \Big\vert \mathbf{u}_{k_{\alpha}}^{\mathcal{C}_2}\right),\\ 
v^{(r)}_{2,c} {}&= C_{U_{\mathcal{C}_{12}} \vert (U_{\mathcal{C}_{11}},\mathbf{U}_{\mathcal{C}_2})}\left(u^{(r)}_{2}(\mathbf{u}_{k_{\alpha}}^{\mathcal{C}_2}) \Big\vert u^{(r)}_{1}(\mathbf{u}_{k_{\alpha}}^{\mathcal{C}_2}),\mathbf{u}_{k_{\alpha}}^{\mathcal{C}_2}\right),
\end{align*}
or using the other order of the variables
\begin{align*}
v^{(r)}_{2,m} &= C_{U_{\mathcal{C}_{12}} \vert \mathbf{U}_{\mathcal{C}_2}}\left(u^{(r)}_{2}(\mathbf{u}_{k_{\alpha}}^{\mathcal{C}_2}) \Big\vert \mathbf{u}_{k_{\alpha}}^{\mathcal{C}_2} \right),\\
v^{(r)}_{1,c} &= C_{U_{\mathcal{C}_{11}} \vert (U_{\mathcal{C}_{12}},\mathbf{U}_{\mathcal{C}_2})}\left(u^{(r)}_{1}(\mathbf{u}_{k_{\alpha}}^{\mathcal{C}_2}) \Big\vert u^{(r)}_{2}(\mathbf{u}_{k_{\alpha}}^{\mathcal{C}_2}),\mathbf{u}_{k_{\alpha}}^{\mathcal{C}_2} \right).
\end{align*}
Then, we test for a uniform distribution of the four quantities $v^{(r)}_{1,m},v^{(r)}_{2,m},v^{(r)}_{2,c}$, and $v^{(r)}_{1,c}$ using the Kolmogorov-Smirnov test. Further, we test for independence of $\{ v^{(r)}_{1,m}, v^{(r)}_{2,c} \}$ and $\{v^{(r)}_{1,c}, v^{(r)}_{2,m}\}$ using a bivariate dependence test based on Kendall's tau \cite{hollanderNonparametricStatisticalMethods2013}.
In total, six tests are performed on one bivariate sample $\mathbf{u}^{(r)}(\mathbf{u}_{k_{\alpha}}^{\mathcal{C}_2})$, $r=1, \dots, R$, which means that we use a Bonferroni correction to avoid the multiple testing problem. Hence, our significance cut-off is now set to $0.05/6=0.0083$. We reject that the sample is from the desired distribution if at least one p-value out of six is below this threshold.

\subsubsection*{Setup 3: R-vine copula, d=5, Bivariate Conditional Density Expressed with Integration}

We show the chosen R-vine in \autoref{fig:bivarsetup3fampar} in the Appendix. We consider the conditional density of $(U_{2}, U_{4} \vert U_{1}= u_{1}, U_{3}= u_{3}, U_{5}= u_{5})$.

\begin{figure*}
     \centering
     \begin{subfigure}[b]{\textwidth}
         \centering
         \includegraphics[width=\linewidth]{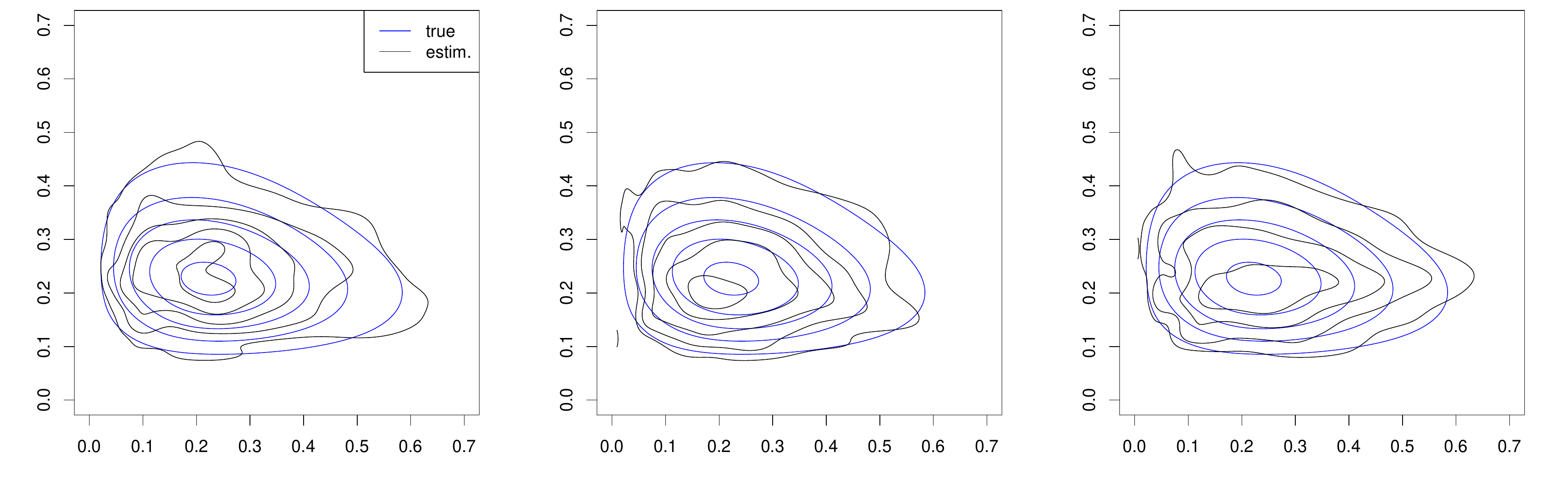}
     \end{subfigure}
     
    \begin{subfigure}[b]{\textwidth}
         \centering
         \includegraphics[width=\linewidth]{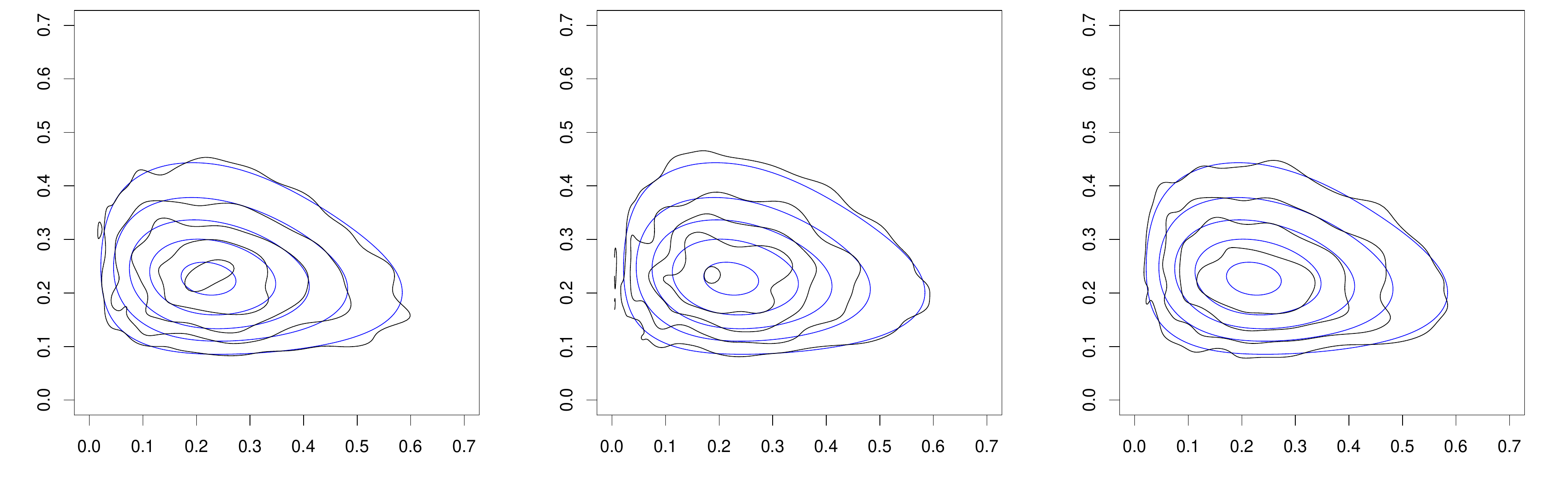}
    \end{subfigure}
         
    \begin{subfigure}[b]{\textwidth}
         \centering
         \includegraphics[width=\linewidth]{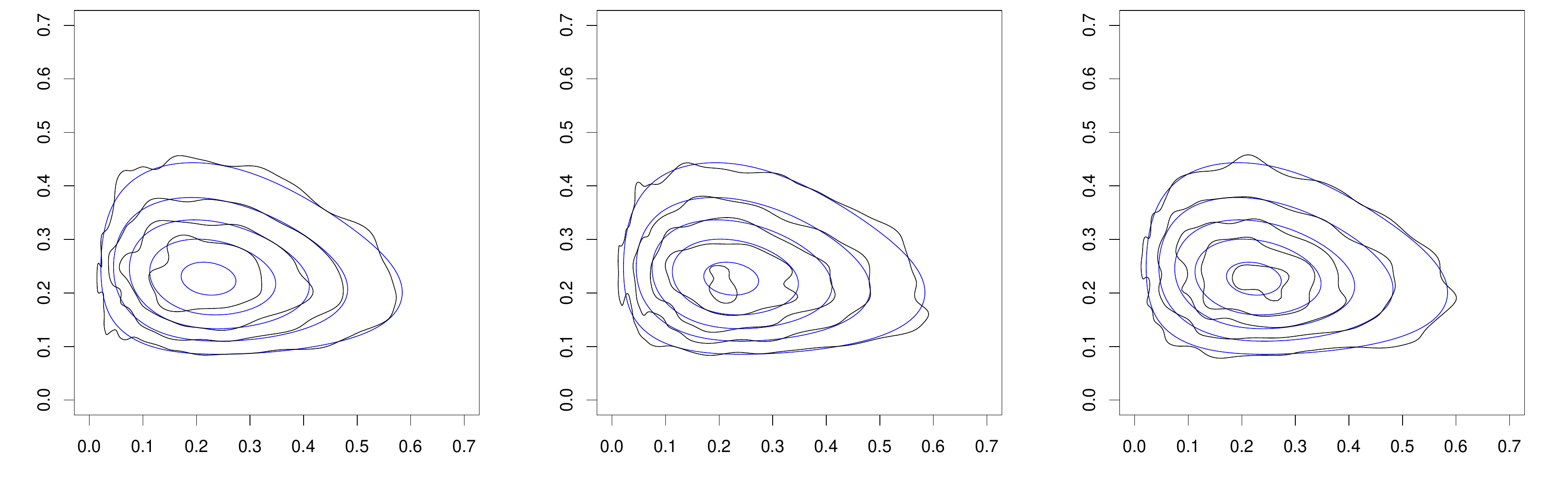}
     \end{subfigure}     
    \caption{Comparison of estimated densities and the true density with \textbf{medium conditioning values} for simulation setup 3. Each column shows the plot for one chosen iteration. The rows correspond to different sample sizes with order $R=1000$, 5000, and 10000.}
    \label{fig:bivar_setup3_cont3}
\end{figure*}

A visual comparison of the kernel density of the sampled values to the theoretical density of
$(\mathbf{U}_{\mathcal{C}_1} \vert \mathbf{U}_{\mathcal{C}_2} = \mathbf{u}_{k_{\alpha}}^{\mathcal{C}_2})=(U_{\mathcal{C}_{11}},U_{\mathcal{C}_{12}} \vert \mathbf{U}_{\mathcal{C}_2} = \mathbf{u}_{k_{\alpha}}^{\mathcal{C}_2})$
is given.
In \cite{havlickovaAnalysisConditionalVine2022}, contour plots for all conditioning values and three out of $100$ different iterations are given. Here, we only show the contour plots for medium conditioning values and three arbitrarily chosen iterations in \autoref{fig:bivar_setup3_cont3}. To avoid boundary issues, a transformation is performed; see Section \ref{Sec:TransTwoDim} in the Appendix.
Again, the estimated densities are close to the theoretical density. Further, the fit improves for a higher value of $R$.

The percentage of iterations for which the six tests are rejected after the Bonferroni correction is given in \autoref{tab:bivar_setup3_results}. The percentage of rejected tests is always lower than 5\%, which is a very goodresult.

\begin{table*}
\centering
\begin{tabular}{|c|c|c|c|}
  \hline
 \multirow{2}{*}{} & \multicolumn{3}{|c|}{Conditioning value} \\
 & Low & Medium & Large \\ 
  \hline
  R=1000 &   2\% &   0\%  &   1\%  \\ \hline
  R=5000 &   3\% &   4\%  &   5\%  \\ \hline
  R=10000 &   5\%  & 4\% &   3\%  \\ 
  \hline
\end{tabular}
\caption{The percentage of iterations for which at least one of the six tests is rejected after the Bonferroni correction.}
\label{tab:bivar_setup3_results}
\end{table*}

\section{Data Application}\label{Sec:Data}
\begin{figure*}
     \centering
         \includegraphics[width=1\linewidth]{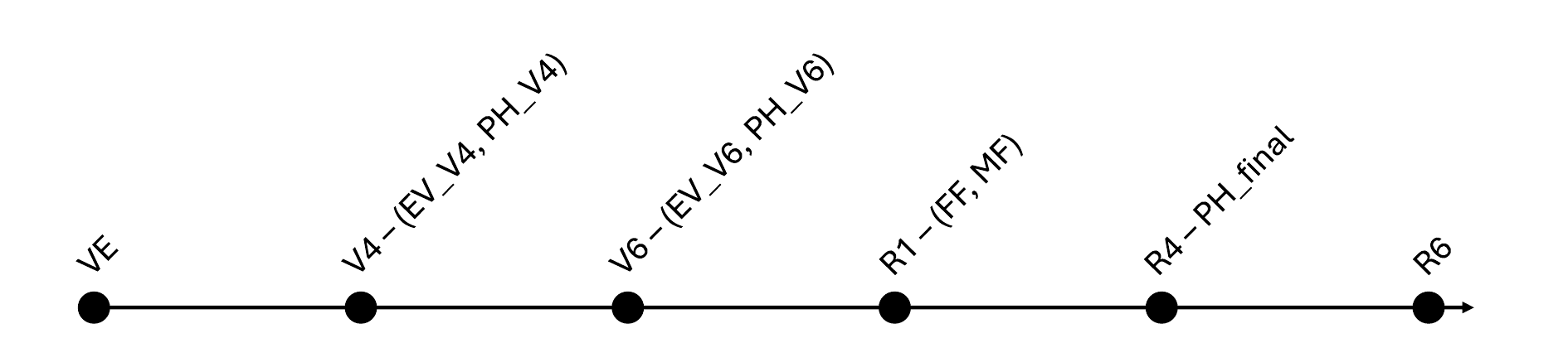}
    \caption{Timeline of some maize growth stages, where the traits in our data set are written next to their associated growth stage.}
    \label{fig:maizetime}
\end{figure*}

Maize has natural development stages as described in \cite{nleya2016corn} and shown in Figure \ref{fig:maizetime}, starting from its vegetative stage VE (emergence) and ending in its reproductive stage R6 (physiological maturity). From the early stages, plant breeders aim to predict the maize's final plant height at growth stage R4 to estimate the height at the end of the breeding season, which might impact its yield \cite{agronomy12040958}. In addition, it is important to estimate the female and male flowering time measured at the growth stage R1 so that breeders can take prevention measures in changing climate environments if they are too early or too late. 

We work with seven phenotypic traits of the Kemater Landmais Gelb landrace of maize: early vigor as a 1-9 score ranging from very small (1) to very vigorous (9) (\texttt{EV\_V4}) and early plant height in centimeters at the growth stage V4 (\texttt{PH\_V4}), early vigor as a score (\texttt{EV\_V6}) and early plant height in centimeters at the growth stage V6 (\texttt{PH\_V6}), female (\texttt{FF}) and male flowering (\texttt{MF}) time in days at the growth stage R1, and the final plant height in centimeters at the growth stage R4 (\texttt{PH\_final}). Thus, we have seven continuous features of 471 observations as analyzed in \cite{holker2019european} and given in the link: \url{https://figshare.com/articles/journal_contribution/Data_from_Mayer_et_al_2020_Nat_Commun_/12137142}. Two extreme observations are removed for the analysis, leaving 469 observations in the data set. 

Our interest is in sampling from (1) the conditional distribution of \texttt{PH\_final} given the other traits, i.e., sampling from a univariate conditional distribution and (2) the conditional distribution of (\texttt{PH\_final}, \texttt{FF}, \texttt{MF}) given the remaining four traits, i.e., sampling from a multivariate conditional distribution. Our algorithm described in Section \ref{Sec:Proposal} is used for the sampling. Given the MCMC samples of the conditional distributions, we obtain point estimates like the marginal mean and median of the traits, credible intervals, and empirical probabilities of having a high final plant height and early flowering times.
Furthermore, we are interested in (3) the four-dimensional analysis of how the conditional Kendall's $\tau$ of (\texttt{EV\_V6}, \texttt{PH\_V6} $\mid$ \texttt{EV\_V4}, \texttt{PH\_V4}) changes with different values of \texttt{EV\_V4} and \texttt{PH\_V4}. Our algorithm from Section \ref{Sec:Proposal} can also be used for this analysis, obtaining an empirical Kendall`s $\hat{\tau}$ from the MCMC samples of \texttt{EV\_V6} and \texttt{PH\_V6} given fixed values of \texttt{EV\_V4} and \texttt{PH\_V4}. Such early-stage maize traits are important to assess maize's development \cite{holker2019european}. Therefore, with our analysis, plant breeders can understand how the maize's traits will depend on each other in later stages, given their values in the early stages.

The data set is transformed to the unit hypercube to have approximate copula data using the estimated distribution functions of the kernel density estimation implemented in the \texttt{R} package \texttt{kde1d} \cite{kde1dpackage}. We remark that we sample from the univariate and multivariate conditional distributions of vine copula models of the traits. However, we can easily convert the samples to their original scale with the inverse of their estimated marginal distribution functions.

We remark that all vines used in this chapter have been tested for the simplifying assumption using the constant conditional correlation test proposed in \cite{kurz2022testing}, implemented in the \texttt{R} package \texttt{pacotest}. The simplifying assumption remains unchallenged for all considered bivariate conditional copulas except for one. Applying the chi-square-type test of independence proposed in \cite{derumignyTestsSimplifyingAssumption2017} on this bivariate conditional copula however does not reject the simplifying assumption. Hence, there is no statistical evidence against the simplifying assumption in any of the vines used for this data analysis.

\subsection{Exploratory Data Analysis}
\autoref{fig:datahist} in the Appendix shows that the univariate marginal distributions of the traits are non-Gaussian. The traits \texttt{EV\_V4}, \texttt{EV\_V6}, \texttt{PH\_V4}, and \texttt{PH\_V6} are left-skewed, while right-skewness is observed for the traits \texttt{PH\_final} and \texttt{MF}. The trait \texttt{FF} looks symmetric.

\begin{figure*}
    \centering
    \includegraphics[width=0.7\linewidth]{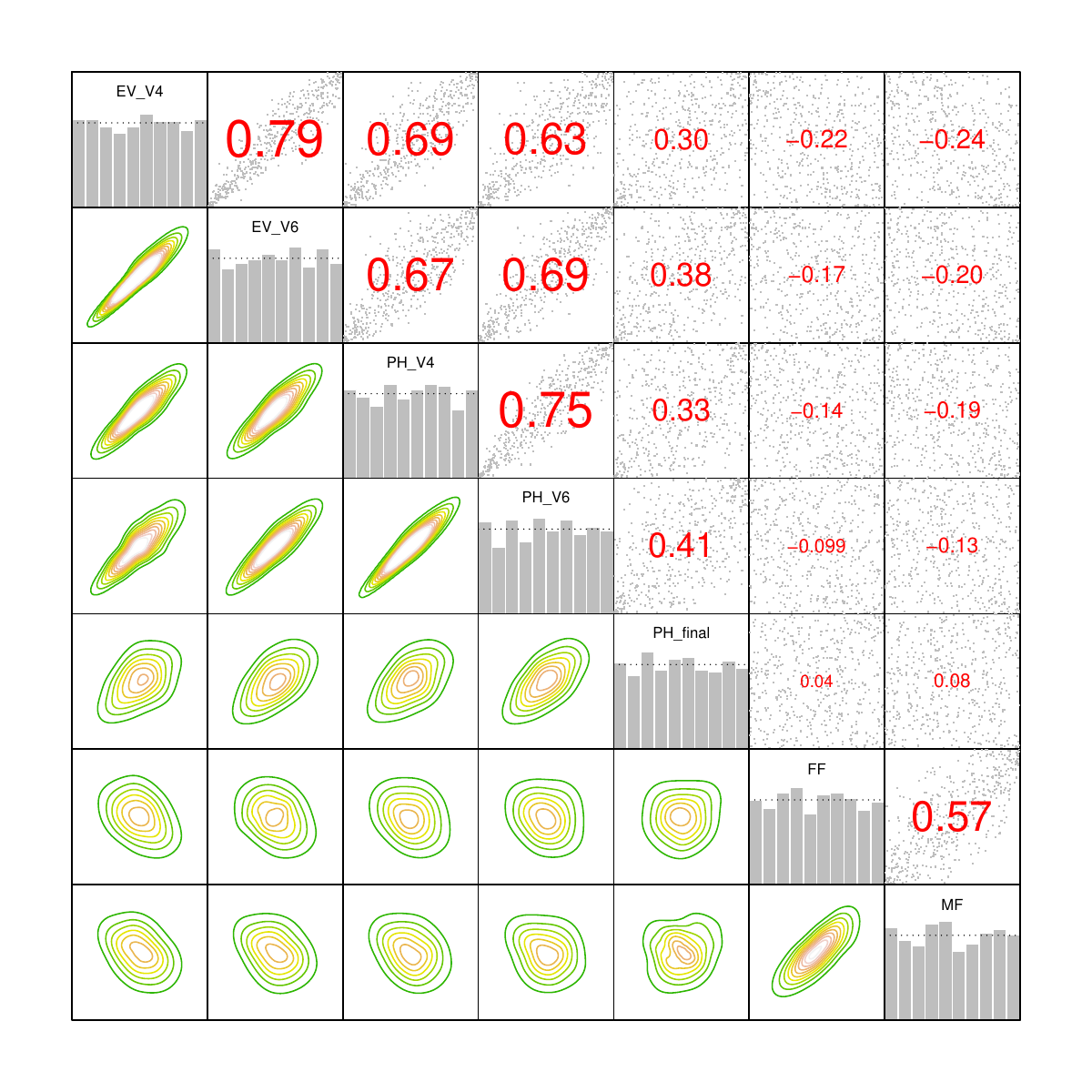}
    \caption{Marginally normalized bivariate contour plots of the copula data.}
    \label{fig:datapairs}
\end{figure*}

\autoref{fig:datapairs} shows the histograms of the approximated copula data on the diagonal, the estimated dependence strength among the pairs of traits measured by Kendall's $\tau$ in the upper diagonal together with the copula data, and its marginally normalized bivariate contour plots in the lower diagonal. A high-sized, positive dependence exists among the traits of the growth stages V4 and V6; their dependence with the final plant height is medium-sized, positive, and that with the female and male flowering time is low-sized, negative. Moreover, all pairwise dependence structures look non-elliptical and hence non-Gaussian in \autoref{fig:datapairs}, except for the pair (\texttt{FF}, \texttt{MF}).

\subsection{Univariate Conditional Distribution Analysis}\label{Sec:UniResp}

Here, we estimate the conditional quantiles and mean of \texttt{PH\_final} given the six early growth stage traits, i.e., a univariate conditional distribution analysis. Given the trait information at the early growth stages, such an analysis helps plant breeders to approximate the final maize yield whose proxy is \texttt{PH\_final}. Hence,  we are interested in the conditional distribution of \texttt{PH\_final} given the other traits based on vine copulas.

In the following, we introduce and compare several ways of selecting a vine tree structure for the univariate conditional distribution estimation. We also compare their in-sample behavior regarding different performance measures. With our sampling approach, we can work with any valid vine tree structure. Thus, we consider two different R-vines, for which we obtain results by sampling using HMC.
Furthermore, we use D-vines, obtained by \texttt{vinereg} and \texttt{sparsevinereg} methods, described later. An expression without integration for the univariate conditional distribution is available in both D-vines.

The heuristic algorithm suggested by \cite{Dissmann2013} finds the maximum spanning tree at each tree level, where the weights are the absolute estimated Kendall's $\tau$ between the tree level's nodes, and AIC is used to select the pair copula families and their parameters. Following this algorithm,
we obtain the vine model shown in \autoref{fig:RVMMaize} in the Appendix. Here, it is enough to show the first two trees since the second tree is a path; thus, all other trees are uniquely determined. Since \texttt{PH\_final} is a leaf node in each tree level, its conditional distribution has a form without integration. Nevertheless, we sample from the conditional distribution of \texttt{PH\_final} given the remaining traits based on this vine model and compare the estimated kernel density estimate of the MCMC samples to the true density. \autoref{fig:Plots_OneDim_Comparison} in the Appendix shows this comparison visually for three observations, where our sampling approximates the true density very well. 

The Dissmann algorithm is not focused on the estimation of the conditional distribution, i.e., prediction, while selecting the vine structure. 
Therefore, as another simple forward-selection heuristic, we propose to maximize the correlations and partial correlations of \texttt{PH\_final} \cite{dissertationSahin} given the other traits after converting them into normal scores \cite{joeDependenceModelingCopulas2014}.
Using their normal scores, we implement this forward-selection heuristic and start by finding the trait with the highest absolute correlation with \texttt{PH\_final}. We compare six correlation values of traits with the response \texttt{PH\_final} and choose the trait with the highest one. Then, we estimate correlations and partial correlations given the first chosen trait of the remaining traits with \texttt{PH\_final}, calculating five correlations with   \texttt{PH\_final} and four partial correlations of remaining traits and \texttt{PH\_final} given the first chosen trait. The trait among the five traits studied with the highest value is added to the vine structure. If all correlation values were always the highest among all the above-studied correlations and partial correlations, the resulting structure in the first tree would be a C-vine. The remaining tree structures are allowed to be chosen by the Dissmann algorithm, preserving the proximity condition. 
Furthermore, we select the pair copula families of the chosen structure by the AIC. We give the details of our selection in Section \ref{Sec:Heuristic} in the Appendix, including the vine tree structure, where \texttt{PH\_final} is not a leaf node (see Figure \ref{fig:RVMMaizeNotLeaf} in the Appendix). We also sample from this vine model.

A similar but different idea is proposed by \cite{chang2019prediction}, abbreviated as \texttt{copreg}, where the correlation among normal scores of predictors is maximized in the first tree, and the remaining trees allowed by the proximity condition are selected by maximizing partial correlations. Later, the response is added to the tree of predictors so that it maximizes the associated (partial) correlations. We remark that such a selection does not consider the impact of the predictors on the response. Furthermore, as seen in Table \ref{tab:datamethods}, the selected and estimated vine model by \texttt{copreg} coincides with the one from the  Dissmann algorithm in our application.

After sampling from the conditional distribution based on the above chosen simplified vine copula models (40,000 MCMC samples for each of  469 conditioning values on the copula scale), we use the empirical quantile function of the estimated kernel density estimate of \texttt{PH\_final} to obtain samples on the original data scale from the conditional distribution of \texttt{PH\_final} given for each value of the remaining traits.

For each conditioning value, a separate simulation is carried out, running four chains with 100,000 iterations, respectively. By using only every $10^{th}$ value, thinning is performed for each chain, leading to 40,000 MCMC samples per conditioning value.
As the chains are parallelized, we consider the maximum computation time of the four chains per conditioning value.
For the conditioning value with the minimum (maximum) computation time, sampling takes 3 minutes (19 hours). The median is 38 minutes.
We run all computations with an AMD 7713 processor, running R version 4.3.3 and \texttt{Stan} version 2.33.0, parallelizing the chains and the conditioning variables.

In the literature, another way of obtaining the conditional distribution of \texttt{PH\_final} given the others is to restrict the vine structure to D-vines and \texttt{PH\_final} to a leaf node in all trees, as mentioned in Section \ref{Sec:IntCopula}. Such an approach provides an expression without integration for the conditional distribution of interest using vines as described in Section \ref{sec:Background}. \cite{krausDVineCopulaBased2017, sahin2022high} follow this approach and develop a forward variable selection tailored to univariate conditional distribution estimation purposes. Their implementation is given in the \texttt{R} packages \texttt{vinereg} \cite{nagler2018vinereg} and \texttt{sparsevinereg} \cite{sahin2022high}, respectively.

Next, we compare the in-sample prediction performance of these vine copula models concerning the root mean-squared-error (RMSE) for the conditional mean estimation and the pinball loss ($PL_\alpha$) \cite{steinwart2011estimating} at the quantile levels  10\%, 50\%, and 90\% for the conditional quantile estimation. Smaller RMSE and $PL_\alpha$ values are better.
For the D-vines, these performance measures are based on predictions arising from the simple expressions without integration. For the R-vines, the empirical mean and quantiles of the  MCMC samples are used for the assessment. However, HMC can also be applied to the D-vines, leading to the same performance measures in Table \ref{tab:datamethods} like the appraoch without integration.

\autoref{tab:datamethods} shows that the D-vine-based methods perform better than the R-vine-based methods in terms of the RMSE, $PL_{0.10}$, and $PL_{0.90}$.  On the other hand, our heuristic is the best regarding the pinball loss at the median level in \autoref{tab:datamethods}.
We remark that the differences in the performance measures in \autoref{tab:datamethods} are due to the selected vine model, not due to the estimation approach.

\autoref{tab:datavinereg} in the Appendix gives the orders of the selected D-vine models by \texttt{vinereg} and \texttt{sparsevinereg}, respectively, while \autoref{tab:vinereg} and \autoref{tab:sparsevinereg} in the Appendix give the selected pair copula families and their parameter estimate based on the fitted vine copulas. Both select all variables in the forward selection procedure to estimate the conditional distribution of \texttt{PH\_final} given the remaining traits.
Furthermore, both identify \texttt{PH\_V6} and \texttt{MF} as the first and second predictors, respectively. Next, \texttt{EV\_V6} is added to the model by \texttt{sparsevinereg}, while \texttt{vinereg} adds \texttt{PH\_V4}. Despite the different D-vine structures chosen by both methods, none dominates the other regarding the in sample prediction accuracy measures given in \autoref{tab:datamethods}.

\begin{table*}[ht]
\centering
\begin{tabular}{llllll}
  \hline
Method & RMSE &$PL_{0.10}$ & $PL_{0.50}$&$PL_{0.90}$ & HMC \\
  \hline
R-vine (Dissmann \& \texttt{copreg}, leaf response) & 291.80 & 20.64 & 23.66 & 20.32 & yes \\ 
  R-vine (our heuristic, non-leaf response) & 294.46  & 20.72 & \hl{20.45} & 19.50 & yes \\ 
  D-vine (\texttt{vinereg}) & 291.30  & \hl{19.67} & 22.60 & 21.39 &no \\ 
  D-vine (\texttt{sparsevinereg}) &\hl{290.62} & 22.55 & 22.31 & \hl{19.03}  & no\\  \hline
\end{tabular}
\caption{Comparison of vine-based univariate conditional distribution estimation (in sample prediction) methods regarding the root mean-squared-error (RMSE) and pinball loss ($PL_\alpha$) at the quantile levels 5\%, 10\%, and 90\% using 469 observations (in-sample). Smaller RMSE and $PL_\alpha$ values are better. The last column shows if the results come from HMC (yes) or not (no).}
\label{tab:datamethods}
\end{table*}

\subsection{Multivariate Conditional Distribution Analysis}\label{Sec:MultiResponse}

Female and male flowering times matter for breeders, so adjustments in sowing dates may be required. Furthermore, the difference in female and male flowering time defines the Anthesis-silking interval (\texttt{ASI}=\texttt{FF}-\texttt{MF}). Researchers have identified final plant height and \texttt{ASI} impacting grain yield, making them two outcomes of interest in maize breeding programs \cite{silva2022grain}. The shorter \texttt{ASI}, the better for yield, especially under drought conditions \cite{silva2022grain}. From \autoref{fig:maizetime}, we see that other traits, except the final plant height, are measured before flowering. Hence, we are interested in the analysis of the multivariate conditional distribution of   (\texttt{PH\_final}, \texttt{FF}, \texttt{MF}) given the other four traits using the vine copula model identified by Dissman's algorithm in Section \ref{Sec:UniResp} (see \autoref{fig:RVMMaize} and \autoref{tab:dissmann} in the Appendix).
We remark that the vine tree structure selection can be tailored to express the multivariate conditional distribution without integration; for instance, \cite{tepegjozovaBivariateVineCopula2023, zhu2021simplified} proposed a heuristic. It is on our agenda to develop further ideas in this direction. Still, since Dissman's algorithm models the strongest (conditional) pairwise dependencies stepwise to select the vine structure, we use it to show our sampling algorithm's multivariate applications further. 
Next, we perform a stress testing: what would be the impact of very low and very high early growth traits  (\texttt{EV\_V6}, \texttt{PH\_V6}, \texttt{EV\_V4}, \texttt{PH\_V4}) on (\texttt{PH\_final}, \texttt{FF}, \texttt{MF}), thereby on the final plant height and \texttt{ASI}.

As the four traits (\texttt{EV\_V6}, \texttt{PH\_V6}, \texttt{EV\_V4}, \texttt{PH\_V4}) are highly positively dependent (see \autoref{fig:datapairs}), it is realistic that they are either all small or all large. Therefore, to allow for two extreme stress scenarios, we set these four traits on the u-scale to $0.01$ (small scenario) and $0.99$ (large scenario). 

We remark that the trivariate conditional density we sample can be expressed without integration. However, integration is needed to estimate the corresponding univariate quantiles or joint conditional probabilities.
Hence, we sample from the joint trivariate conditional distribution using our sampling algorithm and obtain 8000 MCMC samples (4 chains with 2000 MCMC samples, respectively) for each of the two conditioning scenarios. Using these samples, we can estimate univariate quantiles and joint probabilities using Monte Carlo without the need to integrate numerically.

In the following, given the MCMC samples from the joint trivariate conditional distribution, we analyze the univariate and bivariate margins of that distribution and itself.

\subsubsection{Univariate Margins}\label{Sec:UnivMarg}

First, we analyze the three univariate margins of the joint trivariate conditional distribution:
(\texttt{PH\_final} $\vert$\texttt{EV\_V6}, \texttt{PH\_V6}, \texttt{EV\_V4}, \texttt{PH\_V4}),
(\texttt{FF} $\vert$\texttt{EV\_V6}, \texttt{PH\_V6}, \texttt{EV\_V4}, \texttt{PH\_V4}), and
(\texttt{MF}$\vert$\texttt{EV\_V6}, \texttt{PH\_V6}, \texttt{EV\_V4}, \texttt{PH\_V4}) based on the MCMC samples of
(\texttt{PH\_final}, \texttt{FF}, \texttt{MF}$\vert$\texttt{EV\_V6}, \texttt{PH\_V6}, \texttt{EV\_V4}, \texttt{PH\_V4})
in the two stress scenarios (abbreviated as joint analysis in \autoref{tab:stresstest}). Since these univariate conditional distributions can also be estimated by fitting three five-dimensional vine copulas like one for (\texttt{MF}, \texttt{EV\_V6}, \texttt{PH\_V6}, \texttt{EV\_V4}, \texttt{PH\_V4}), we also fit and analyze such models (abbreviated as separate analysis in \autoref{tab:stresstest}). However, we point out that the univariate margins of the joint trivariate conditional distribution include the joint impact of the conditioning variables on (\texttt{PH\_final}, \texttt{FF},  \texttt{MF}). Hence, univariate separate analyses might result in losing joint information and do not assess the importance of all predictors on the multivariate response. Furthermore, despite being out of the scope of this article, if the variable selection was tailored for the estimation of the joint trivariate conditional distribution  (instead of using the vine model given in \autoref{tab:dissmann} in the Appendix), the selection of the covariates from such estimation compared to three univariate conditional distribution estimations could be different.

The summaries of our samples on the u-scale are given in \autoref{tab:stressuquant} in the Appendix. To interpret them better, we convert them to the original scale as explained in Section \ref{Sec:Data}. Furthermore, we obtain samples of \texttt{ASI} as the difference of \texttt{FF} from \texttt{MF} for each scenario and analysis and the observed values of the covariates. We remark that we do not regard \texttt{ASI} as a variable or response in our vine copula models because we can determine inferences about \texttt{ASI} from the MCMC samples of \texttt{FF} and \texttt{MF}.

\autoref{tab:stresstest} lists the estimated  univariate (conditional) quantiles of (1) the four variables 
\texttt{PH\_final}, \texttt{FF},  \texttt{MF}, and \texttt{ASI}, respectively, from the univariate margins of the joint trivariate conditional distribution, (2) univariate conditional distributions from three five-dimensional vine copula models separately, and (3) the observed values for \texttt{PH\_final}, \texttt{FF},  \texttt{MF} and \texttt{ASI} ignoring the covariates. The estimated values are close to each other in the joint or separate analysis, except for \texttt{ASI}.
The differences result from the fact that the separate analysis assumes conditional independence between \texttt{FF} and \texttt{MF}, compared to the joint analysis.
This shows that inference for \texttt{ASI} should be based on the joint analysis approach, which we discuss in more detail now.

For the joint analysis, \autoref{tab:stresstest} shows that the estimated conditional median of the final plant height is smaller than its unconditional sample median if it is not vigorous and very small in the growth stages V4 and V6 (small scenario). In addition,  the plant takes longer to flower in the conditional median than the unconditional one. However, the median anthesis-silking interval is still around two to three days both in the conditional small scenario and unconditional setup.  On the other hand, when the plant is already vigorous and large in the growth stages V4 and V6 (large scenario), the final plant height is also large, and the female and male flowering is earlier than the unconditional scenario. \texttt{ASI} varies between zero and seven days for the different quantile levels. The small conditioning values have a more extreme effect than the large ones on \texttt{PH\_final} for the estimated median, where unconditional and conditional medians differ around  11\% for the large scenario and 32\%  for the small scenario in absolute values, respectively.

\begin{table*}[ht]
\centering
\begin{tabular}{lcrrrr}
  \hline
Scenario &$\alpha$& \texttt{PH\_final} & \texttt{FF}  & \texttt{MF}  & \texttt{ASI}  \\ 
  \hline
\multirow{3}{*}{Small scenario - joint analysis}& 0.10 & 98.62 & 77.44 & 74.87 & -0.55 \\ 
 & 0.50 & 106.58 & 82.11 & 79.11 & 2.96 \\ 
&  0.90 & 124.59 & 86.93 & 83.52 & 6.27 \\ \hline
\multirow{3}{*}{Small scenario - separate analysis} & 0.10 & 98.58 & 77.40 & 74.93 & -3.72 \\ 
  &0.50  & 106.70 & 81.87 & 79.17 & 2.85 \\ 
&  0.90  & 124.48 & 86.69 & 83.48 & 9.08 \\  \hline
\multirow{3}{*}{Large scenario - joint analysis} &  0.10 & 138.49 & 72.61 & 69.72 & 0.29 \\ 
 & 0.50  & 157.78 & 77.31 & 73.72 & 3.51 \\ 
& 0.90  & 177.75 & 81.91 & 77.88 & 6.82 \\ \hline
\multirow{3}{*}{Large scenario - separate analysis}&  0.10& 138.83 & 72.08 & 69.70 & -2.99 \\ 
&  0.50  & 157.65 & 77.13 & 73.65 & 3.38 \\ 
&  0.90  & 177.97 & 81.72 & 77.87 & 9.55 \\  \hline
\multirow{3}{*}{unconditional} & 0.10 & 116.79 & 74.35 & 71.42 & -0.35 \\ 
&  0.50  & 140.09 & 79.71 & 76.30 & 2.99 \\ 
& 0.90  & 165.46 & 84.82 & 81.38 & 6.66 \\ 
   \hline
\end{tabular}
\caption{Estimated univariate conditional quantiles of \texttt{PH\_final}, \texttt{FF}, \texttt{MF} and \texttt{ASI}, respectively, based on a joint/separate HMC analysis approach for the small/large scenario together with empirical quantiles using the observed values of  \texttt{PH\_final}, \texttt{FF}, \texttt{MF} and \texttt{ASI} for quantile levels $\alpha=(0.10, 0.50, 0.90)$ 
(\texttt{ASI} is calculated as the difference of \texttt{FF} and \texttt{MF}).}
\label{tab:stresstest}
\end{table*}

\subsubsection{Bivariate Margins}

\begin{figure*}[t]
    \centering
 \includegraphics[width=\linewidth]{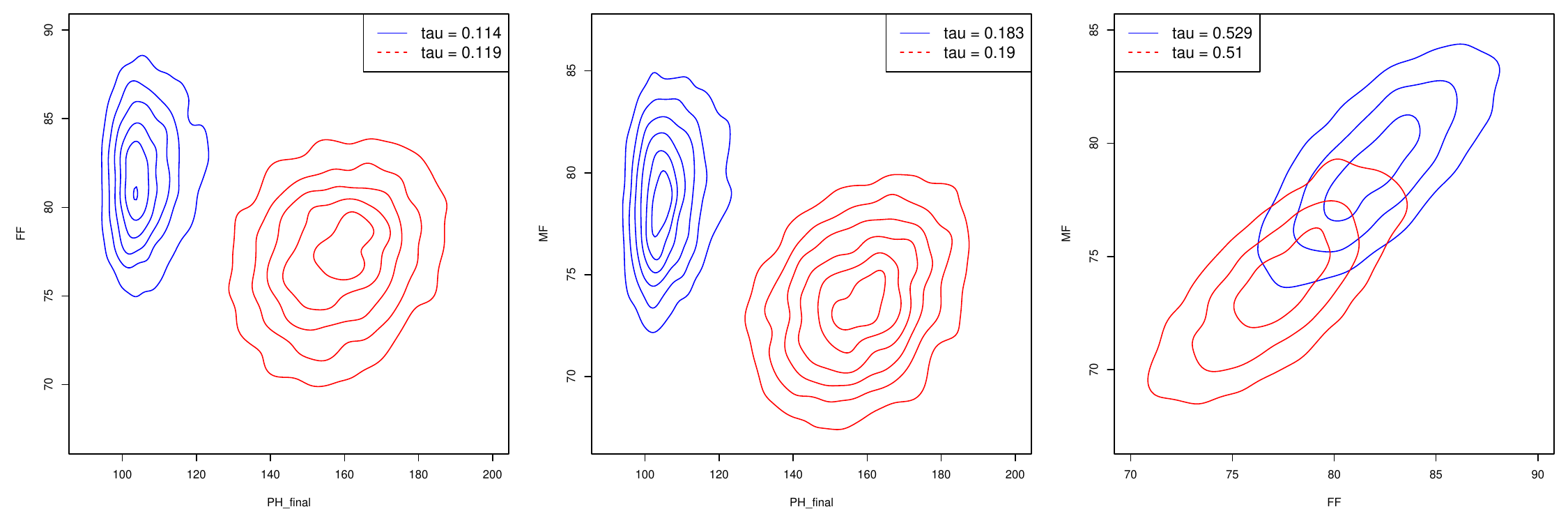}

\caption{Estimated bivariate contours based on the joint analysis for the small (blue) and large (red) scenario for the three bivariate conditional distributions (\texttt{PH\_final}, \texttt{FF}$\vert$\texttt{EV\_V6}, \texttt{PH\_V6}, \texttt{EV\_V4}, \texttt{PH\_V4}) (left), (\texttt{PH\_final},  \texttt{MF}$\vert$\texttt{EV\_V6}, \texttt{PH\_V6}, \texttt{EV\_V4}, \texttt{PH\_V4}) (middle), and (\texttt{FF}, \texttt{MF}$\vert$\texttt{EV\_V6}, \texttt{PH\_V6}, \texttt{EV\_V4}, \texttt{PH\_V4}) (right).}
\label{fig:Contours}
\end{figure*}

Given the joint MCMC samples of (\texttt{PH\_final}, \texttt{FF}, \texttt{MF}$\vert$\texttt{EV\_V6}, \texttt{PH\_V6}, \texttt{EV\_V4}, \texttt{PH\_V4}), we can also consider the three bivariate marginal conditional distributions:
(\texttt{PH\_final}, \texttt{FF}$\vert$\texttt{EV\_V6}, \texttt{PH\_V6}, \texttt{EV\_V4}, \texttt{PH\_V4}),
(\texttt{PH\_final},  \texttt{MF}$\vert$\texttt{EV\_V6}, \texttt{PH\_V6}, \texttt{EV\_V4}, \texttt{PH\_V4}), and
(\texttt{FF}, \texttt{MF}$\vert$\texttt{EV\_V6}, \texttt{PH\_V6}, \texttt{EV\_V4}, \texttt{PH\_V4}).
We remark that a vine copula could be constructed in such a way that the associated bivariate conditional density can be estimated without integration (see, for example, \cite{zhu2021simplified} or \cite{tepegjozovaBivariateVineCopula2023}). Yet, using  the same
vine copula model from the first pair would not necessarily give expressions  without integration for the other two bivariate conditional distributions. 

In our application, for each of the three bivariate conditional densities of interest, we show MCMC-based estimated contour plots of the small scenario (blue) and the large scenario (red) in \autoref{fig:Contours}. The change of the univariate empirical conditional quantiles observed in Section \ref{Sec:UnivMarg} between the two scenarios is also visible here. Furthermore, we can see that the shape changes for the conditional distributions (\texttt{PH\_final}, \texttt{FF}$\vert$\texttt{EV\_V6}, \texttt{PH\_V6}, \texttt{EV\_V4}, \texttt{PH\_V4}) and (\texttt{PH\_final}, \texttt{MF}$\vert$\texttt{EV\_V6}, \texttt{PH\_V6}, \texttt{EV\_V4}, \texttt{PH\_V4}), allowing for a smaller range of \texttt{PH\_final} in the small scenario compared to the large scenario.

Moreover, we give the conditional empirical Kendall's $\tau$ for the three pairs of traits in \autoref{fig:Contours}. The differences between the two scenarios are small for all three bivariate conditional distributions.
The Kendall's $\tau$ values are non-zero, meaning that all three variables are conditionally dependent.
This indicates that a joint conditional analysis is more reasonable than univariate conditional analyses since the identified bivariate conditional dependencies in \autoref{fig:Contours} are not taken into account.
In the unconditional case, the empirical Kendall's $\tau$ between \texttt{PH\_final} and \texttt{FF} (\texttt{MF}) is 0.04 (0.08), see \autoref{fig:datapairs}, showing that there is no evidence of monotonic dependence. For the conditional case, this increases to 0.114 (0.183) for the small scenario and 0.119 (0.19) for the large scenario, respectively. 
Between \texttt{MF} and \texttt{FF}, the empirical unconditional Kendall's $\tau$ is medium-sized at 0.57. For the conditional case, this decreases to 0.529 (small scenario) and 0.51 (large scenario), but it is still medium-sized.

\subsubsection{Trivariate Analysis}

In this analysis, the aim is to obtain joint probabilities
\begin{align}
\label{eq: triprob}
P({}&\texttt{PH\_final}>c_{\texttt{PH}}, \texttt{FF}<c_{\texttt{FF}}, \texttt{MF}<c_{\texttt{MF}}\vert\texttt{EV\_V6}=c_{\texttt{EV\_V6}}, \nonumber\\
&\texttt{PH\_V6}=c_{\texttt{PH\_V6}}, \texttt{EV\_V4}=c_{\texttt{EV\_V4}}, \texttt{PH\_V4}=c_{\texttt{PH\_V4}})
\end{align}
under the extreme scenarios. Given that the observed early growth traits are very low or very high, we aim to guide plant breeders on the joint probability of the final plant height still being larger than a threshold ($c_{\texttt{PH}}$) and flowering times being not later than the desired ones ($c_{\texttt{FF}}$ and $c_{\texttt{MF}}$). 
Again, both scenarios are considered by setting
$c_{\texttt{EV\_V6}}, c_{\texttt{PH\_V6}}, c_{\texttt{EV\_V4}}, c_{\texttt{PH\_V4}}$
to the values of 0.01 (small scenario) and 0.99 (large scenario), transformed to the original scale using the marginal distribution functions. The probabilities are estimated by relative frequencies using the available MCMC samples.

We consider two setups: in the first setup, we want to validate that the three responses are conditionally dependent.
For this, we set $c_{\texttt{PH}}$ to the univariate conditional $(1-\alpha)$-quantile of (\texttt{PH\_final}$\vert$\texttt{EV\_V6}, \texttt{PH\_V6}, \texttt{EV\_V4}, \texttt{PH\_V4}) and $c_{\texttt{FF}}$ and $c_{\texttt{MF}}$ to the univariate conditional $\alpha$-quantiles of \texttt{FF} and \texttt{MF}, respectively. These quantiles are obtained by the empirical quantiles of the respective MCMC samples obtained from the univariate separate sampling for both scenarios in Section \ref{Sec:UnivMarg}. Hence, the values $c_{\texttt{PH}}$, $c_{\texttt{FF}}$, and $c_{\texttt{MF}}$ are different for the small and large scenario, using the univariate MCMC samples from the small and large scenario, respectively. The values of $\alpha$ are set to $0.5$ and $0.1$.
If the three responses were conditionally independent, this joint probability would be $\alpha^3$, since in this case we have
\begin{align*}
    P(&\texttt{PH\_final}>c_{\texttt{PH}}, \texttt{FF}<c_{\texttt{FF}}, \texttt{MF}<c_{\texttt{MF}} \vert\texttt{EV\_V6}=c_{\texttt{EV\_V6}},\\
    & \texttt{PH\_V6}=c_{\texttt{PH\_V6}}, \texttt{EV\_V4}=c_{\texttt{EV\_V4}}, \texttt{PH\_V4}=c_{\texttt{PH\_V4}}) \\
    =&P(\texttt{PH\_final}>c_{\texttt{PH}}\vert\texttt{EV\_V6}=c_{\texttt{EV\_V6}}, \texttt{PH\_V6}=c_{\texttt{PH\_V6}},\\
    &\texttt{EV\_V4}=c_{\texttt{EV\_V4}}, \texttt{PH\_V4}=c_{\texttt{PH\_V4}})\\
    \cdot &P(\texttt{FF}<c_{\texttt{FF}}\vert\texttt{EV\_V6}=c_{\texttt{EV\_V6}}, \texttt{PH\_V6}=c_{\texttt{PH\_V6}},\\
    &\texttt{EV\_V4}=c_{\texttt{EV\_V4}}, \texttt{PH\_V4}=c_{\texttt{PH\_V4}})\\
    \cdot &P(\texttt{MF}<c_{\texttt{MF}} \vert\texttt{EV\_V6}=c_{\texttt{EV\_V6}}, \texttt{PH\_V6}=c_{\texttt{PH\_V6}},\\
    &\texttt{EV\_V4}=c_{\texttt{EV\_V4}}, \texttt{PH\_V4}=c_{\texttt{PH\_V4}})
\end{align*}
and the values of $c_{\texttt{PH}}$, $c_{\texttt{FF}}$, and $c_{\texttt{MF}}$ are chosen to be the quantiles of the univariate conditional distributions, leading to a value of $\alpha$ for all three univariate conditional probabilities.
From \autoref{tab:Probs}, it is clear that this is not the case. The values of $0.1525$ and $0.1435$ are different from $0.5^3=0.125$ for $\alpha=.5$, and the values $0.0016$ and $0.0014$ are different from $0.1^3=0.001$ for $\alpha=.1$.  Still, the conditional dependence among the three responses is not strong.

In the second setup, the probabilities of interest given in \autoref{eq: triprob} are compared for the small and large scenarios, respectively. For this, we set $c_{\texttt{PH}}$ to the empirical (1-$\alpha$)-quantile of the observed values of \texttt{PH\_final}. The values $c_{\texttt{FF}}$ and $c_{\texttt{MF}}$ are set to the $\alpha$-quantiles of the observed traits \texttt{FF} and \texttt{MF}, respectively. Hence, the values $c_{\texttt{PH}}$, $c_{\texttt{FF}}$, and $c_{\texttt{MF}}$ are the same for both the small and the large scenario, however the MCMC samples for the large/small scenario differ. We obtain estimates of the probability of a plant having a high final height with early male and female flowering times conditioned on \texttt{EV\_V6}, \texttt{PH\_V6}, \texttt{EV\_V4}, and \texttt{PH\_V4} having large/small values. From \autoref{tab:Probs}, it is clear that for both values of $\alpha=0.5$ and $\alpha=0.1$, this probability is much higher for the large scenario compared to the small scenario. This is expected as we assume a high and vigorous plant for this scenario's growth stages V4 and V6.

\begin{table*}[ht]
\centering
\begin{tabular}{r|rr|rr}
  & \multicolumn{2}{c|}{\shortstack{\textbf{First Setup:} thresholding 
  values \\ are alpha quantiles of MCMC samples}} & \multicolumn{2}{c}{\shortstack{\textbf{Second Setup:} thresholding  values \\  are alpha quantiles of observations}} \\ 
  \hline
  & $\alpha=0.5$ & $\alpha=0.1$ & $\alpha=0.5$ & $\alpha=0.1$ \\ 
  \hline
  Small scenario & 0.1525 & 0.0016 & 0.0010 & 0.0000 \\ 
  Large scenario & 0.1435 & 0.0014 & 0.5707 & 0.0243 \\ 
  Unconditional &  & & 0.0120 & 0.0002 \\ 
\end{tabular}
\caption{Estimated (conditional) probabilities given in \autoref{eq: triprob} for the first and second setup for different values of $\alpha$ and both the small (first row)  and large scenario (second row).
For the first setup, the thresholding values for the small scenario are ($c_{\texttt{PH}}=107.7$, $c_{\texttt{FF}}=81.9$, $c_{\texttt{MF}}=79.2$) for $\alpha=0.5$ and ($c_{\texttt{PH}}=124.5$, $c_{\texttt{FF}}=77.4$, $c_{\texttt{MF}}=74.9$) for $\alpha=0.1$, respectively.
For the large scenario, they are ($c_{\texttt{PH}}=157.7$, $c_{\texttt{FF}}=77.1$, $c_{\texttt{MF}}=73.6$) for $\alpha=0.5$ and ($c_{\texttt{PH}}=177.9$, $c_{\texttt{FF}}=72.1$, $c_{\texttt{MF}}=69.7$) for $\alpha=0.1$, respectively.
For the second setup, the thresholding  values are the same for both scenarios with ($c_{\texttt{PH}}=140.1$, $c_{\texttt{FF}}=79.7$, $c_{\texttt{MF}}=76.3$) for $\alpha=0.5$ and ($c_{\texttt{PH}}=165.5$, $c_{\texttt{FF}}=74.4$, $c_{\texttt{MF}}=71.4$) for $\alpha=0.1$. The last row  estimates $P(\texttt{PH\_final}>c_{\texttt{PH}}, \texttt{FF}<c_{\texttt{FF}}, \texttt{MF}<c_{\texttt{MF}})$ using the same thresholding values as for the second setup. }
\label{tab:Probs}
\end{table*}
The difference between the first and second setup is due to the different values of $c_{\texttt{PH}}$, $c_{\texttt{FF}}$, and $c_{\texttt{MF}}$ that are chosen. In the first setup, these values are more extreme as they are sampled within the two scenarios of small and large. Furthermore, for the small and large scenario, two different sets of sampled values are used. On the other hand, for the second setup, the same conditioning values are used for both scenarios, resulting in a bigger difference between the small and the large scenario. The exact values utilized are given in the caption of \autoref{tab:Probs}.

For the second setup, we also obtain the empirical unconditional joint probability  $P(\texttt{PH\_final}>c_{\texttt{PH}}, \texttt{FF}<c_{\texttt{FF}}, \texttt{MF}<c_{\texttt{MF}})$  using the empirical probabilities based on the observed values for \texttt{PH\_final}, \texttt{FF}, and \texttt{MF} as thresholds. This unconditional probability lies between the conditional probabilities of the small and the large scenarios, showing a behavior that is less extreme than the two extreme scenarios.

\subsection{Conditional Kendall's \texorpdfstring{$\tau$ } oof (\texttt{EV\_V6}, \texttt{PH\_V6} $\vert$ \texttt{EV\_V4}, \texttt{PH\_V4})} \label{Sec:Kendall}

In a final analysis, we are interested in how the growth stage V4 influences the later growth stage V6. This means we study the dependence between early vigor and plant height at the growth stage V6, given different values at the growth stage V4.
Hence, we want to evaluate how the conditional Kendall's $\tau$ of (\texttt{EV\_V6}, \texttt{PH\_V6} $\mid$ \texttt{EV\_V4}, \texttt{PH\_V4}) changes as the values of \texttt{EV\_V4} and \texttt{PH\_V4} are changing.
For this, we consider a grid of 100 bivariate tuples
$$(\text{\texttt{EV\_V4}}^g, \text{\texttt{PH\_V4}}^h), g=1,\ldots,10,h=1,\ldots,10.$$
The ten grid points $\text{\texttt{EV\_V4}}^g$ are equally spaced in the observed range of \texttt{EV\_V4} and similarly for $\text{\texttt{PH\_V4}}^h$.
They are transformed to the u-scale using the kernel estimated distribution functions of \texttt{EV\_V4} and \texttt{PH\_V4}.

Since we are only interested in the four variables \texttt{EV\_V4}, \texttt{PH\_V4}, \texttt{EV\_V6}, and \texttt{PH\_V6} for this part of the analysis, we consider the corresponding four-dimensional vine (see \autoref{tab:FourDimTree}), which is a subtree of the Dissmann-based seven-dimensional R-vine tree from Section \ref{Sec:UniResp} given in \autoref{fig:RVMMaize}. Note that the density corresponding to the desired conditional distribution is not available without integration. 

For every grid point $(\text{\texttt{EV\_V4}}^g, \text{\texttt{PH\_V4}}^h), g,h=1,\ldots,10$, we perform MCMC sampling on the u-scale 100 times with 8000 iterations, respectively, to obtain samples 
\begin{align*}
    &(\text{\texttt{EV\_V6}}^{g,(r)}_k, \text{\texttt{PH\_V6}}^{h,(r)}_k),\\
    &g,h=1,\ldots,10, r=1,\ldots,8000, k=1,\ldots,100,
\end{align*}
from (\texttt{EV\_V6}, \texttt{PH\_V6} $\mid$ $\text{\texttt{EV\_V4}}^g$, $\text{\texttt{PH\_V4}}^h$).
For each grid point $(g,h)$, we obtain 100 empirical conditional Kendall's $\tau$ estimates
$$\hat{\tau}^k_{\text{\texttt{EV\_V6}, \texttt{PH\_V6} $\mid$ \texttt{EV\_V4}$^g$, \texttt{PH\_V4}$^h$}}, k=1,\ldots,100,$$
of the MCMC samples $(\text{\texttt{EV\_V6}}^{g,(r)}_k, \text{\texttt{PH\_V6}}^{h,(r)}_k), r=1,\ldots,8000$.
From these 100 values, we can determine the average and the empirical $5$\% and $95$\% quantiles of the conditional Kendall's $\tau$ values.

\begin{figure*}
         \includegraphics[width=0.5\linewidth]{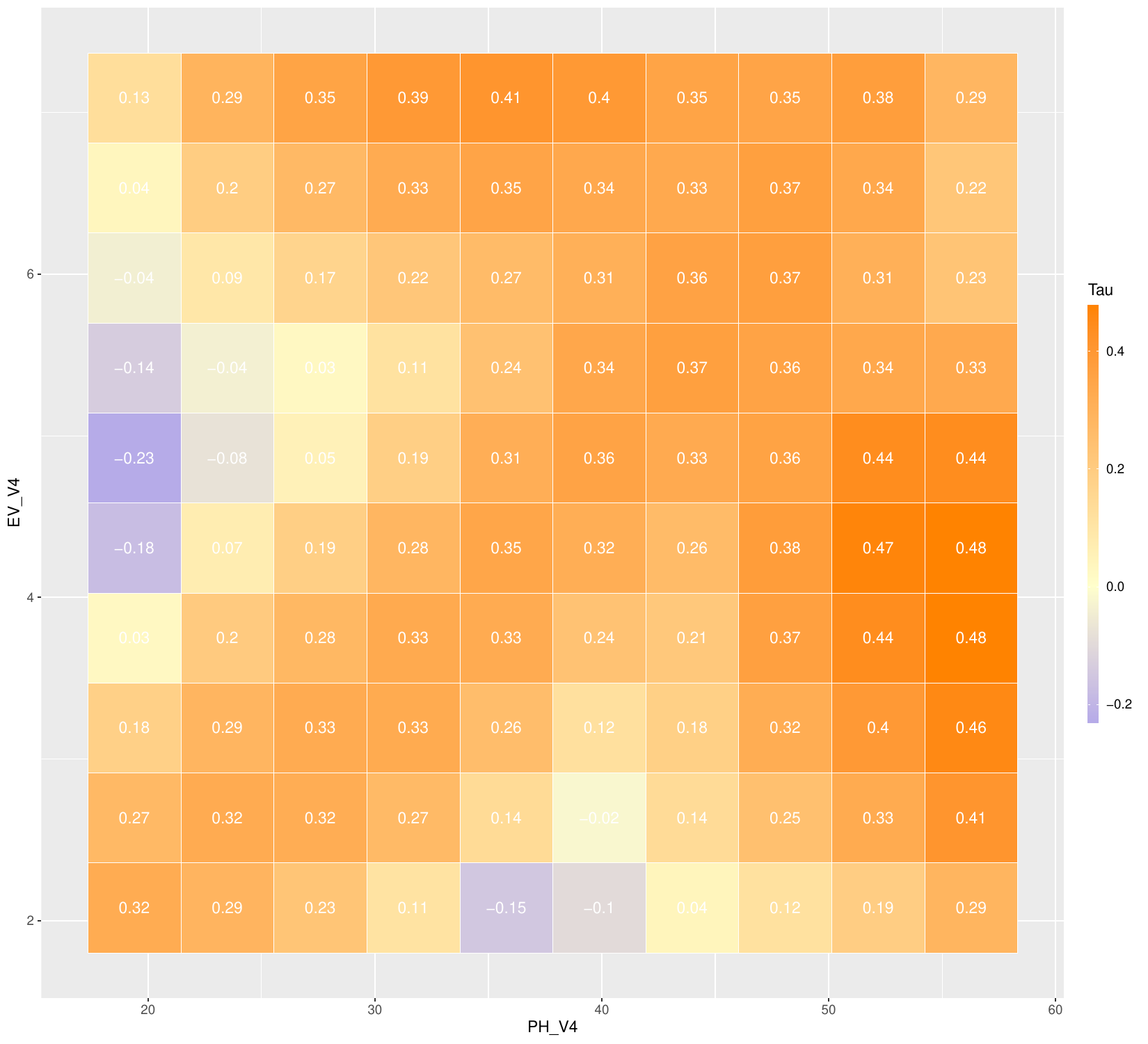}
         \includegraphics[width=0.5\linewidth]{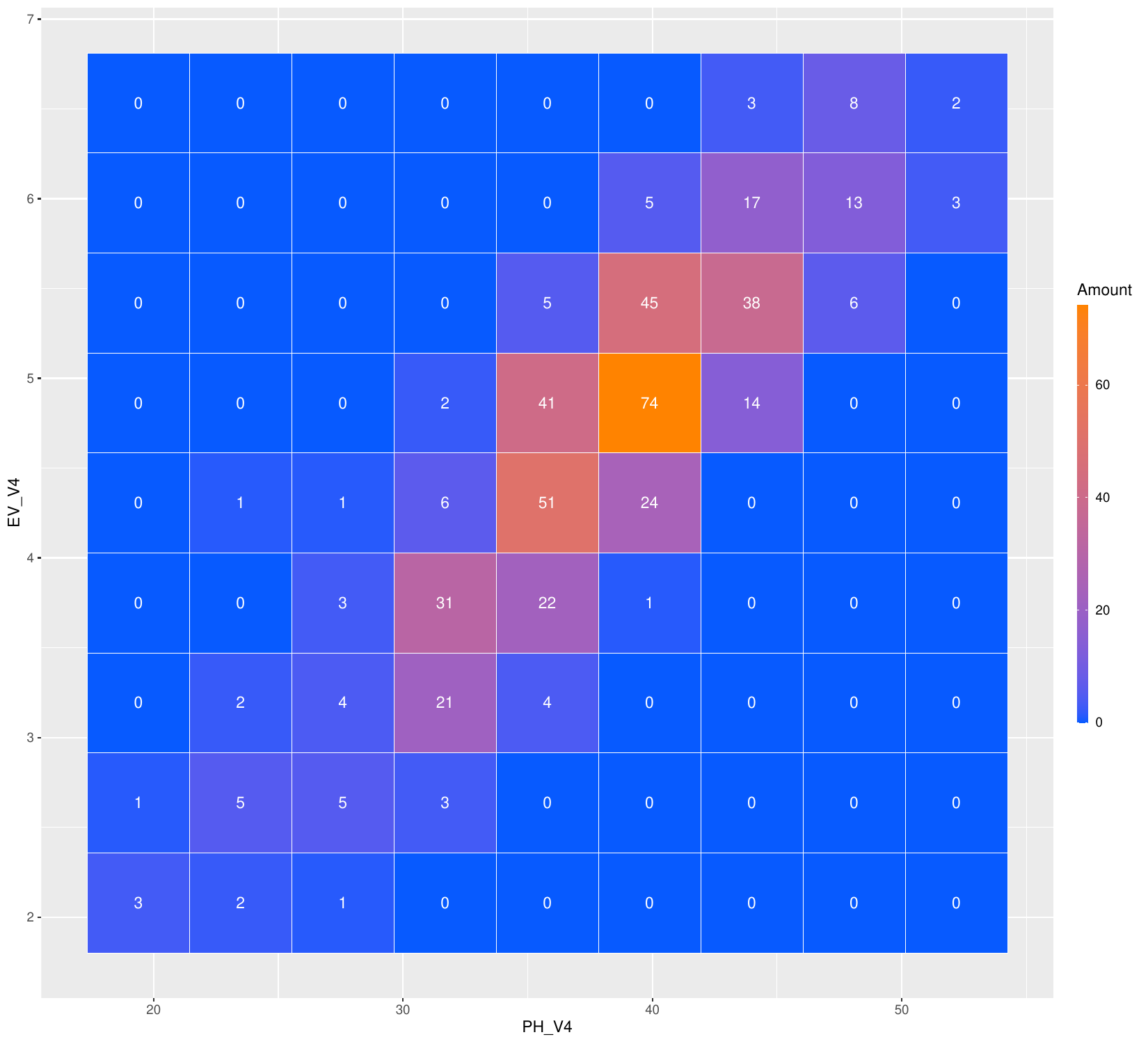}
        \caption{\textbf{Left:} The average Kendall's $\tau$ values of (\texttt{EV\_V6}, \texttt{PH\_V6} $\mid$ \texttt{EV\_V4}, \texttt{PH\_V4}) for different grid points of \texttt{EV\_V4} and \texttt{PH\_V4}. \textbf{Right:} The amount of data points for different grid points of \texttt{\texttt{EV\_V4}} and \texttt{\texttt{PH\_V4}}.}
     \label{fig:Tiles}
\end{figure*}

The left side of \autoref{fig:Tiles} shows the estimated average conditional Kendall's $\tau$ value for each grid point.
The conditional Kendall's $\tau$ values range from $-0.23$ to $0.48$.
The corresponding $5$\% and $95$\% quantiles show a very similar behavior and are not given here.

The right side of \autoref{fig:Tiles} shows the number of data points falling into
\begin{align*}
    &(\text{\texttt{EV\_V4}}^g,\text{\texttt{EV\_V4}}^{g+1}) \times (\text{\texttt{PH\_V4}}^h,\text{\texttt{PH\_V4}}^{h+1}),\\
    &g=1,\ldots,9,h=1,\ldots,9.
\end{align*}
It is clear that most of the data lie on the anti-diagonal of the $10\times 10$ matrix in the right panel of \autoref{fig:Tiles}, where height and vigor at the growth stage V4 are both small, medium, or high. Hence, the findings can be separated into scenarios that are more or less observed in the data.

\begin{itemize}
    \item \textbf{Similar values for \texttt{PH\_V4} and \texttt{EV\_V4} (Most often observed in the data)}: The scenario, which is mostly observed in the data (see right side of \autoref{fig:Tiles}), is the case of being on the anti-diagonal part of the matrix. Here, a positive conditional Kendall's $\tau \in (0.3,0.37)$ between \texttt{PH\_V6} and \texttt{EV\_V6} is given, showing no big changes for the different conditioning values of \texttt{PH\_V4} and \texttt{EV\_V4} . The positive values of $\tau$ mean that these plants are vigorous (non-vigorous) at stage V6 if they are high (low) at stage V6 and vice versa.
    \item \textbf{Small/medium value of
    \texttt{PH\_V4}/ \texttt{EV\_V4} or medium/small value 
    of \texttt{PH\_V4}/ \texttt{EV\_V4} (Some observations in the data):} In the two cases of (1) having a small value of \texttt{PH\_V4} combined with a medium value of \texttt{EV\_V4} and (2) having a small value of \texttt{EV\_V4} combined with a medium value of \texttt{PH\_V4}, there is a negative conditional Kendall's $\tau$ between \texttt{PH\_V6} and \texttt{EV\_V6}. This means, for example, that for small plants with medium vigor at stage V4, being high at stage V6 corresponds to not being vigorous at stage V6. This is intuitive, as these plants must use their energy to grow.
    \item \textbf{Large/medium  value of \texttt{PH\_V4}/ \texttt{EV\_V4} or  medium/large value of \texttt{PH\_V4}/ \texttt{EV\_V4} (Some observations in the data):} On the other hand, for the two cases of (1) having a large value of \texttt{PH\_V4} combined with a medium value of \texttt{EV\_V4} and of (2) having a large value of \texttt{EV\_V4} combined with a medium value of \texttt{PH\_V4}, the strongest positive conditional Kendall's $\tau > 0.4$ between \texttt{PH\_V6} and \texttt{EV\_V6} is observed. Hence, for example, very high plants with medium vigor at stage V4 will be vigorous if they are also high at stage V6. They do not have to grow so much anymore, so they can use the energy to become more vigorous.
    \item \textbf{Large/small  value of \texttt{PH\_V4}/ \texttt{EV\_V4} or  small/large value of \texttt{PH\_V4}/ \texttt{EV\_V4} (No observations in the data):} The scenarios of (1) being very high but very weak and (2) being very strong but very low at stage V4 are not interesting as these plants rarely existed in the data set.
\end{itemize}

\section{Discussion and Conclusion}\label{Sec:Conc}
We propose a straightforward way of sampling from any conditional distribution of a simplified vine with an arbitrary vine tree structure using the programming language \texttt{Stan}.
Extensive simulation studies show very good results of this sampling algorithm for univariate and bivariate conditional distributions.
Since Markov Chain Monte Carlo sampling provides information on the full conditional distribution, we can obtain further quantities of interest, like the mean and median, credible intervals, and conditional dependence measures like Kendall's $\tau$.

Our data application, using phenotypic traits of maize based on simplified vine models, which have been tested for the simplifying assumption, points to several possible application areas of our sampling approach.
When sampling from a univariate conditional distribution and obtaining the corresponding point estimate, i.e., predicting a response given the other variables, our algorithm performs similarly to the existing D-vine-based approaches of \texttt{vinereg} and \texttt{sparsevinereg}.
For multivariate responses, the associated conditional density cannot be obtained without integration for arbitrary vine copula based regression models. However, our approach allows us to sample from the desired conditional distribution. If several conditional distributions are of interest, our approach gives consistent answers since it is based on a single underlying joint model. We harvest this ability in our application.

For example, we show conditional dependence for the three responses of interest (final plant height, female flowering time, and male flowering time), given the maize plant's height and vigour at earlier growth stages by stress testing. Moreover, we can estimate conditional Kendall's $\tau$ values where the dimension of the conditioning variable is higher than one, including confidence intervals. This quantifies how the conditional dependence between plant height and vigor at a later growth stage (V6) changes, conditioned on these values at an earlier growth stage (V4).

While our current applications include a prediction of univariate and trivariate responses and estimation of conditional Kendall's $\tau$ in the analysis of maize traits, further interesting applications of our proposal are classification and learning graphical structures with conditional independence tests. Other potential application domains are finance and climatology, e.g., through stress testing.

Moreover, our real data analysis shows the importance of tailoring the vine structure learning to conditional distribution estimation. This has also been noted in \cite{zhu2021simplified}. For the univariate response case, we propose a simple heuristic. Still, selecting flexible vine structures to estimate multivariate conditional distributions is a future research area to extend the heuristics of \cite{chang2019prediction} and \cite{zhu2021simplified} to a conditional copula likelihood based one. 

Finally, since most real data sets include discrete or mixed discrete-continuous variables, developing a sampling approach for such cases with vines is on the agenda. 

As a final remark, we note that the proposed conditional sampling approach can be utilized for specified joint densities, which are not necessarily vine copula based.

\backmatter

\begin{appendices}
\section{Simulation Study}

\subsection{Transformation One-Dimensional Case}\label{Sec:TransOneDim}

To avoid issues at the boundaries \cite{naglerNonparametricEstimationSimplified2017} of the support $[0,1]$ of
$c_{U_{\mathcal{C}_1} \vert \mathbf{U}_{\mathcal{C}_2}}(\cdot \vert \mathbf{u}_{k_{\alpha}}^{\mathcal{C}_2})$,
we transform it to
$f_{Z_{\mathcal{C}_1} \vert \mathbf{Z}_{\mathcal{C}_2}}(\cdot \vert \mathbf{z}_{k_{\alpha}}^{\mathcal{C}_2})$
on $(-\infty,\infty)$.
Here, $Z_{\mathcal{C}_1}=\Phi^{-1}(U_{\mathcal{C}_1})$,
$Z_{\mathcal{C}_{2,i}}=\Phi^{-1}(U_{\mathcal{C}_{2,i}})$, and
$z_{k_{\alpha}}^{\mathcal{C}_{2,i}}=\Phi^{-1}(u_{k_{\alpha}}^{\mathcal{C}_{2,i}}), i=1,\ldots,\ell$.
The density $c_{U_{\mathcal{C}_1} \vert \mathbf{U}_{\mathcal{C}_2}}(\cdot \vert \mathbf{u}_{k_{\alpha}}^{\mathcal{C}_2})$ is then given by
\begin{equation} \label{eq:transformationzscale}
    c_{U_{\mathcal{C}_1} \vert \mathbf{U}_{\mathcal{C}_2}}(u \vert \mathbf{u}_{k_{\alpha}}^{\mathcal{C}_2})=
    \frac{f_{Z_{\mathcal{C}_1} \vert \mathbf{Z}_{\mathcal{C}_2}}(\Phi^{-1}(u) \vert \mathbf{z}_{k_{\alpha}}^{\mathcal{C}_2})}{\phi(\Phi^{-1}(u))}.
\end{equation}
To obtain a kernel density estimate
$\hat{f}_{Z_{\mathcal{C}_1} \vert \mathbf{Z}_{\mathcal{C}_2}}(\cdot \vert \mathbf{z}_{k_{\alpha}}^{\mathcal{C}_2})$
of
$f_{Z_{\mathcal{C}_1} \vert \mathbf{Z}_{\mathcal{C}_2}}(\cdot \vert \mathbf{z}_{k_{\alpha}}^{\mathcal{C}_2})$ on $(-\infty,\infty)$,
we need to obtain samples $z^{(r)}(\mathbf{u}_{k_{\alpha}}^{\mathcal{C}_2})$ of the corresponding distribution function
$F_{Z_{\mathcal{C}_1} \vert \mathbf{Z}_{\mathcal{C}_2}}(\cdot \vert \mathbf{z}_{k_{\alpha}}^{\mathcal{C}_2})$.
Using the MCMC samples $u^{(r)}(\mathbf{u}_{k_{\alpha}}^{\mathcal{C}_2})$ of
$C_{U_{\mathcal{C}_1} \vert \mathbf{U}_{\mathcal{C}_2}}(\cdot \vert \mathbf{u}_{k_{\alpha}}^{\mathcal{C}_2})$,
we obtain the desired samples using the probability integral transform and \cite[Eq. (3.4)]{krausDVineCopulaBased2017} by
{\tiny
\begin{align*}
    &z^{(r)}(\mathbf{u}_{k_{\alpha}}^{\mathcal{C}_2})=
    F^{-1}_{Z_{\mathcal{C}_1} \vert \mathbf{Z}_{\mathcal{C}_2}}\left(C_{U_{\mathcal{C}_1} \vert \mathbf{U}_{\mathcal{C}_2}}\left(u^{(r)}(\mathbf{u}_{k_{\alpha}}^{\mathcal{C}_2}) \vert \mathbf{u}_{k_{\alpha}}^{\mathcal{C}_2}\right) \Big\vert \mathbf{z}_{k_{\alpha}}^{\mathcal{C}_2}\right)\\
    &=
    \Phi^{-1}\left(C^{-1}_{U_{\mathcal{C}_1} \vert \mathbf{U}_{\mathcal{C}_2}}\left(C_{U_{\mathcal{C}_1} \vert \mathbf{U}_{\mathcal{C}_2}}\left(u^{(r)}(\mathbf{u}_{k_{\alpha}}^{\mathcal{C}_2}) \vert \mathbf{u}_{k_{\alpha}}^{\mathcal{C}_2} \right) \Big\vert \mathbf{u}_{k_{\alpha}}^{\mathcal{C}_2} \right)\right)\\
    &=\Phi^{-1}\left(u^{(r)}(\mathbf{u}_{k_{\alpha}}^{\mathcal{C}_2}) \right).
\end{align*}}
Then, \autoref{eq:transformationzscale} can be used to transform $\hat{f}_{Z_{\mathcal{C}_1} \vert \mathbf{Z}_{\mathcal{C}_2}}(\cdot \vert \mathbf{z}_{k_{\alpha}}^{\mathcal{C}_2})$ back to $[0,1]$ to obtain an estimated $\hat{c}_{U_{\mathcal{C}_1} \vert \mathbf{U}_{\mathcal{C}_2}}(\cdot \vert \mathbf{u}_{k_{\alpha}}^{\mathcal{C}_2})$.

\subsection{The Vine Specifications of Simulation Setup 6 - Case 1}

\begin{figure*}
     \centering
     \begin{subfigure}[b]{0.4\textwidth}
         \centering
         \includegraphics[width=\linewidth]{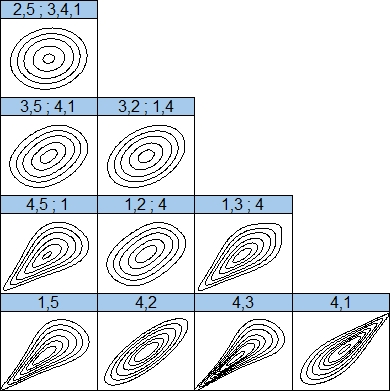}
     \end{subfigure}
     \hfill
     \begin{subfigure}[b]{0.4\textwidth}
         \centering
         \includegraphics[width=\linewidth]{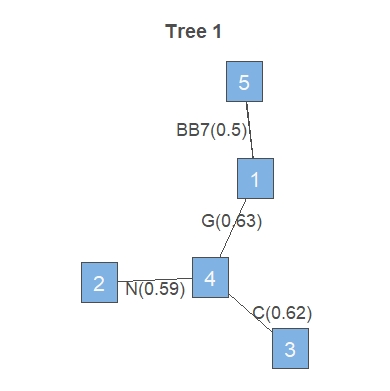}
     \end{subfigure}
     
     \medskip
     \begin{subfigure}[b]{0.3\textwidth}
         \centering
         \includegraphics[width=\linewidth]{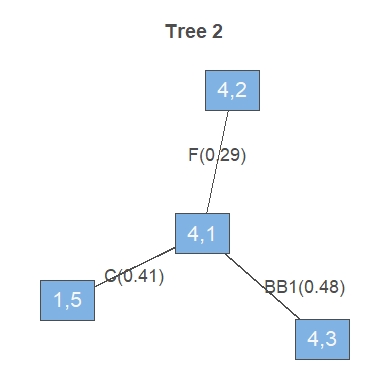}
     \end{subfigure}
     \hfill
     \begin{subfigure}[b]{0.3\textwidth}
         \centering
         \includegraphics[width=\linewidth]{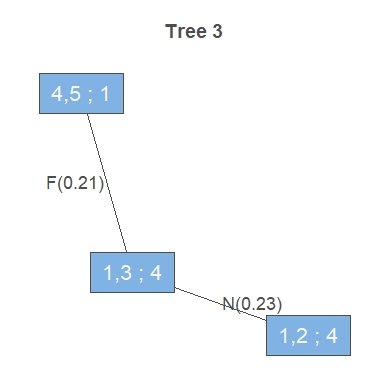}
     \end{subfigure}
     \hfill
     \begin{subfigure}[b]{0.3\textwidth}
         \centering
         \includegraphics[width=\linewidth]{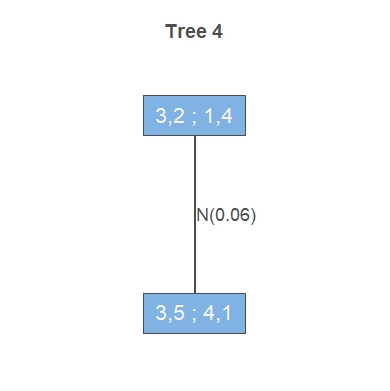}
     \end{subfigure}    
    \caption{Marginal normalized pair contour plots (top) and R-vine tree structure plots of the chosen R-vine specification in simulation setup 6 together with chosen pair copula families (N=Gauss, C=Clayton, F=Frank, G=Gumbel, BB1=biparameter bivariate 1, BB7=biparameter bivariate 7) and their associated Kendall's $\tau$ in parentheses.}
    \label{fig:univarsetup6fampar}
\end{figure*}

These vine specifications are given in \autoref{fig:univarsetup6fampar}.

\subsection{Transformation Two-Dimensional Case} \label{Sec:TransTwoDim}

For this, we transform
$c_{\mathbf{U}_{\mathcal{C}_1} \vert \mathbf{U}_{\mathcal{C}_2}}(\cdot \vert \mathbf{u}_{k_{\alpha}}^{\mathcal{C}_2})$ on $[0,1]^2$
to $f_{\mathbf{Z}_{\mathcal{C}_1} \vert \mathbf{Z}_{\mathcal{C}_2}}(\cdot \vert \mathbf{z}_{k_{\alpha}}^{\mathcal{C}_2})$,
where
$Z_{\mathcal{C}_{1,1}}=\Phi^{-1}(U_{\mathcal{C}_{1,1}})$,$Z_{\mathcal{C}_{1,2}}=\Phi^{-1}(U_{\mathcal{C}_{1,2}})$, $Z_{\mathcal{C}_{2,i}}=\Phi^{-1}(U_{\mathcal{C}_{2,i}})$, and $z_{k_{\alpha}}^{\mathcal{C}_{2,i}}=\Phi^{-1}(u_{k_{\alpha}}^{\mathcal{C}_{2,i}}), i=1,\ldots,\ell$.
This leads to
\begin{align}
    \label{eq:transformationzscalemulti}
    &c_{\mathbf{U}_{\mathcal{C}_1} \vert \mathbf{U}_{\mathcal{C}_2}}(u_1, u_2 \vert \mathbf{u}_{k_{\alpha}}^{\mathcal{C}_2})\nonumber\\
    &:= \frac{f_{\mathbf{Z}_{\mathcal{C}_1} \vert \mathbf{Z}_{\mathcal{C}_2}}(\Phi^{-1}(u_1), \Phi^{-1}(u_2) \vert \mathbf{z}_{k_{\alpha}}^{\mathcal{C}_2})}{\phi(\Phi^{-1}(u_1)) \phi(\Phi^{-1}(u_2))}.
\end{align}
In order to obtain samples
$\mathbf{z}^{(r)}(\mathbf{u}_{k_{\alpha}}^{\mathcal{C}_2})$
from
$F_{\mathbf{Z}_{\mathcal{C}_1} \vert \mathbf{Z}_{\mathcal{C}_2}}(\cdot \vert \mathbf{z}_{k_{\alpha}}^{\mathcal{C}_2})$,
we use the given MCMC samples
$(u^{(r)}_{1}(\mathbf{u}_{k_{\alpha}}^{\mathcal{C}_2}), u^{(r)}_{2}(\mathbf{u}_{k_{\alpha}}^{\mathcal{C}_2}))$
from
$C_{\mathbf{U}_{\mathcal{C}_1} \vert \mathbf{U}_{\mathcal{C}_2}}(\cdot \vert \mathbf{u}_{k_{\alpha}}^{\mathcal{C}_2})$.
Using the inverse Rosenblatt transform, the probability integral transform, and Eq. (3.4) in \cite {krausDVineCopulaBased2017}, we obtain
{\tiny
\begin{align*}
    &z^{(r)}_{2}(\mathbf{u}_{k_{\alpha}}^{\mathcal{C}_2}) =
    F^{-1}_{Z_{\mathcal{C}_{12}} \vert \mathbf{Z}_{\mathcal{C}_2}}\left(C_{U_{\mathcal{C}_{12}} \vert \mathbf{U}_{\mathcal{C}_2}}\left(u^{(r)}_{2}\left(\mathbf{u}_{k_{\alpha}}^{\mathcal{C}_2}\right) \vert \mathbf{u}_{k_{\alpha}}^{\mathcal{C}_2}\right) \Big\vert \mathbf{z}_{k_{\alpha}}^{\mathcal{C}_2}\right)\\
    &=\Phi^{-1}\left(C^{-1}_{U_{\mathcal{C}_{12}} \vert \mathbf{U}_{\mathcal{C}_2}}\left(C_{U_{\mathcal{C}_{12}} \vert \mathbf{U}_{\mathcal{C}_2}}\left(u^{(r)}_{2}\left(\mathbf{u}_{k_{\alpha}}^{\mathcal{C}_2}\right) \vert \mathbf{u}_{k_{\alpha}}^{\mathcal{C}_2} \right) \Big\vert \mathbf{u}_{k_{\alpha}}^{\mathcal{C}_2}\right)\right)\\
    &=\Phi^{-1}\left(u^{(r)}_{2}(\mathbf{u}_{k_{\alpha}}^{\mathcal{C}_2})\right),\\
    &z^{(r)}_{1}(\mathbf{u}_{k_{\alpha}}^{\mathcal{C}_2})= F^{-1}_{Z_{\mathcal{C}_{11}} \vert (Z_{\mathcal{C}_{12}},\mathbf{Z}_{\mathcal{C}_2})}\\
    & \left(C_{U_{\mathcal{C}_{11}} \vert (U_{\mathcal{C}_{12}},\mathbf{U}_{\mathcal{C}_2})}\left(u^{(r)}_{1}(\mathbf{u}_{k_{\alpha}}^{\mathcal{C}_2}) \vert u^{(r)}_{2}(\mathbf{u}_{k_{\alpha}}^{\mathcal{C}_2}),\mathbf{u}_{k_{\alpha}}^{\mathcal{C}_2}\right)\Big\vert z^{(r)}_{2}(\mathbf{u}_{k_{\alpha}}^{\mathcal{C}_2}),\mathbf{z}_{k_{\alpha}}^{\mathcal{C}_2}\right)\\
    &=\Phi^{-1}\left(u^{(r)}_{1}(\mathbf{u}_{k_{\alpha}}^{\mathcal{C}_2})\right).
\end{align*}}
Using these samples, we obtain an estimated $\hat{f}_{\mathbf{Z}_{\mathcal{C}_1} \vert \mathbf{Z}_{\mathcal{C}_2}}(\cdot \vert \mathbf{z}_{k_{\alpha}}^{\mathcal{C}_2})$ and with \autoref{eq:transformationzscalemulti} an estimated $\hat{c}_{\mathbf{U}_{\mathcal{C}_1} \vert \mathbf{U}_{\mathcal{C}_2}}(\cdot \vert \mathbf{u}_{k_{\alpha}}^{\mathcal{C}_2})$.

\subsection{The Vine Specifications of Simulation Setup 3 - Case 2}

\begin{figure*}
     \centering
     \begin{subfigure}[b]{0.4\textwidth}
         \centering
         \includegraphics[width=\linewidth]{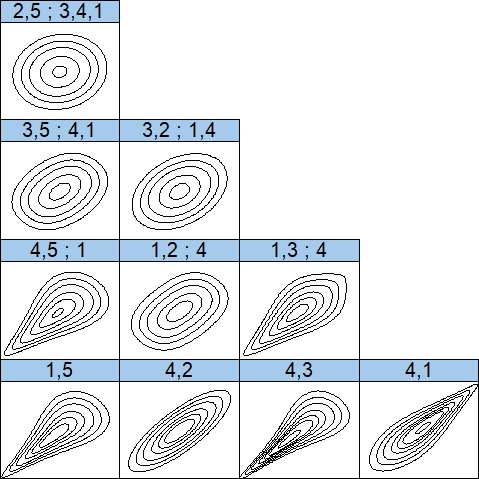}
     \end{subfigure}
     \hfill
     \begin{subfigure}[b]{0.4\textwidth}
         \centering
         \includegraphics[width=\linewidth]{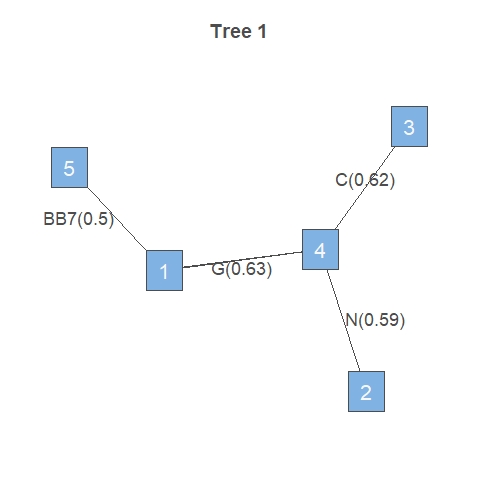}
     \end{subfigure}
     
     \medskip
     \begin{subfigure}[b]{0.3\textwidth}
         \centering
         \includegraphics[width=\linewidth]{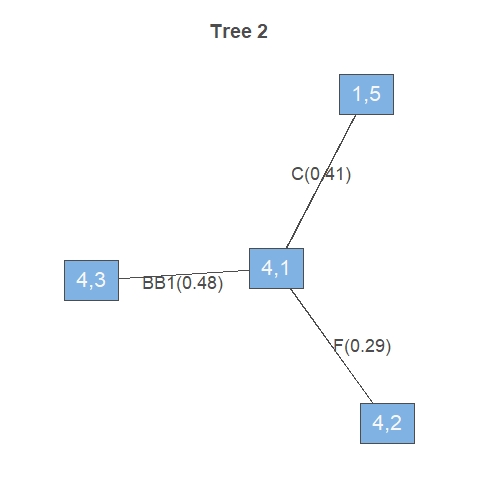}
     \end{subfigure}
     \hfill
     \begin{subfigure}[b]{0.3\textwidth}
         \centering
         \includegraphics[width=\linewidth]{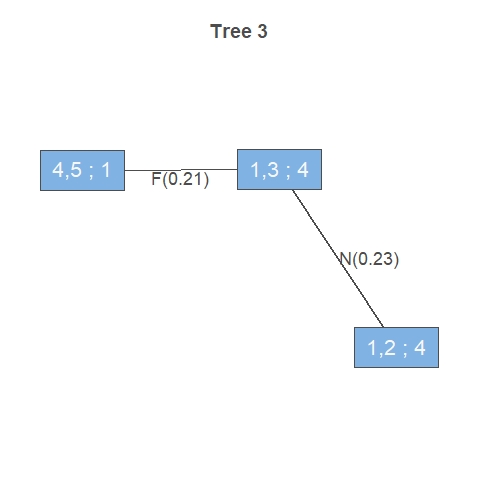}
     \end{subfigure}
     \hfill
     \begin{subfigure}[b]{0.3\textwidth}
         \centering
         \includegraphics[width=\linewidth]{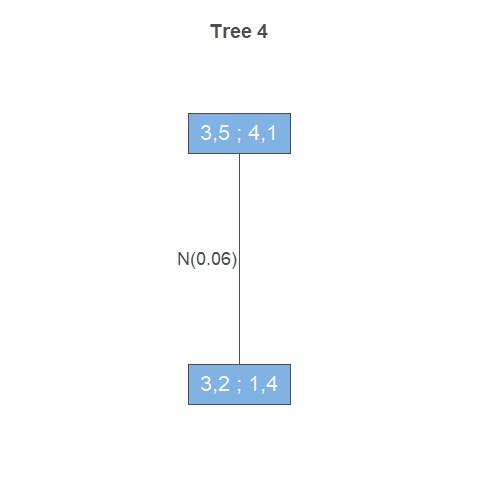}
     \end{subfigure}    
    \caption{Marginal normalized pair contour plots (top) and R-vine tree structure plots of the chosen R-vine specification in the bivariate simulation setup 3.}
    \label{fig:bivarsetup3fampar}
\end{figure*}

These vine specifications are given in \autoref{fig:bivarsetup3fampar}.

\section{Maize Data Set}

\subsection{Exploratory Data Analysis of the Maize Data Set}

\begin{figure*}
     \centering
     \begin{subfigure}[b]{0.49\textwidth}
         \centering
         \includegraphics[width=\linewidth]{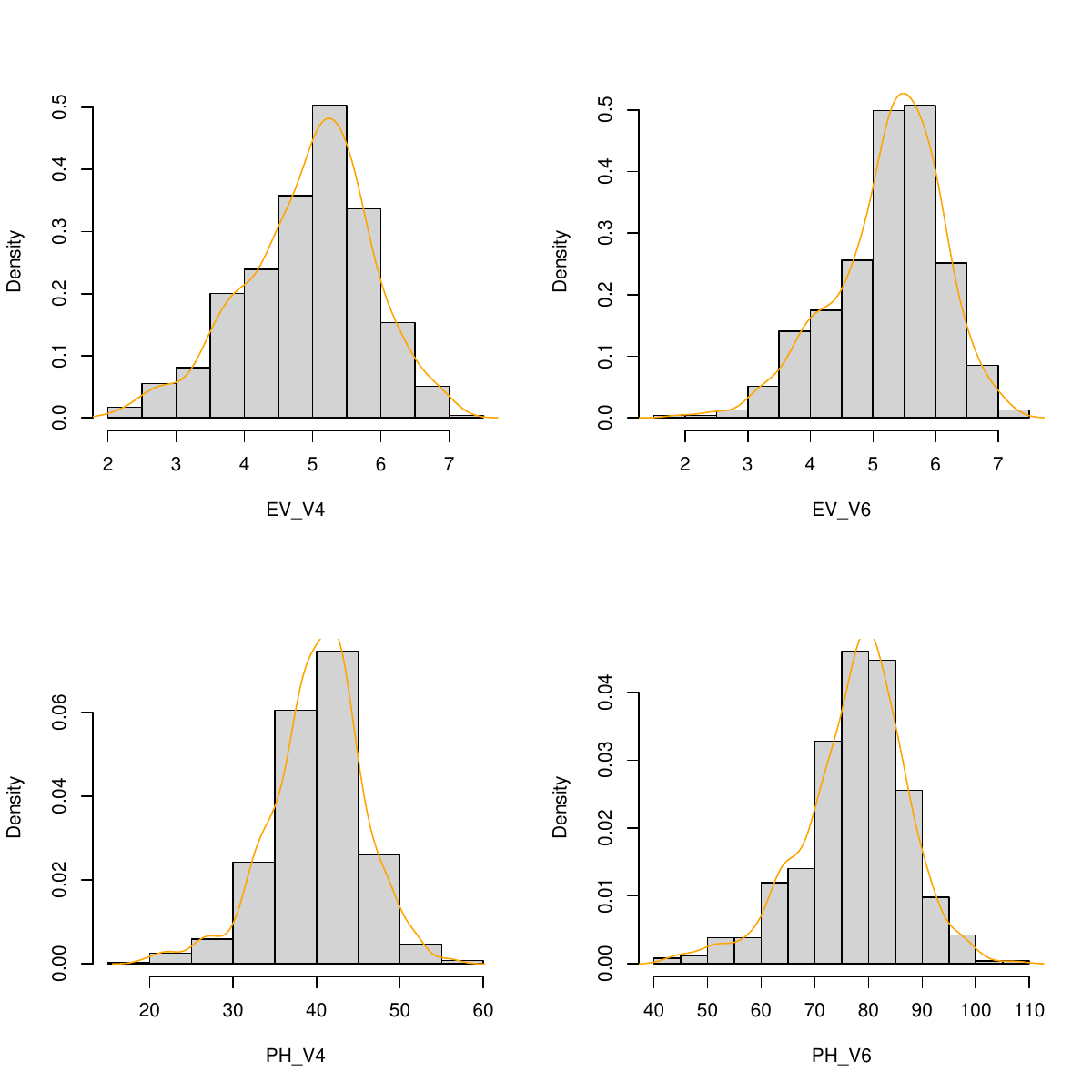}
     \end{subfigure}
     \hfill
     \begin{subfigure}[b]{0.49\textwidth}
         \centering
         \includegraphics[width=\linewidth]{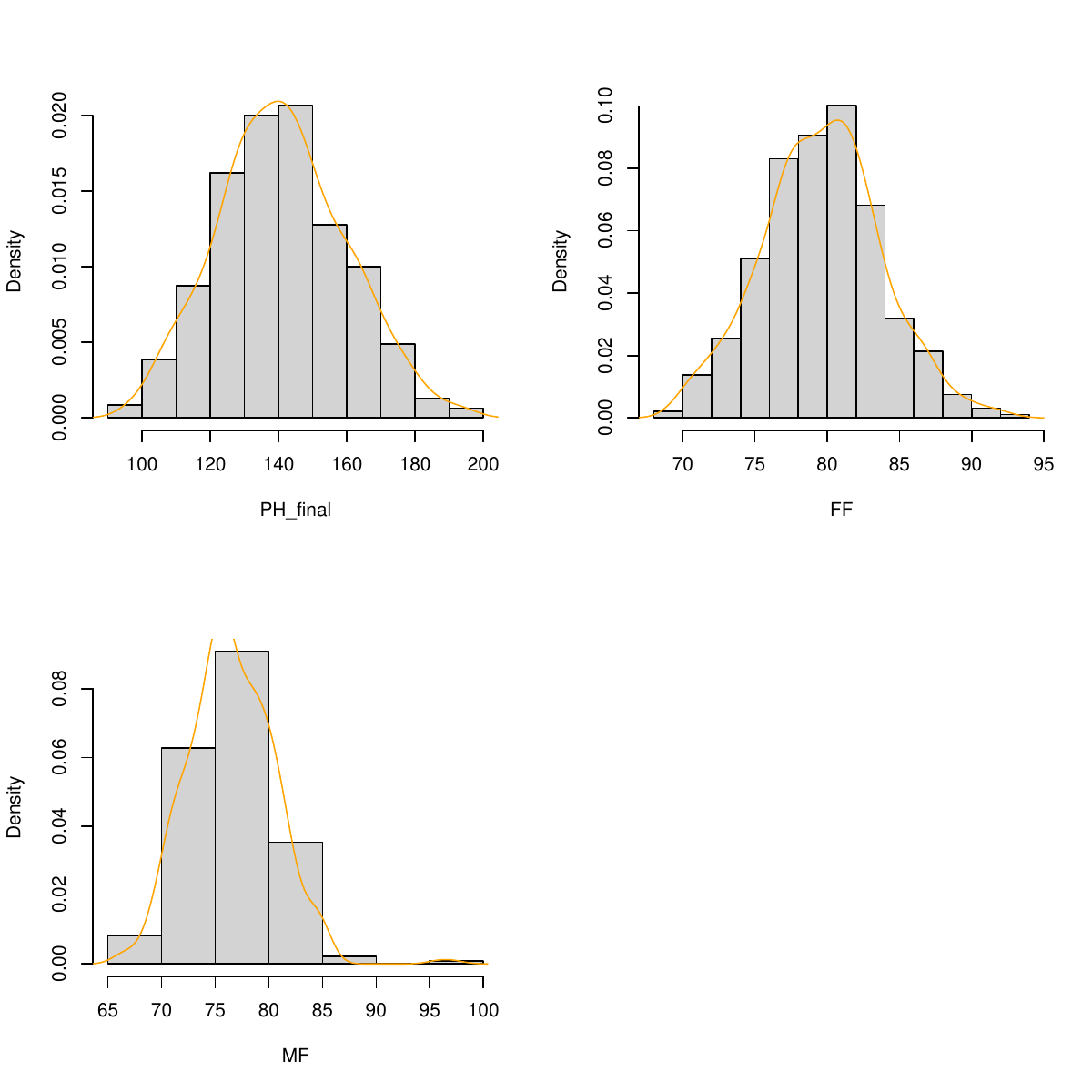}
     \end{subfigure}
        \caption{Histogram of the traits (variables) in Section \ref{Sec:Data}, where the orange curve corresponds to the associated univariate kernel density estimate.}
     \label{fig:datahist}
\end{figure*}

The histogram of the traits is given in \autoref{fig:datahist}.

\subsection{Summaries of 7-dimensional R-vine obtained with Dissmann Algorithm}

\begin{figure*}
         \centering
         \includegraphics[width=.8\linewidth]{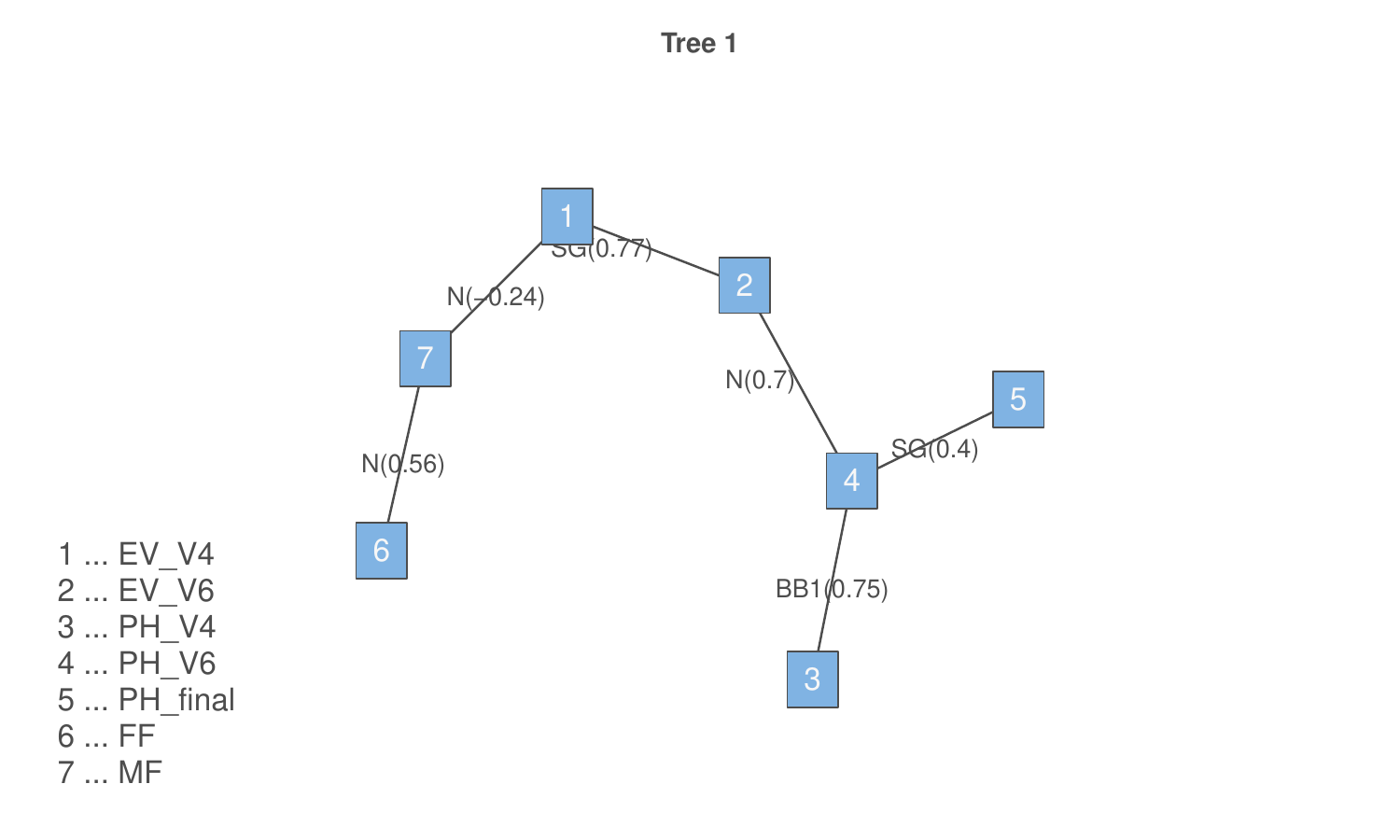}
     \vfill
         \centering
         \includegraphics[width=.8\linewidth]{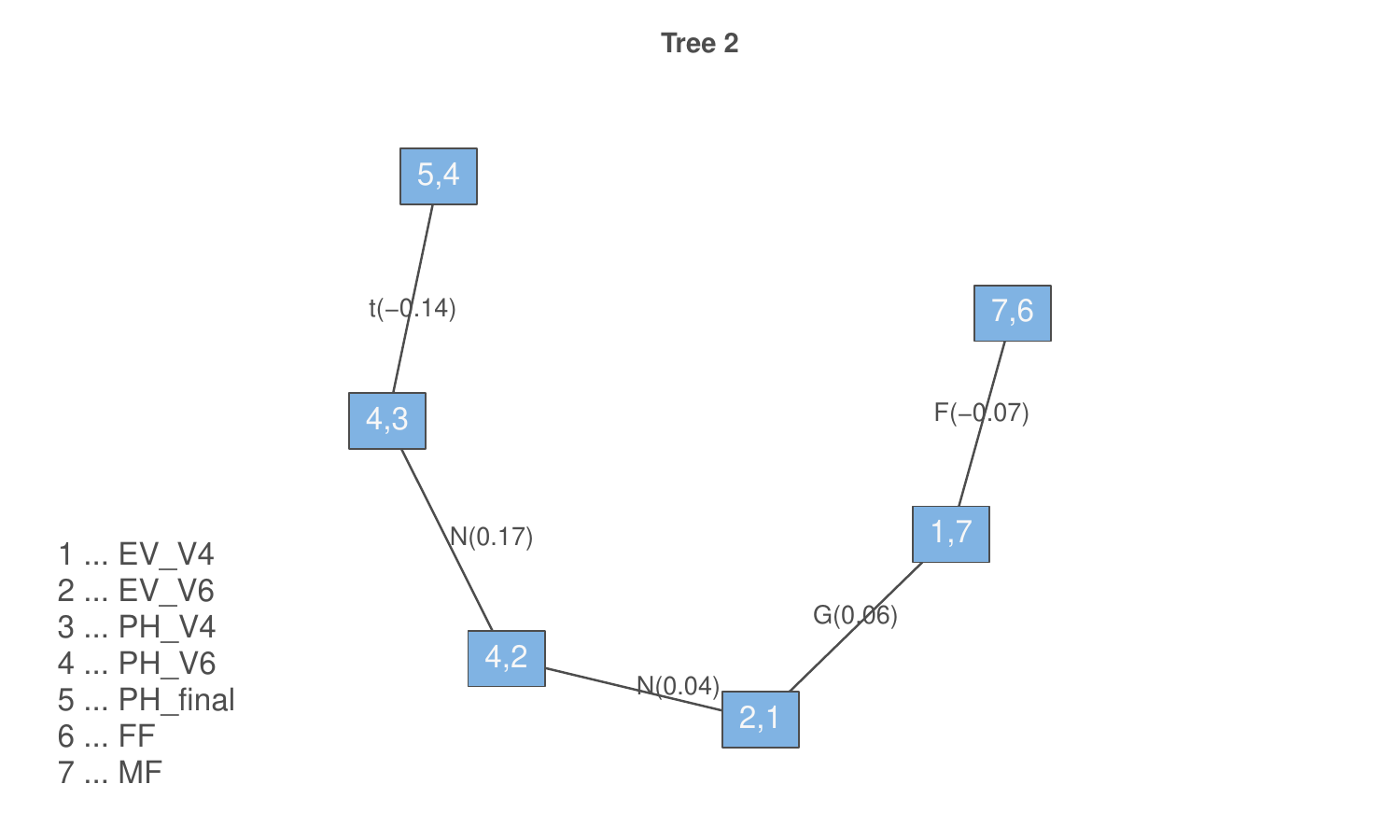}
        \caption{The first (top) and second (bottom) tree of the vine model obtained with the Dissmann algorithm. A letter at an edge with numbers inside the parenthesis refers to its bivariate copula family with its Kendall's $\hat{\tau}$.}
     \label{fig:RVMMaize}
\end{figure*}

The summaries of this R-vine are given in \autoref{fig:RVMMaize} and \autoref{tab:dissmann}.

\subsection{Comparison of True and Estimated Densities}

\begin{figure*}
    \centering
    \includegraphics[width=0.5\textwidth]{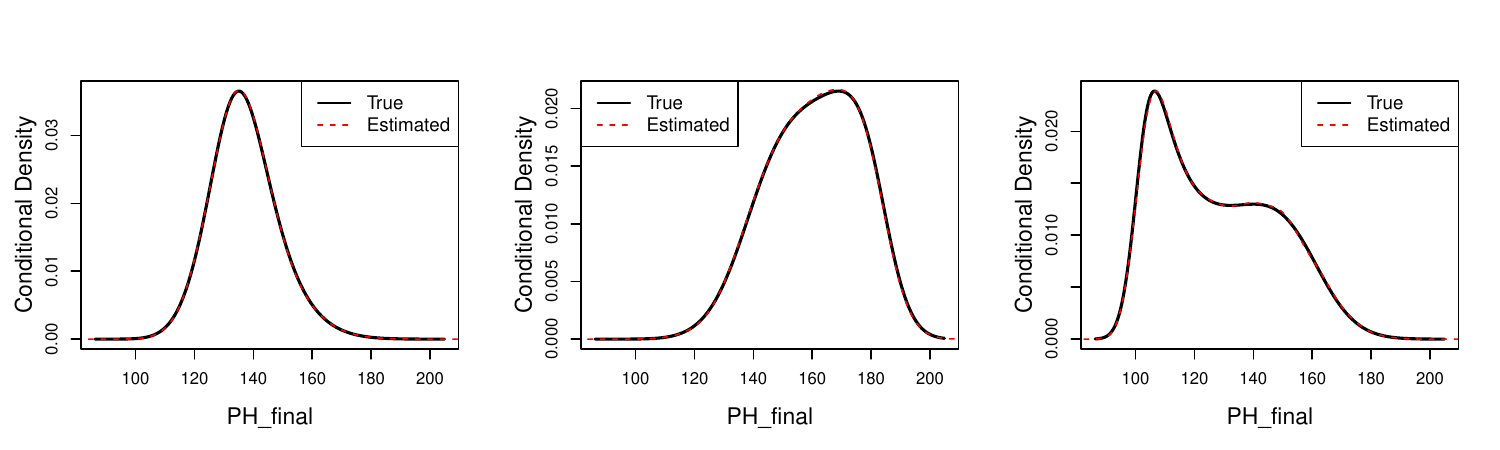}
    \caption{Comparison of the true density with the estimated kernel density using MCMC samples on the original scale in Section \ref{Sec:Data}. The observations are chosen to have the minimum (left), median (middle), and maximum (right) Euclidean distance of the seven traits on the u-scale to $\mathbf{u_c}$. }
  	\label{fig:Plots_OneDim_Comparison}
\end{figure*}

The true and estimated densities are compared in \autoref{fig:Plots_OneDim_Comparison}.

\subsection{Summaries of 7-dimensional R-vine selected by Maximizing (Partial) Correlations}

The summaries of this tree are given in \autoref{fig:RVMMaizeNotLeaf} and \autoref{tab:RVMMaizeNotLeaf}.

\subsection{Summaries of 7-dimensional D-vines obtained by \texttt{vinereg} and \texttt{sparsevinereg}}

These summaries are given in \autoref{tab:datavinereg}, \autoref{tab:vinereg}, \autoref{tab:sparsevinereg}.

\subsection{Summaries of 4-dimensional R-vine obtained with Dissmann Algorithm}

These summaries are given in \autoref{tab:FourDimTree}.

\subsection{Empirical Conditional Quantiles on Copula Scale for Section \ref{Sec:MultiResponse}}

These quantiles are given in \autoref{tab:stressuquant}.

\subsection{Details on Heuristic in Section \ref{Sec:Data}}\label{Sec:Heuristic}
\begin{figure*}
    \centering
    \includegraphics[width=1\linewidth]{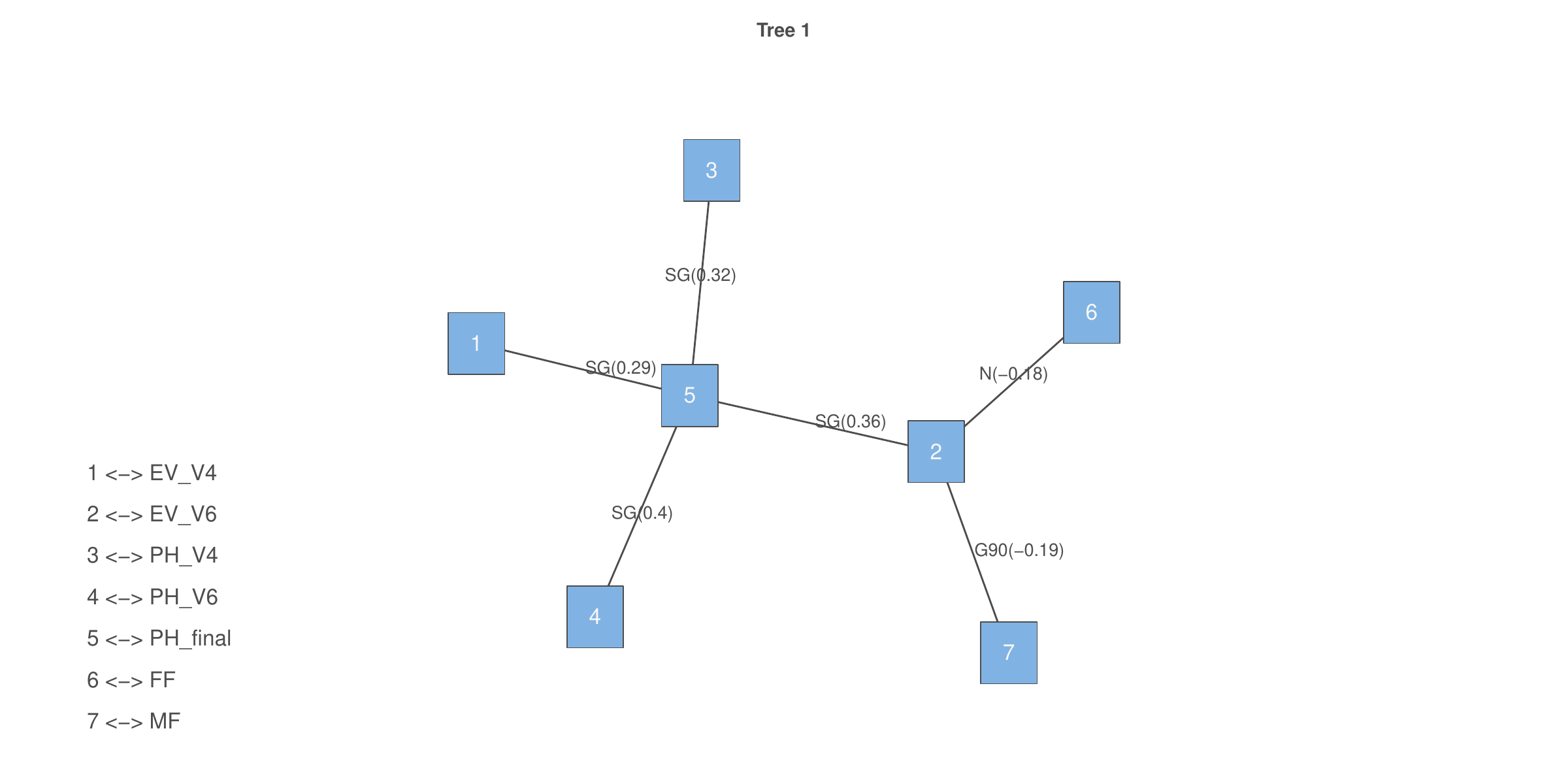}
    \includegraphics[width=1\linewidth]{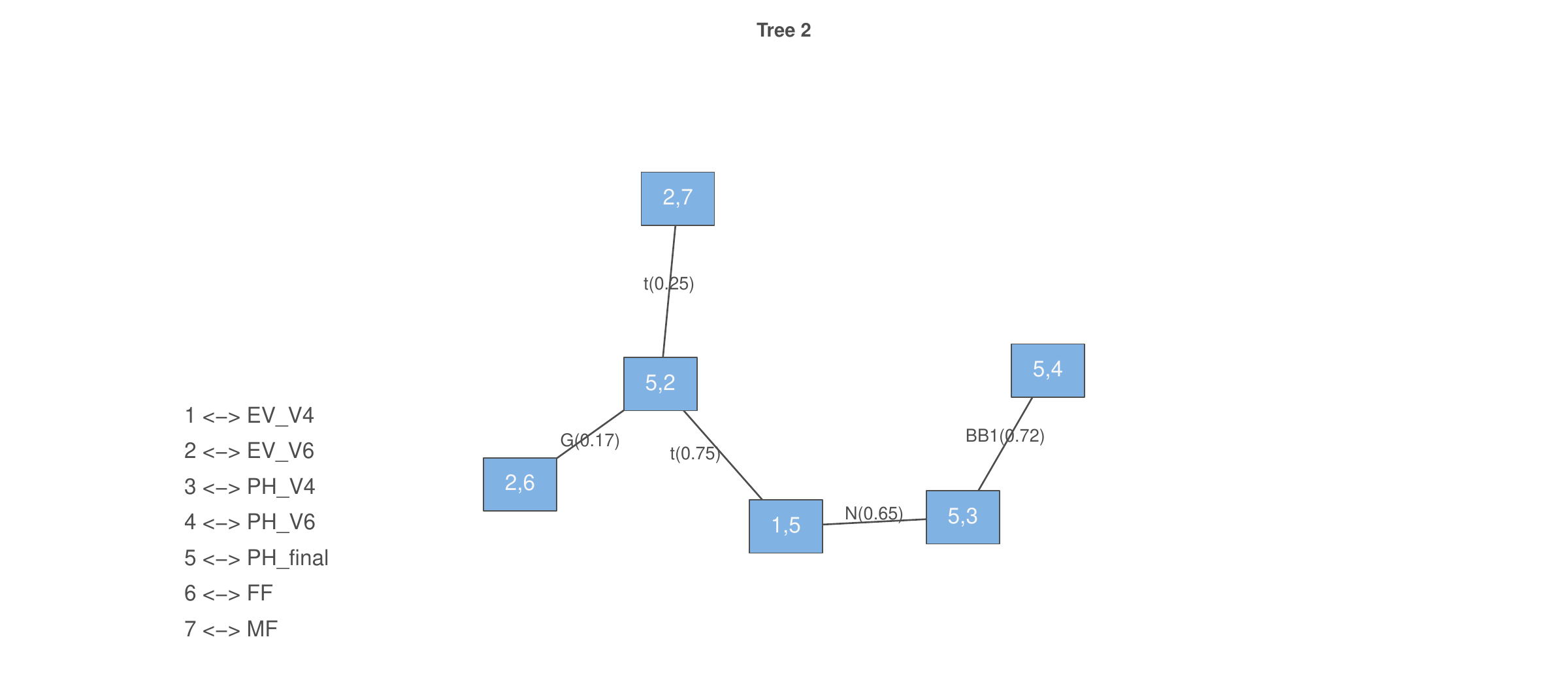}
    \caption{The vine model selected and estimated by maximizing (partial) correlations. A letter at an edge with numbers inside the parenthesis refers to its bivariate copula family with its Kendall's $\hat{\tau}$.}
    \label{fig:RVMMaizeNotLeaf}
\end{figure*}

In the following, we give the details on a heuristic to find the vine tree structure tailored to conditional density estimation using the data in Section \ref{Sec:Data}.
\begin{itemize}
    \item[1.] After converting $trait\in \{\texttt{EV\_V4}, \ldots, \texttt{MF}\}$ to normal scores ($\bm{z}_{trait}$), we estimate the Pearson's correlation of the response with other traits as follows. Due to the highest estimated value,  \texttt{PH\_V6} is connected to \texttt{PH\_final} in the first tree.

\begin{table}[!h]
\resizebox{\linewidth}{!}{%
\begin{tabular}{lrrrrrrr}
  \hline
 & \texttt{EV\_V4}& \texttt{EV\_V6}& \texttt{PH\_V4} & \texttt{PH\_V6} & \texttt{PH\_f.} & \texttt{FF} & \texttt{MF} \\ 
  \hline
 \texttt{EV\_V4}& 1.00 & 0.93 & 0.88 & 0.83 & 0.45 & -0.33 & -0.36 \\ 
\texttt{EV\_V6}& 0.93 & 1.00 & 0.87 & 0.88 & 0.55 & -0.27 & -0.31 \\ 
\texttt{PH\_V4}& 0.88 & 0.87 & 1.00 & 0.93 & 0.50 & -0.22 & -0.30 \\ 
 \texttt{PH\_V6}& 0.83 & 0.88 & 0.93 & 1.00 & 0.60 & -0.16 & -0.22 \\ 
\texttt{PH\_f.}& 0.45 & 0.55 & 0.50 & 0.60 & 1.00 & 0.07 & 0.12 \\ 
 \texttt{FF}& -0.33 & -0.27 & -0.22 & -0.16 & 0.07 & 1.00 & 0.77 \\ 
 \texttt{MF}& -0.36 & -0.31 & -0.30 & -0.22 & 0.12 & 0.77 & 1.00 \\ 
   \hline
\end{tabular}}
\end{table}
 \item[2.] Next, we estimate the partial correlations of the remaining traits with \texttt{PH\_final} given \texttt{PH\_V6} as follows. The highest estimated correlation of the response (without \texttt{PH\_V6}) is 0.55 with \texttt{EV\_V6} which is higher than the highest estimated partial correlation of \texttt{PH\_final} and \texttt{MF} given \texttt{PH\_V6} of 0.32. Thus, we connect \texttt{EV\_V6} to \texttt{PH\_final} in the first tree. As a result, \texttt{PH\_final} is not a leaf node in the first tree (see \autoref{fig:RVMMaizeNotLeaf}).
\begin{table}[ht]
    \centering
    \begin{tabular}{ccccccc}
    \hline
        Est. partial corr.  & \multicolumn{5}{c}{trait}  \\ \hline
         &\texttt{EV\_V4} & \texttt{EV\_V6}&\texttt{PH\_V4} & \texttt{FF}& \texttt{MF}\\ \hline
       $\hat{\rho}_{\texttt{PH\_final}, trait; \texttt{PH\_V6}}$  &0.11& 0.05& 0.20&0.21& 0.32\\\hline
    \end{tabular}
\end{table}
\item[3.] We iterate similarly and estimate the partial correlations of the remaining traits with \texttt{PH\_final} given \texttt{EV\_V6} as follows. The highest estimated correlation of the response is 0.50 with \texttt{PH\_V4} after \texttt{EV\_V6} and \texttt{PH\_V6} are already chosen for the first tree, but the highest estimated partial correlation of \texttt{PH\_final} and \texttt{MF} given \texttt{PH\_V6} is 0.32 and that of \texttt{PH\_final} and \texttt{MF} given \texttt{EV\_V6} is 0.37. Hence, we connect \texttt{PH\_V4} to \texttt{PH\_final} in the first tree. We remark that $\hat{\rho}_{\texttt{PH\_final}, trait; \texttt{PH\_V4}, \texttt{EV\_V6}}$ is not reasonable to consider since \texttt{PH\_final} is not a leaf in the first tree, thus these edges cannot occur in this pair copula construction.
\begin{table}[ht]
    \centering
    \begin{tabular}{cccccc}
    \hline
    Est. partial corr.  & \multicolumn{4}{c}{trait}  \\ \hline
         &\texttt{EV\_V4} &\texttt{PH\_V4} & \texttt{FF}& \texttt{MF}\\ \hline
       $\hat{\rho}_{\texttt{PH\_final}, trait; \texttt{EV\_V6}}$  &0.21& 0.05&0.28& 0.37\\\hline
    \end{tabular}
\end{table}

\item[4.] We iterate similarly and estimate the partial correlations of the remaining traits with \texttt{PH\_final} given \texttt{PH\_V4} as follows. Still, the highest estimated correlation of the response (without \texttt{PH\_V6},  \texttt{EV\_V6}, \texttt{PH\_V4}) is 0.45 with \texttt{EV\_V4} which is higher than the highest estimated partial correlation of \texttt{PH\_final} and \texttt{MF} given \texttt{PH\_V6} of 0.32, that of \texttt{PH\_final} and \texttt{MF} given \texttt{EV\_V6} of 0.37, and  that of \texttt{PH\_final} and \texttt{MF} given \texttt{PH\_V4} of 0.33. Hence, connect \texttt{EV\_V4}  to \texttt{PH\_final} in the first tree.
\begin{table}[ht]
    \centering
    \begin{tabular}{ccccc}
    \hline
    Est. partial corr.  & \multicolumn{3}{c}{trait}  \\ \hline
         &\texttt{EV\_V4}  & \texttt{FF}& \texttt{MF}\\ \hline
       $\hat{\rho}_{\texttt{PH\_final}, trait; \texttt{PH\_V4}}$  &0.03&0.22& 0.33\\\hline
    \end{tabular}
\end{table}

\item[5.] Next, the highest estimated partial correlation of \texttt{PH\_final} and \texttt{MF} given \texttt{EV\_V6} of 0.37 is higher than the highest estimated correlation of the response with \texttt{FF} and \texttt{MF}. Thus, we connect  \texttt{MF} to \texttt{EV\_V6} in the first tree so that \texttt{PH\_final} and \texttt{MF} given \texttt{EV\_V6} is modeled in the second tree. Likewise, \texttt{FF} is connected to \texttt{EV\_V6} in the first tree, resulting in the final structure given in \autoref{fig:RVMMaizeNotLeaf}.  The remaining tree structures are then chosen by the Dissmann algorithm.
\begin{table}[ht]
    \centering
    \begin{tabular}{cccc}
    \hline
        Estimated partial correlation  & \multicolumn{2}{c}{trait}  \\ \hline
          & \texttt{FF}& \texttt{MF}\\ \hline
       $\hat{\rho}_{\texttt{PH\_final}, trait; \texttt{EV\_V4}}$  &0.26& 0.34\\\hline
    \end{tabular}
\end{table}
\end{itemize}

\begin{table*}
\centering
\begin{tabular}{rrrrrrr}
  \hline
tree & conditioned & conditioning & family & rotation & parameters & tau \\ 
  \hline
1 & 5, 4 &  & gumbel   & 180 & 1.7 &  0.396 \\ 
  1 & 3, 4 &  & bb1      &   0 & 1.3, 2.4 &  0.750 \\ 
  1 & 4, 2 &  & gaussian &   0 & 0.89 &  0.696 \\ 
  1 & 2, 1 &  & gumbel   & 180 & 4.4 &  0.773 \\ 
  1 & 1, 7 &  & gaussian &   0 & -0.36 & -0.236 \\ 
  1 & 6, 7 &  & gaussian &   0 & 0.77 &  0.555 \\ 
  2 & 5, 3 & 4 & t        &   0 & -0.22, 8.19 & -0.139 \\ 
  2 & 3, 2 & 4 & gaussian &   0 & 0.26 &  0.169 \\ 
  2 & 4, 1 & 2 & gaussian &   0 & 0.055 &  0.035 \\ 
  2 & 2, 7 & 1 & gumbel   &   0 & 1.1 &  0.062 \\ 
  2 & 1, 6 & 7 & frank    &   0 & -0.63 & -0.070 \\ 
  3 & 5, 2 & 3, 4 & clayton  &   0 & 0.17 &  0.077 \\ 
  3 & 3, 1 & 2, 4 & t        &   0 & 0.49, 8.35 &  0.329 \\ 
  3 & 4, 7 & 1, 2 & frank    &   0 & 0.92 &  0.102 \\ 
  3 & 2, 6 & 7, 1 & clayton  &   0 & 0.056 &  0.027 \\ 
  4 & 5, 1 & 2, 3, 4 & frank    &   0 & -1.3 & -0.138 \\ 
  4 & 3, 7 & 1, 2, 4 & t        &   0 & -0.1, 9.8 & -0.064 \\ 
  4 & 4, 6 & 7, 1, 2 & bb8      &   0 & 1.27, 0.93 &  0.094 \\ 
  5 & 5, 7 & 1, 2, 3, 4 & gumbel   &   0 & 1.2 &  0.190 \\ 
  5 & 3, 6 & 7, 1, 2, 4 & bb8      & 180 & 1.39, 0.81 &  0.085 \\ 
  6 & 5, 6 & 7, 1, 2, 3, 4 & clayton  & 270 & 0.077 & -0.037 \\ 
   \hline
\end{tabular}
\caption{Summary of the seven-dimensional vine model obtained with the Dissmann algorithm. The variables are 1=\texttt{EV\_V4}, 2=\texttt{EV\_V6}, 3=\texttt{PH\_V4}, 4=\texttt{PH\_V6}, 5=\texttt{PH\_final}, 6=\texttt{FF}, 7=\texttt{MF}.}
\label{tab:dissmann}
\end{table*}

\begin{table*}
\centering
\begin{tabular}{rrrrrrr}
  \hline
tree & conditioned & conditioning & family & rotation & parameters & tau \\ 
  \hline
1 & 7, 2 &  & gumbel   & 270 & 1.2 & -0.186 \\ 
  1 & 4, 5 &  & gumbel   & 180 & 1.7 &  0.396 \\ 
  1 & 6, 2 &  & gaussian &   0 & -0.27 & -0.176 \\ 
  1 & 3, 5 &  & gumbel   & 180 & 1.5 &  0.320 \\ 
  1 & 2, 5 &  & gumbel   & 180 & 1.6 &  0.361 \\ 
  1 & 5, 1 &  & gumbel   & 180 & 1.4 &  0.292 \\ 
  2 & 7, 5 & 2 & t        &   0 & 0.38, 13.78 &  0.248 \\ 
  2 & 4, 3 & 5 & bb1      &   0 & 0.63, 2.70 &  0.718 \\ 
  2 & 6, 5 & 2 & gumbel   &   0 & 1.2 &  0.167 \\ 
  2 & 3, 1 & 5 & gaussian &   0 & 0.85 &  0.645 \\ 
  2 & 2, 1 & 5 & t        &   0 & 0.92, 6.57 &  0.752 \\ 
  3 & 7, 6 & 5, 2 & gaussian &   0 & 0.71 &  0.507 \\ 
  3 & 4, 1 & 3, 5 & t        &   0 & 0.12, 7.80 &  0.075 \\ 
  3 & 6, 1 & 5, 2 & gaussian &   0 & -0.16 & -0.105 \\ 
  3 & 3, 2 & 1, 5 & bb7      & 180 & 1.04, 0.19 &  0.106 \\ 
  4 & 7, 1 & 6, 5, 2 & indep    &   0 &  &  0.000 \\ 
  4 & 4, 2 & 1, 3, 5 & t        &   0 & 0.43, 9.90 &  0.282 \\ 
  4 & 6, 3 & 1, 5, 2 & frank    &   0 & 0.72 &  0.080 \\ 
  5 & 7, 3 & 1, 6, 5, 2 & gaussian &   0 & -0.13 & -0.080 \\ 
  5 & 4, 6 & 2, 1, 3, 5 & bb8      &   0 & 1.21, 0.97 &  0.084 \\ 
  6 & 7, 4 & 3, 1, 6, 5, 2 & t        &   0 & 0.057, 10.285 &  0.036 \\ 
   \hline
\end{tabular}
\caption{Summary of the vine selected and estimated by maximizing (partial) correlations. The variables are 1=\texttt{EV\_V4}, 2=\texttt{EV\_V6}, 3=\texttt{PH\_V4}, 4=\texttt{PH\_V6}, 5=\texttt{PH\_final}, 6=\texttt{FF}, 7=\texttt{MF}.}
\label{tab:RVMMaizeNotLeaf}
\end{table*}

\begin{table*}
\centering
\begin{tabular}{|c|c|c|c|c|c|c|c|}
\hline
Model & 1 & 2 & 3 & 4 & 5 & 6& 7\\
  \hline
 \texttt{vinereg} & \texttt{PH\_final} & \texttt{PH\_V6} & \texttt{MF} & \texttt{PH\_V4} & \texttt{EV\_V6} & \texttt{EV\_V4} & \texttt{FF}\\
 \hline
\texttt{sparsevinereg} & \texttt{PH\_final} & \texttt{PH\_V6} & \texttt{MF} & \texttt{EV\_V6} & \texttt{EV\_V4} & \texttt{PH\_V4} & \texttt{FF}\\
   \hline
\end{tabular}
\caption{Order of the D-vine models selected and estimated by \texttt{vinereg} (top) and \texttt{sparsevinereg} (bottom) in Section \ref{Sec:Data}.}
\label{tab:datavinereg}
\end{table*}

\begin{table*}
\centering
\begin{tabular}{rrrrrrr}
  \hline
tree & conditioned & conditioning & family & rotation & parameters & tau \\ 
  \hline
1 & 1, 5 &  & gumbel   & 180 & 1.7 &  0.396 \\ 
  1 & 5, 7 &  & bb7      & 270 & 1.05, 0.27 & -0.141 \\ 
  1 & 7, 4 &  & gaussian &   0 & -0.3 & -0.193 \\ 
  1 & 4, 3 &  & gaussian &   0 & 0.87 &  0.675 \\ 
  1 & 3, 2 &  & gumbel   & 180 & 4.4 &  0.773 \\ 
  1 & 2, 6 &  & gaussian &   0 & -0.33 & -0.215 \\ 
  2 & 1, 7 & 5 & bb1      & 180 & 0.34, 1.09 &  0.218 \\ 
  2 & 5, 4 & 7 & bb1      &   0 & 1.3, 2.4 &  0.751 \\ 
  2 & 7, 3 & 4 & joe      & 270 & 1.2 & -0.085 \\ 
  2 & 4, 2 & 3 & gaussian &   0 & 0.39 &  0.256 \\ 
  2 & 3, 6 & 2 & frank    &   0 & 0.58 &  0.064 \\ 
  3 & 1, 4 & 7, 5 & frank    &   0 & -0.89 & -0.098 \\ 
  3 & 5, 3 & 4, 7 & bb8      &   0 & 3.2, 0.7 &  0.292 \\ 
  3 & 7, 2 & 3, 4 & frank    &   0 & -1.2 & -0.136 \\ 
  3 & 4, 6 & 2, 3 & frank    &   0 & 0.92 &  0.102 \\ 
  4 & 1, 3 & 4, 7, 5 & clayton  &   0 & 0.23 &  0.103 \\ 
  4 & 5, 2 & 3, 4, 7 & bb8      &  90 & 1.62, 0.93 & -0.197 \\ 
  4 & 7, 6 & 2, 3, 4 & gaussian &   0 & 0.73 &  0.523 \\ 
  5 & 1, 2 & 3, 4, 7, 5 & bb8      &  90 & 1.41, 0.88 & -0.112 \\ 
  5 & 5, 6 & 2, 3, 4, 7 & indep    &   0 &  &  0.000 \\ 
  6 & 1, 6 & 2, 3, 4, 7, 5 & clayton  & 270 & 0.077 & -0.037 \\ 
   \hline
\end{tabular}
\caption{Summary of the vine selected and estimated by \texttt{vinereg}. The variables are 1=\texttt{PH\_final}, 2=\texttt{EV\_V4}, 3=\texttt{EV\_V6}, 4=\texttt{PH\_V4}, 5=\texttt{PH\_V6}, 6=\texttt{FF}, 7=\texttt{MF}.}
\label{tab:vinereg}
\end{table*}

\begin{table*}
\centering
\begin{tabular}{rrrrrrr}
  \hline
tree & conditioned & conditioning & family & rotation & parameters & tau \\ 
  \hline
1 & 1, 2 &  & gumbel   & 180 & 1.7 &  0.396 \\ 
  1 & 2, 3 &  & bb7      & 270 & 1.05, 0.27 & -0.141 \\ 
  1 & 3, 4 &  & gumbel   & 270 & 1.2 & -0.186 \\ 
  1 & 4, 5 &  & gumbel   & 180 & 4.4 &  0.773 \\ 
  1 & 5, 6 &  & gaussian &   0 & 0.88 &  0.691 \\ 
  1 & 6, 7 &  & gaussian &   0 & -0.22 & -0.143 \\ 
  2 & 1, 3 & 2 & bb1      & 180 & 0.34, 1.09 &  0.218 \\ 
  2 & 2, 4 & 3 & gaussian &   0 & 0.88 &  0.687 \\ 
  2 & 3, 5 & 4 & frank    &   0 & -1.2 & -0.136 \\ 
  2 & 4, 6 & 5 & gaussian &   0 & 0.24 &  0.151 \\ 
  2 & 5, 7 & 6 & frank    &   0 & -2 & -0.211 \\ 
  3 & 1, 4 & 3, 2 & clayton  &   0 & 0.14 &  0.067 \\ 
  3 & 2, 5 & 4, 3 & gaussian &   0 & 0.081 &  0.052 \\ 
  3 & 3, 6 & 5, 4 & frank    &   0 & 0.44 &  0.049 \\ 
  3 & 4, 7 & 6, 5 & clayton  & 180 & 0.08 &  0.038 \\ 
  4 & 1, 5 & 4, 3, 2 & frank    &   0 & -1.4 & -0.155 \\ 
  4 & 2, 6 & 5, 4, 3 & bb1      & 180 & 0.24, 1.83 &  0.511 \\ 
  4 & 3, 7 & 6, 5, 4 & gaussian &   0 & 0.74 &  0.526 \\ 
  5 & 1, 6 & 5, 4, 3, 2 & gaussian &   0 & -0.11 & -0.071 \\ 
  5 & 2, 7 & 6, 5, 4, 3 & indep    &   0 &  &  0.000 \\ 
  6 & 1, 7 & 6, 5, 4, 3, 2 & clayton  & 270 & 0.077 & -0.037 \\ 
   \hline
\end{tabular}
\caption{Summary of the vine selected and estimated by \texttt{sparsevinereg}. The variables are 1=\texttt{PH\_final}, 2=\texttt{PH\_V6},3=\texttt{MF}, 4=\texttt{EV\_V6}, 5=\texttt{EV\_V4}, 6=\texttt{PH\_V4}, 7=\texttt{FF}.}
\label{tab:sparsevinereg}
\end{table*}

\begin{table*}
\centering
\begin{tabular}{rrrrrrr}
  \hline
tree & conditioned & conditioning & family & rotation & parameters & tau \\ 
  \hline
1 & 1, 2 &  & gumbel   & 180 & 4.4 & 0.773 \\ 
  1 & 2, 4 &  & gaussian &   0 & 0.89 & 0.696 \\ 
  1 & 3, 4 &  & bb1      &   0 & 1.3, 2.4 & 0.750 \\ 
  2 & 1, 4 & 2 & gaussian &   0 & 0.055 & 0.035 \\ 
  2 & 2, 3 & 4 & gaussian &   0 & 0.26 & 0.169 \\ 
  3 & 1, 3 & 4, 2 & t        &   0 & 0.49, 8.35 & 0.329 \\ 
   \hline
\end{tabular}
\caption{Summary of the four-dimensional vine model obtained with the Dissmann algorithm. The variables are 1=\texttt{EV\_V4}, 2=\texttt{EV\_V6}, 3=\texttt{PH\_V4}, 4=\texttt{PH\_V6}.}
\label{tab:FourDimTree}
\end{table*}

\begin{table*}
\centering
\begin{tabular}{rrrrrrr}
  \hline
 & 10\% & 30\% & 50\% & 70\% & 90\% & 95\% \\ 
  \hline 
  quantile\_\texttt{PH\_final}\_small & 0.01 & 0.01 & 0.03 & 0.05 & 0.19 & 0.35 \\ 
  quantile\_\texttt{FF}\_small & 0.29 & 0.55 & 0.73 & 0.86 & 0.96 & 0.98 \\ 
  quantile\_\texttt{MF}\_small & 0.35 & 0.59 & 0.75 & 0.87 & 0.96 & 0.98 \\ 
  \hline
  quantile\_\texttt{PH\_final}\_large & 0.46 & 0.68 & 0.81 & 0.90 & 0.97 & 0.99 \\ 
  quantile\_\texttt{FF}\_large & 0.04 & 0.15 & 0.28 & 0.46 & 0.71 & 0.81 \\ 
  quantile\_\texttt{MF}\_large & 0.04 & 0.13 & 0.25 & 0.40 & 0.65 & 0.75 \\ 
   \hline
\end{tabular}
\caption{Empirical univariate conditional quantiles of  \texttt{FF}, \texttt{MF}, and \texttt{PH\_final}, respectively on the copula scale for different stress scenarios (small with the four conditioning values set to 0.01, large with that of 0.99) based on HMC samples using the vine copula model \autoref{tab:dissmann}.}
\label{tab:stressuquant}
\end{table*}

\end{appendices}

\clearpage
\bibliographystyle{apalike} 
\bibliography{References_New}

\begin{thebibliography}{}

\bibitem[Aas et~al., 2009]{Aas2009}
Aas, K., Czado, C., Frigessi, A., and Bakken, H. (2009).
\newblock {Pair-Copula Constructions of Multiple Dependence}.
\newblock {\em Insurance: Mathematics and Economics}, 44(2):182--198.

\bibitem[Acar et~al., 2012]{acar2012beyond}
Acar, E.~F., Genest, C., and Ne{\v{s}}lehov{\'a}, J. (2012).
\newblock {Beyond Simplified Pair-Copula Constructions}.
\newblock {\em Journal of Multivariate Analysis}, 110:74--90.

\bibitem[Bedford and Cooke, 2001]{bedfordProbabilityDensityDecomposition2001}
Bedford, T. and Cooke, R.~M. (2001).
\newblock Probability {{Density Decomposition}} for {{Conditionally Dependent Random Variables Modeled}} by {{Vines}}.
\newblock {\em Annals of Mathematics and Artificial Intelligence}, 32(1):245--268.

\bibitem[Betancourt, 2018]{betancourtConceptualIntroductionHamiltonian2018}
Betancourt, M. (2018).
\newblock A {{Conceptual Introduction}} to {{Hamiltonian Monte Carlo}}.
\newblock {\em arXiv:1701.02434 [stat]}.

\bibitem[Brechmann et~al., 2013]{brechmannConditionalCopulaSimulation2013}
Brechmann, E.~C., Hendrich, K., and Czado, C. (2013).
\newblock Conditional {{Copula Simulation}} for {{Systemic Risk Stress Testing}}.
\newblock {\em Insurance / Mathematics \& economics}, 53(3).

\bibitem[Carpenter et~al., 2017]{carpenterStanProbabilisticProgramming2017}
Carpenter, B., Gelman, A., Hoffman, M.~D., Lee, D., Goodrich, B., Betancourt, M., Brubaker, M., Guo, J., Li, P., and Riddell, A. (2017).
\newblock Stan: {{A Probabilistic Programming Language}}.
\newblock {\em Journal of Statistical Software}, 76:1--32.

\bibitem[Chang and Joe, 2019]{chang2019prediction}
Chang, B. and Joe, H. (2019).
\newblock {Prediction Based on Conditional Fistributions of Vine Copulas}.
\newblock {\em Computational Statistics \& Data Analysis}, 139:45--63.

\bibitem[Czado, 2019]{czadoAnalyzingDependentData2019}
Czado, C. (2019).
\newblock {\em Analyzing {{Dependent Data}} with {{Vine Copulas}}: {{A Practical Guide With R}}}.
\newblock Lecture {{Notes}} in {{Statistics}}. {Springer International Publishing}.

\bibitem[Czado and Nagler, 2022]{czado2022vine}
Czado, C. and Nagler, T. (2022).
\newblock {Vine Copula Based Modeling}.
\newblock {\em Annual Review of Statistics and Its Application}, 9:453--477.

\bibitem[Daul et~al., 2003]{daul_grouped_2003}
Daul, S., De~Giorgi, E.~G., Lindskog, F., and {McNeil}, A. (2003).
\newblock {The Grouped t-Copula with an Application to Credit Risk}.
\newblock {\em Available at {SSRN} 1358956}.

\bibitem[Derumigny and Fermanian, 2017]{derumignyTestsSimplifyingAssumption2017}
Derumigny, A. and Fermanian, J.-D. (2017).
\newblock About {{Tests}} of the ``{{Simplifying}}'' {{Assumption}} for {{Conditional Copulas}}.
\newblock {\em Dependence Modeling}, 5(1):154--197.

\bibitem[Di{\ss}mann et~al., 2013]{Dissmann2013}
Di{\ss}mann, J., Brechmann, E.~C., Czado, C., and Kurowicka, D. (2013).
\newblock {Selecting and Estimating Regular Vine Copulae and Application to Financial Returns}.
\newblock {\em Computational Statistics and Data Analysis}, 59:52--69.

\bibitem[Gelman et~al., 2013]{gelmanBayesianDataAnalysis2013}
Gelman, A., Carlin, J.~B., Stern, H.~S., Dunson, D.~B., Vehtari, A., and Rubin, D.~B. (2013).
\newblock {\em Bayesian {{Data Analysis}}}.
\newblock {Chapman and Hall/CRC}, {New York}, 3 edition.

\bibitem[Hastie et~al., 2009]{hastie2009elements}
Hastie, T., Tibshirani, R., and Friedman, J.~H. (2009).
\newblock {\em {The Elements of Statistical Learning: Data Mining, Inference, and Prediction}}, volume~2.
\newblock Springer, New York, NY.

\bibitem[Havl{\'i}ckov{\'a}, 2022]{havlickovaAnalysisConditionalVine2022}
Havl{\'i}ckov{\'a}, P. (2022).
\newblock Analysis of {{Conditional Vine Copula Distributions Using Hamiltonian Monte Carlo}}.
\newblock Master's thesis, Technische Universitaet Muenchen.

\bibitem[Herrmann, 2018]{herrmann2018regular}
Herrmann, J. (2018).
\newblock {Regular Vine Copula Based Quantile Regression}.
\newblock Master's thesis, Technical University of Munich.

\bibitem[Hoffman et~al., 2014]{hoffmanNoUTurnSamplerAdaptively2014}
Hoffman, M.~D., Gelman, A., et~al. (2014).
\newblock The {{No-U-Turn Sampler}}: {{Adaptively Setting Path Lengths}} in {{Hamiltonian Monte Carlo}}.
\newblock {\em J. Mach. Learn. Res.}, 15(1):1593--1623.

\bibitem[H{\"o}lker et~al., 2019]{holker2019european}
H{\"o}lker, A.~C., Mayer, M., Presterl, T., Bolduan, T., Bauer, E., Ordas, B., Brauner, P.~C., Ouzunova, M., Melchinger, A.~E., and Sch{\"o}n, C.-C. (2019).
\newblock {European Maize Landraces made Accessible for Plant Breeding and Genome-Based Studies}.
\newblock {\em Theoretical and Applied Genetics}, 132(12):3333--3345.

\bibitem[Hollander et~al., 2013]{hollanderNonparametricStatisticalMethods2013}
Hollander, M., Wolfe, D.~A., and Chicken, E. (2013).
\newblock Nonparametric {{Statistical Methods}}, 3rd {{Edition}} | {{Wiley}}.

\bibitem[Joe, 1993]{joe_parametric_1993}
Joe, H. (1993).
\newblock {Parametric Families of Multivariate Distributions with Given Margins}.
\newblock {\em Journal of Multivariate Analysis}, 46(2):262--282.

\bibitem[Joe, 2014]{joeDependenceModelingCopulas2014}
Joe, H. (2014).
\newblock {\em Dependence {{Modeling}} with {{Copulas}}}.
\newblock {CRC press}.

\bibitem[K{\"a}hm, 2014]{kahmAssessingSystemRelevance2014}
K{\"a}hm, O. (2014).
\newblock Assessing {{System Relevance}} of {{Financial Institutions Using Pair-Copula Constructions}} for {{Modeling}}.

\bibitem[Kendall, 1938]{kendall1938new}
Kendall, M.~G. (1938).
\newblock {A New Measure of Rank Correlation}.
\newblock {\em Biometrika}, 30(1-2):81--93.

\bibitem[Klein, 2024]{klein2024distributional}
Klein, N. (2024).
\newblock {Distributional Regression for Data Analysis}.
\newblock {\em Annual Review of Statistics and Its Application}, 11.

\bibitem[Kraus and Czado, 2017]{krausDVineCopulaBased2017}
Kraus, D. and Czado, C. (2017).
\newblock {D-Vine Copula Based Quantile Regression}.
\newblock {\em Computational Statistics \& Data Analysis}, 110:1--18.

\bibitem[Kurz and Spanhel, 2022]{kurz2022testing}
Kurz, M.~S. and Spanhel, F. (2022).
\newblock {Testing the Simplifying Assumption in High-Dimensional Vine Copulas}.
\newblock {\em Electronic Journal of Statistics}, 16(2):5226--5276.

\bibitem[McNeil, 2008]{mcneil_sampling_2008}
McNeil, A.~J. (2008).
\newblock {Sampling Nested Archimedean Copulas}.
\newblock {\em Journal of Statistical Computation and Simulation}, 78(6):567--581.

\bibitem[Mroz et~al., 2021]{mroz2021simplifying}
Mroz, T., Fuchs, S., and Trutschnig, W. (2021).
\newblock {How Simplifying and Flexible is the Simplifying Assumption in Pair-Copula Constructions--Analytic Answers in Dimension Three and a Glimpse Beyond}.
\newblock {\em Electron. J. Stat.}, 15:1951--1992.

\bibitem[Nagler, 2022]{nagler2018vinereg}
Nagler, T. (2022).
\newblock {\em {vinereg: D-Vine Quantile Regression}}.
\newblock R package version 0.8.1.

\bibitem[Nagler, 2024]{nagler2024simplified}
Nagler, T. (2024).
\newblock {Simplified Vine Cpula Models: State of Science and Affairs}.
\newblock {\em arXiv preprint arXiv:2410.16806}.

\bibitem[Nagler et~al., 2017]{naglerNonparametricEstimationSimplified2017}
Nagler, T., Schellhase, C., and Czado, C. (2017).
\newblock Nonparametric {{Estimation}} of {{Simplified Vine Copula Models}}: {{Comparison}} of {{Methods}}.
\newblock {\em Dependence Modeling}, 5(1):99--120.

\bibitem[Nagler et~al., 2023]{naglerVineCopulaStatisticalInference2023}
Nagler, T., Schepsmeier, U., Stoeber, J., Brechmann, E.~C., Graeler, B., Erhardt, T., Almeida, C., Min, A., Czado, C., Hofmann, M., Killiches, M., Joe, H., and Vatter, T. (2023).
\newblock {\em {{VineCopula}}: {{Statistical Inference}} of {{Vine Copulas}}}.
\newblock R package version 2.6.1.

\bibitem[Nagler and Vatter, 2022a]{kde1dpackage}
Nagler, T. and Vatter, T. (2022a).
\newblock {\em kde1d: Univariate Kernel Density Estimation}.
\newblock R package version 1.0.4.

\bibitem[Nagler and Vatter, 2022b]{nagler2022rvinecopulib}
Nagler, T. and Vatter, T. (2022b).
\newblock {\em rvinecopulib: High Performance Algorithms for Vine Copula Modeling}.
\newblock R package version 0.7.2.1.0.

\bibitem[Neal et~al., 2011]{nealMcmcUsingHamiltonian2012}
Neal, R.~M. et~al. (2011).
\newblock {MCMC using Hamiltonian Dynamics}.
\newblock {\em {Handbook of Markov Chain Monte Carlo}}, 2(11):2.

\bibitem[Nesterov, 2009]{nesterovPrimaldualSubgradientMethods2009}
Nesterov, Y. (2009).
\newblock {Primal-Dual Subgradient Methods for Convex Problems}.
\newblock {\em Math. Program.}, 120(1):221--259.

\bibitem[Nleya et~al., 2016]{nleya2016corn}
Nleya, T., Chungu, C., and Kleinjan, J. (2016).
\newblock {Corn Growth and Development}.
\newblock {\em Grow Corn Best Manag. Pract}.

\bibitem[Oehme et~al., 2022]{agronomy12040958}
Oehme, L.~H., Reineke, A.-J., Weiß, T.~M., Würschum, T., He, X., and Müller, J. (2022).
\newblock {Remote Sensing of Maize Plant Height at Different Growth Stages Using UAV-Based Digital Surface Models (DSM)}.
\newblock {\em Agronomy}, 12(4).

\bibitem[Rosenblatt, 1956]{rosenblattRemarksNonparametricEstimates1956}
Rosenblatt, M. (1956).
\newblock Remarks on {{Some Nonparametric Estimates}} of a {{Density Function}}.
\newblock {\em Ann. Math. Statist.}, 27(3):832--837.

\bibitem[Sahin, 2023]{dissertationSahin}
Sahin, {\"O}. (2023).
\newblock {\em {Statistical Learning Based on Vine Copulas with Societal Applications}}.
\newblock PhD thesis, Technische Universitaet Muenchen.

\bibitem[Sahin and Czado, 2024]{sahin2022high}
Sahin, {\"O}. and Czado, C. (2024).
\newblock {High-Dimensional Sparse Vine Copula Regression with Application to Genomic Prediction}.
\newblock {\em Biometrics}.

\bibitem[Schellhase and Spanhel, 2018]{schellhase2018estimating}
Schellhase, C. and Spanhel, F. (2018).
\newblock {Estimating Non-Simplified Vine Copulas using Penalized Splines}.
\newblock {\em Statistics and Computing}, 28:387--409.

\bibitem[Silva et~al., 2022]{silva2022grain}
Silva, P.~C., S{\'a}nchez, A.~C., Opazo, M.~A., Mardones, L.~A., and Acevedo, E.~A. (2022).
\newblock {Grain Yield, Anthesis-Silking Interval, and Phenotypic Plasticity in Response to Changing Environments: Evaluation in Temperate Maize Hybrids}.
\newblock {\em Field Crops Research}, 285:108583.

\bibitem[Sklar, 1959]{sklarFonctionsRepartitionDimensions1959}
Sklar, A. (1959).
\newblock {Fonctions de R\'epartition \`a n Dimensions et Leurs Marges}.
\newblock {\em Publications de l'Institut de Statistique de l'Universit\'e de Paris}, (8):229--231.

\bibitem[StanDevelopmentTeam, 2023]{standevelopmentteamStanModelingLanguage2023}
StanDevelopmentTeam (2023).
\newblock Stan {{Modeling Language Users Guide}} and {{Reference Manual}}, 2.32. {{https://mc-stan.org}}.

\bibitem[Steinwart and Christmann, 2011]{steinwart2011estimating}
Steinwart, I. and Christmann, A. (2011).
\newblock {Estimating Conditional Quantiles with the Help of the Pinball Loss}.
\newblock {\em Bernoulli}, 17(1):211--225.

\bibitem[Stoeber et~al., 2013]{stoeber2013simplified}
Stoeber, J., Joe, H., and Czado, C. (2013).
\newblock {Simplified Pair Copula Constructions—Limitations and Extensions}.
\newblock {\em Journal of Multivariate Analysis}, 119:101--118.

\bibitem[Tepegjozova and Czado, 2022]{tepegjozovaBivariateVineCopula2023}
Tepegjozova, M. and Czado, C. (2022).
\newblock {Bivariate Vine Copula Based Regression, Bivariate Level and Quantile Curves}.
\newblock {\em arXiv preprint arXiv:2205.02557}.

\bibitem[Tepegjozova et~al., 2022]{tepegjozova2022nonparametric}
Tepegjozova, M., Zhou, J., Claeskens, G., and Czado, C. (2022).
\newblock {Nonparametric C-and D-Vine-Based Quantile Regression}.
\newblock {\em Dependence Modeling}, 10(1):1--21.

\bibitem[Thomas and Tu, 2021]{thomasLearningHamiltonianMonte2020}
Thomas, S. and Tu, W. (2021).
\newblock Learning {{Hamiltonian Monte Carlo}} in {{R}}.
\newblock {\em The American Statistician}, 75(4):403--413.

\bibitem[Vatter and Chavez-Demoulin, 2015]{vatter2015generalized}
Vatter, T. and Chavez-Demoulin, V. (2015).
\newblock {Generalized Additive Models for Conditional Dependence Structures}.
\newblock {\em Journal of Multivariate Analysis}, 141:147--167.

\bibitem[Vatter and Nagler, 2018]{vatter2018generalized}
Vatter, T. and Nagler, T. (2018).
\newblock {Generalized Additive Models for Pair-Copula Constructions}.
\newblock {\em Journal of Computational and Graphical Statistics}, 27(4):715--727.

\bibitem[Vehtari et~al., 2021]{vehtariRankNormalizationFoldingLocalization2021}
Vehtari, A., Gelman, A., Simpson, D., Carpenter, B., and B{\"u}rkner, P.-C. (2021).
\newblock Rank-{{Normalization}}, {{Folding}}, and {{Localization}}: {{An Improved R\textasciicircum}} for {{Assessing Convergence}} of {{MCMC}} (with {{Discussion}}).
\newblock {\em Bayesian Analysis}, 16(2):667--718.

\bibitem[Zhu et~al., 2021]{zhu2021simplified}
Zhu, K., Kurowicka, D., and Nane, G.~F. (2021).
\newblock {Simplified R-vine Based Forward Regression}.
\newblock {\em Computational Statistics \& Data Analysis}, 155:107091.

\end{thebibliography}

\end{document}